\documentclass[twoside,12pt]{report} 
\usepackage{epsf,graphicx,amsmath,epsfig,latexsym,amssymb,axodraw,color,cite,a4wide,amsfonts,amsbsy,fancyhdr,setspace} 

\usepackage{vmargin} 
\setpapersize{A4} 
\setmargins{32mm}{20mm}{154mm}{240mm}{12pt}{20pt}{12pt}{36pt}

\def\ltsim{\lower3pt\hbox{$\, \buildrel < \over \sim \, $}}  
\def\gtsim{\lower3pt\hbox{$\, \buildrel > \over \sim \, $}}  
\def\ga{\mathrel{\raise.3ex\hbox{$>$\kern-.75em\lower1ex\hbox{$\sim$}}}}  
\def\la{\mathrel{\raise.3ex\hbox{$<$\kern-.75em\lower1ex\hbox{$\sim$}}}}  
\newcommand{\sect}[1]{\section{#1}\setcounter{equation}{0}}  
\def\thesection{\arabic{section}}  
  
\newcommand{\de}{\partial}  
\def\bo{ { \sqcup\llap{ $\sqcap$} } }  
\newcommand{\no}{\noindent} 
\begin{document}

\newpage 
\thispagestyle{empty} 


\vspace*{1cm} 
\begin{center} 
{\Huge\bf Brane-World Multigravity\\} 
\vspace{2cm} 
{\Large\bf  
Antonios  Papazoglou\\} 
{\large Keble College\\ 
\vspace{1.5cm} 
Theoretical Physics\\
Department of Physics\\  
University of Oxford} 
\end{center} 
 
\vspace{2cm} 

\begin{center} 
{\Large\font\oxcrest=oxcrest40 \oxcrest\char'01}~~~~~~~~~~~~~~\epsfig{file=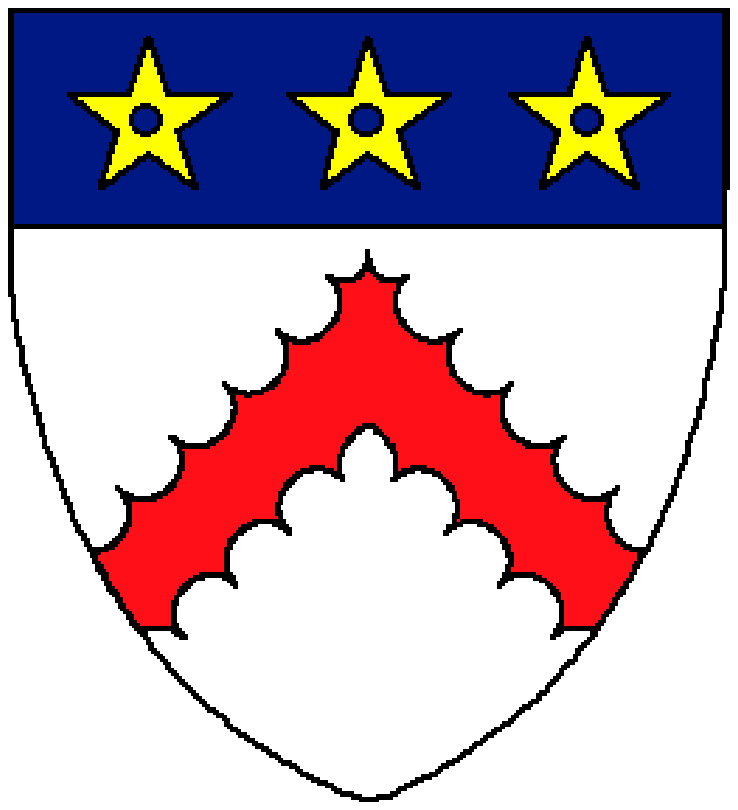,width=1.3cm}
\end{center} 
\vspace{4cm} 
\begin{center} 
\large   
Thesis submitted for the degree of Doctor of Philosophy\\ 
in the University of Oxford\\ 
\vspace{0.3cm} 
$\cdot$ September 2001 $\cdot$
\end{center} 
 
\pagebreak 
\thispagestyle{empty} 
\begin{center} 
{\Large\bf Brane-World Multigravity\\}
\vspace{0.5cm} 
{ {\bf
Antonios  Papazoglou}\\ 
Keble College\\ 
\vspace{0.5cm} 
Theoretical Physics\\
Department of Physics\\  
University of Oxford} 
\vspace{1cm} 
\singlespacing 
\bf Abstract 
\vspace{0.5cm} 
\end{center}          
\setlength{\parindent}{0mm} 
\setlength{\parskip}{0.5\baselineskip} 

In this thesis, we discuss the idea of multigravity, namely the
possibility that a significant component of gravity that we feel at
intermediate distances ($1{\rm mm}<r<10^{26}{\rm cm}$) is due to
massive but ultralight gravitons. We demonstrate how this phenomenon can be realized
in brane-world models in a spacetime with more than four
dimensions and discuss how modifications of gravity at cosmological
scales emerge as a consequence. Firstly, we consider five dimensional multigravity models with flat 
branes. We see how the existence of freely moving negative tension
branes gives rise to ultralight graviton Kaluza-Klein
states. Secondly, we study the
moduli corresponding to the position of the branes and the size of the
extra dimension, the radions and the dilaton respectively. We show that the radions corresponding to
negative tension branes have wrong sign kinetic term. We also derive a 
stabilization condition for the dilaton in a brane model with
general bulk scalar 
field dynamics. Thirdly, we show how we can reproduce the  five
dimensional multigravity models with Anti-de Sitter branes of positive 
tension only. Moreover, we demonstrate how the van Dam-Veltman-Zakharov discontinuity can be circumvented at tree level in Anti-de
Sitter space. Finally, we discuss how the above multigravity models
can be replicated in six dimensions with flat positive tension
branes. This last possibility is the first theoretically and phenomenologically
viable realization of the multigravity scenario.

\setlength{\parindent}{6mm} 
\setlength{\parskip}{0\baselineskip}

\vspace{1cm} 
\begin{center} 
Thesis submitted for the Degree of Doctor of Philosophy \\
in the University of Oxford\\ 
\vspace{0.3cm} 
$\cdot$ September 2001 $\cdot$ 
\end{center} 

\pagebreak

\chapter*{} 
\thispagestyle{empty} 
~
\vspace{4cm} 

\begin{flushright}
{\Large {\textit{
To my parents, Eleni and Yiannis.}}}
\end{flushright}

\pagebreak 

\chapter*{} 
\thispagestyle{empty} 
~
\vspace{8cm} 

{\textit{
``Leave the beaten track behind occasionally and dive into the woods. Every
time you do you will be certain to find something that you have never seen
before. Follow it up, explore all around  it, and before you know it, you
will have something worth thinking about to occupy your mind. All really big
discoveries are the result of thought.''\\
\begin{flushright}
Alexander Graham Bell
\end{flushright}}}

\pagebreak

\doublespacing

\pagenumbering{roman} 
\setcounter{page}{1} \pagestyle{plain} 
 
\tableofcontents 
\pagebreak 

\singlespacing 

\chapter*{} 
~
\vspace{4cm}

\begin{center} 
\singlespacing 
{\bf Publications} 
\vspace{0.5cm} 
\end{center} 
The second chapter is  based on works published in Nucl.\ Phys.\ B {\bf
584}, 313 (2000) and in Nucl.\ Phys.\ B {\bf 595}, 225 (2001). The
third chapter has been published in Phys.\ Lett.\ B {\bf 505},
231 (2001) and in Nucl.\ Phys.\ B {\bf 625}, 179 (2002). The fourth chapter is based on works published in Phys.\
Lett.\ B {\bf 501}, 140 (2001) and in Phys.\ Lett.\ B {\bf 503}, 173
(2001). Finally, the contents of the fifth chapter are published in Phys.\ Rev. \ D {\bf 64}, 124014 (2001).

\pagebreak

~
\vskip5mm

\begin{center} 
\Large \bf {Acknowledgements} 
\end{center} 
\vspace{0.5cm}

First and foremost, I would like to express my deepest gratitude to
the people that have guided me in the paths of Theoretical Physics the 
past three years. I am indebted to my supervisor, collaborator and
friend Ian I. Kogan for our countless stimulating discussions, his
continuous encouragement and his limitless enthusiasm about physics
that inspired me during my stay in Oxford. I am equally grateful to
Graham G. Ross for giving me the privilege to collaborate with him the 
past two years and for generously sharing with me time and ideas that
have stamped my short academic life. Without any doubt I was more than
fortunate to have such teachers. 

Secondly, I would like to thank all the people with whom I have had
rewarding collaborations and discussions. In particular, I thank Thibault Damour for very illuminating  discussions and for crucial suggestions on the manuscript of this thesis. I would like to thank Stavros Mouslopoulos for collaborating
the past years and for our numerous joint publications. I am also
grateful to Luigi Pilo and Jos\'e Santiago for our common work and for very
fruitful discussions.  Moreover, I would like to thank Panagiota Kanti for
valuable correspondence during the past six months. Furthermore, I
want to thank all the people from Theoretical Physics for our endless
conversations about physics and not only, especially 
my colleagues in my offices 1.12 and 6.1.

On the practical side of the things, I would like to thank the
Hellenic State Scholarship Foundation (IKY) for supporting my studies.
I am also grateful to the Department of Theoretical Physics and to
Keble College for the travel grands that made it possible for me to
participate to various schools, conferences and workshops the last
three years. I also would like to thank the coordinators of the ``Beyond 
the Standard Model'' and ``String Phenomenology'' UK networks for
giving me the opportunity to attend and talk in the meetings they organized.

During these three years in Oxford I have met a lot of good friends that
have made my stay particularly enjoyable. Thanks a lot Alejandro,
Alex, Austin, Bayram, Dan, Eleni, Emiliano, Eric, Eva, Fabien, Geza, Katerina,
Liliana, Marianna, M\'ario, Mark, Martin, Martin, Michael,
Nuno, Peter, Peter, Ramon, Riccardo, Robert and Shinsuke.

At last but not least, I am grateful to my parents Eleni and Yiannis,
my sister Maria and my brother-in-law Tassos for their continuous
encouragement, support and above all their love.

\pagebreak

\pagestyle{fancy}

\newpage

\pagenumbering{arabic} 
\setcounter{page}{1} \pagestyle{fancy} 
\renewcommand{\chaptermark}[1]{\markboth{\chaptername%
\ \thechapter:\,\ #1}{}} 
\renewcommand{\sectionmark}[1]{\markright{\thesection\,\ #1}}

\addtolength{\headheight}{3pt} 
\fancyhead{} 
\fancyhead[RE]{\sl\leftmark} 
\fancyhead[RO,LE]{\rm\thepage} 
\fancyhead[LO]{\sl\rightmark} 
\fancyfoot[C,L,E]{}

\singlespacing

\chapter{Introduction}

There has been a lot of attention over the past few years in
theories where the Standard Model (SM) fields are confined on a three
dimensional dynamical hypersurface, a three-brane, in a higher than
four dimensional spacetime. This scenario is widely known as the
brane-world conjecture. This idea dates back to the early eighties
when  independently Akama \cite{Akama:1982jy} and Rubakov and Shaposhnikov \cite{Rubakov:1983bb} suggested
models of our Universe as a topological defect and later Visser
\cite{Visser:1985qm} and Squires \cite{Squires:1986aq} described how
particles can be  gravitationally trapped on the brane. This
conjecture has been subject of renewed interest in the recent years
with the realization that such structures are common in string theory
where the presence of D-branes \cite{Polchinski:1996na} where open stings end is ubiquitous.

We can generally divide the recent brane-world models into two
classes. The first one pioneered by Antoniadis, Arkani-Hamed,
Dimopoulos and Dvali
\cite{Arkani-Hamed:1998rs,Antoniadis:1998ig,Arkani-Hamed:1999nn}
assumes that we live on a 3-brane in a $(3+1+n)$-dimensional spacetime with factorizable 
geometry along the extra dimensions. The higher dimensional Planck scale $M$ is then related to the
4D Planck scale by $M_{\rm Pl}^2=M^{n+2}V_n$ where $V_n$ is the
compactification volume. Taking the size of
the new $n$ dimensions to be sufficiently large and  identifying the
$(4+n)$-Planck scale with the
 electroweak scale (TeV), a hierarchy between the electroweak  and
the Planck scale is generated. This gives an alternative to
technicolour or low energy supersymmetry resolution to 
the hierarchy problem. Experimental and
astrophysical constraints demand that $n\geq2$, $M\gtsim 30\rm TeV$ and allow 
new dimensions
even of submillimeter size.  However, a new hierarchy must now be
explained, namely the ratio of the large compactification radius to the 
electroweak
scale. Even for six extra dimensions this must be greater than
$10^5$. 

The second class of models which attracted equally considerable
attention were the ones involving non-factorizable geometry  along the
extra dimensions and have quite distinct phenomenology from the
factorizable geometry case. These models come under the name of warped
brane  constructions and  are  motivated by heterotic M-theory 
\cite{Horava:1996qa,Witten:1996mz,Horava:1996ma} and its five 
dimensional reduction \cite{Lukas:1999yy,Lukas:1999qs}. In the context  
of these models, one can localize gravity on the brane world 
having four dimensional gravity even with one extra dimension of 
infinite extent \cite{Gogberashvili:1999tb,Randall:1999vf} (RS2), or can 
generate an exponential scale hierarchy with a compact two brane model
of Planck scale volume as it 
was done in the Randall-Sundrum (RS1) model \cite{Randall:1999ee},  
providing a novel geometrical resolution of the  hierarchy 
problem. The scale hierarchy is generated by an exponential with
respect to the extra dimension  warp factor multiplying the Minkowski
metric. We will briefly review in more details this model at the end of this
introduction since we will concentrate on warped brane-world models.

\begin{figure}
\begin{center}
\begin{picture}(320,80)(0,0)
\SetScale{1.4}
\SetOffset(-60,-40)
\LongArrow(0,55)(300,55)

\Text(420,95)[l]{$r$}
\Text(-5,95)[l]{$0$}
\Vertex(0,55){2}

\Text(45,97)[l]{$0.2 {\rm mm}$}

\Text(342,97)[l]{$10^{26} {\rm cm}$}

\SetScale{2}
\C2Text(105,40){Blue}{Red}{Region where Newton's Law}{has been tested}
\SetScale{1.4}
\Vertex(59,55){2}
\Vertex(241,55){2}

\end{picture}
\end{center}
\vskip-5mm
\caption{Our current limit of knowledge of gravity.}
\label{introf}
\end{figure}
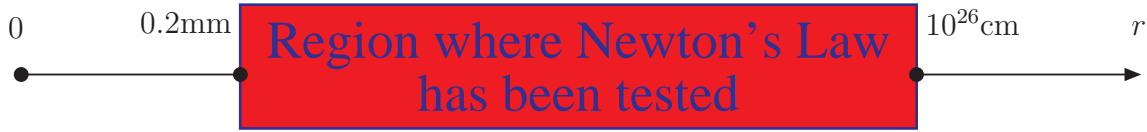

In this thesis we will
discuss another interesting possibility which 
arises from warped brane-world constructions, the scenario of
multigravity. The proposal of multigravity is that the gravitational
interaction that we feel in our Universe is the result of the
combination of massless and massive gravitons or even of massive
gravitons only. The extra dimensional models give us an infinite
tower of massive Kaluza-Klein (KK) gravitons after the process of dimensional
reduction. However, one should keep in mind that  Newton's inverse
square gravitational law is very well tested for distances that range
from the submillimeter regime up to at  most $10^{26}$cm  that
corresponds to $1\%$ of the size of the observable Universe (see Fig.\ref{introf}). Therefore,
the massive KK gravitons that will give rise to multigravity should
have Compton wavelengths bigger than  $10^{26}$cm. Additionally, the
remaining of the KK tower should either have wavelengths  entirely
below the submillimeter scale or if they span the whole region they
should  have very suppressed couplings. It is more than obvious that
these requirements are non trivial and cannot be realized in simple KK
models since the equidistant spectrum that arises in the
typical case will be in contradiction with observations. The aim of
this thesis is to demonstrate how one could take such patterns in
warped  brane-world models. The interesting feature of these
multigravity models is that gravity is modified at large scales where
one starts to be sensitive to the graviton's mass. We should note at
this point that this is the first time that modification of gravity at cosmological scales is
triggered by microphysics considerations which have to do with the
structure of the extra dimensional space of usually Planck scale
volume.  These scenarios
are in principle testable by  the Cosmic Microwave Background (CMB) power
spectrum \cite{Binetruy:2000xv,Bastero-Gil:2001rv}, gravitational lensing
\cite{Uzan:2000mz} and  gravitational wave \cite{Will:1997bb}
observations.

We will firstly present multigravity models which involve only flat
branes in five dimensions. We will begin our presentation by the $''+-+''$ bigravity model \cite{Kogan:2000wc} where, apart
from the graviton zero mode, an anomalously light graviton KK state is
present. This gives the possibility that gravity at intermediate scales 
($1{\rm{mm}}<r<10^{26}{\rm{cm}}$) is mediated by \textit{both} the massless graviton
\textit{and} the ultralight state, leading to the terminology
``bigravity''. At large scales were the Yukawa suppression of the massive 
state turns on, the gravitational constant decreases to a value which
for example in the symmetric configuration is  half of its value at intermediate scales. Similar constructions of freely moving negative tension branes can also
give qualitatively different modifications of gravity at large
scales. We will briefly review the quasi-localized gravity model of Gregory, Rubakov and
Sibiryakov (GRS) \cite{Gregory:2000jc} which has no normalizable zero mode but gravity at intermediate
distances is generated by a ``resonance''-like
\cite{Csaki:2000pp,Dvali:2000rv} coupling of the lowest
part of the continuum of the KK spectrum. In this construction, the
gravitational law at large scales changes from four dimensional to
five dimensional. Although the bigravity and the quasi-localized
gravity models look quite different, they are connected through 
a more general $''+--+''$ model \cite{Kogan:2000cv,Kogan:2000xc}.  We
will explicitly show how the spectrum and the gravitational law
interpolate between the ones obtained in the above cases. The change of the nature of the
gravitational law is observed also in the the crystalline universe
model \cite{Kogan:2000xc} which we will discuss afterwards. In this case gravity at intermediate distances is generated by the
first low lying band of KK states and at large scales gives again five
dimensional law of gravity. For an alternative way of modifying
gravity at ultralarge distances see \cite{Dvali:2000hr,Dvali:2001xg}. At this point we will present a potential
danger arising from the fact that  massive gravitons have extra
polarization states in comparison with the massless one. This is
clearly seen by the difference in the tensor structure of the massive
and massless spin-2 propagators in flat space (van
Dan-Veltman-Zakharov discontinuity) \cite{vanDam:1970vg,Zakharov}. Observationally this
implies that the massive graviton proposal disagrees with well tested
predictions of  Einsteinian gravity \cite{Dvali:2000rv}.

So far our discussion has considered the gravitational force 
due to tensorial gravitational modes.  However, in the gravity
sector there exist moduli corresponding to the fluctuation(s) of the
brane(s) or the size of the extra dimension, the radion(s) \cite{Pilo:2000et,Kogan:2001qx} and the
dilaton\footnote{This is not to be confused with the dilaton field that 
is present in the spectrum of closed strings.} \cite{Charmousis:2000rg,Binetruy:2001tc} respectively, which change the phenomenology of these
models. This will be the subject of chapter 3 where we will derive the
effective action describing these excitations \cite{Kogan:2001qx}. For this purpose we
will firstly consider a general three three-brane model with
arbitrary bulk cosmological constants between the branes. We will
calculate the  kinetic energy for the moduli of the system and see
that the freely moving (\textit{i.e.} not on an orbifold fixed point)
negative tension branes have wrong sign kinetic term
\cite{Dvali:2000km,Pilo:2000et,Kogan:2001qx}. This plays an important role in the context of multigravity models where this ghost
radion(s) cancel the troublesome extra polarization states of the
massive gravitons restoring the phenomenologically favoured tensorial
structure of the graviton propagator
\cite{Pilo:2000et,Gregory:2000iu,Csaki:2000ei}. At large distances the 
ghost radion(s) dominate giving scalar antigravity \cite{Gregory:2000iu}  in all the above constructions 
instead of a reduction of the Newton's constant or a change of the nature
of the gravitational law.  However, the very fact that there  are
physical ghost(s) in the spectrum makes these models unacceptable. Note
that the essential  characteristic of
all these models is the bounce of the warp factor generated by the
presence of the moving negative tension branes. This association
between the bounce in the warp factor and presence of  negative tension branes 
is an unavoidable feature in flat (five dimensional) brane models and
is possibly linked to the
fact that the weaker energy condition is violated at the position of
the bounce \cite{Freedman:1999gp,Witten:2000zk}. In the remaining of
the chapter we will discuss the dilaton for backgrounds with identical 
four-geometries and with bulk dynamics involving arbitrary scalar
fields minimally coupled to gravity  \cite{Papazoglou:2001ed}. In order to do this we will
briefly review the brane-world sum rules derived by Gibbons, Kallosh and
Linde  \cite{Gibbons:2001tf}. Then we will find a generic formula
relating the dilaton mass with the sum of the tensions of the
branes and the curvature of the four-geometries, from which we can see 
immediately the  necessary condition for
stabilizing the overall size of the system.

The flat multigravity models as we have already said suffer from the
presence of ghost scalar mode(s) which renders the theory unacceptable.
In chapter 4 we will present a way out of this  difficulty by
considering $AdS_4$ branes instead of flat branes. In this case the warp factor has a $''\cosh''$ form which naturally
generates a bounce without any need of floating negative tension
branes. Indeed, it is straightforward to replicate the bigravity   and
the crystalline model in this framework. The bigravity model is converted to a
$''++''$ model \cite{Kogan:2001vb} and the crystal model to an infinite array of $''+''$
branes. By
sending the second brane in the $''++''$ model to infinity, one obtains 
the locally localized model of \cite{Karch:2001ct,Schwartz:2001ip,Karch:2001cw} where gravity is
mediated only by the ultralight state since the zero mode becomes
non-normalizable. It is interesting to note how these models confront
the relativistic predictions of Einsteinian gravity since there are no
ghost radion(s) to restore the graviton propagator structure to the
phenomenologically favoured one. We will show that in $(A)dS$ space the
van Dam-Veltman-Zakharov  no-go theorem can be evaded at tree level
for the case that $m/H \to 0$, where $m$ is the mass of the graviton
and $H$ the ``Hubble'' of the $(A)dS$ space, as
it was shown in
\cite{Higuchi:1987py,Higuchi:1989gz,Kogan:2001uy,Porrati:2001cp}. The
extra polarizations of the massive gravitons in this case are very
weakly coupled and decouple in the massless flat limit. A precise
realization of this case is the $''++''$ bigravity model.  The $AdS$ brane models, even though they solve
the theoretical difficulty of the ghost radion(s) of the flat brane
models, face phenomenological difficulties. The presence of a remnant 
negative cosmological constant is at odds with observations, and
furthermore it turns out that all large scale modifications of
gravity predicted by these models, are hidden  behind the 
$AdS$ horizon. Furthermore, as we will discuss at the end of the
chapter an old result by Vainshtein \cite{Vain} suggests that the
discontinuity of the graviton propagator in flat space is not directly
observable by  the bending of the light by the Sun as it was claimed
in \cite{vanDam:1970vg,Zakharov} because non-perturbative effects
become important  and make 
the massless limit continuous.

In the light of this encouraging result it seems appropriate to look for 
multigravity  models of flat branes without ghost fields. As we have
discussed, in five dimensions it is
impossible to have flat brane multigravity models without negative
tension branes. For this reason, in chapter 5 we will consider models
in six dimensions \cite{Kogan:2001yr}. We explicitly show that the difficulties of the five
dimensional models can be evaded in six dimensions. It is possible to construct
multigravity models
with only flat positive tension branes. The branes which localize gravity in
this setup are four-branes, but one of their dimensions is compact,
unwarped and of Planck length. Thus, in the low energy limit the
spacetime on the brane appears four dimensional. In order that these
constructions are realized it is crucial that the tensions and/or the bulk cosmological constant are anisotropic. The five dimensional constructions in this
setup come into two types. One of them involves conical singularities 
at finite proper distance from the four branes. These conical singularities
support three-branes which can be of
positive tension if one has an
angle deficit, of zero tension if one has no angle deficit and of
negative tension if one has an angle excess. The other type of the
multigravity models has no conical singularity at all. It is crucial
to note that there is no five dimensional effective theory for these
constructions, otherwise one would get all the problems faced in the
five dimensional constructions. The low energy effective theory is
directly four dimensional. These six dimensional models give for the
first time a theoretically and phenomenologically viable realization
of multigravity.

In the final chapter, we summarize the multigravity models that we
have presented in this thesis and discuss questions for future
investigation.

Throughout  this thesis we use the mostly plus metric convention
with signature being $(-,+,+,+,\dots)$ and we use the following convention
for the Riemann tensor $R^K_{~\Lambda MN}=\de_M \Gamma^K_{\Lambda
N}-\de_N \Gamma^K_{\Lambda M}+\Gamma^H_{\Lambda
N}\Gamma^K_{MH}-\Gamma^H_{\Lambda M}\Gamma^K_{NH}$ where the
Christoffel symbols are
$\Gamma^K_{MN}={1 \over 2}g^{K \Lambda}\left(\de_Mg_{\Lambda N}+\de_Ng_{\Lambda M}-\de_{\Lambda}g_{MN}\right)$. We work in the units where $\hbar=c=1$. The four dimensional Planck mass and Newton's constant are related as $G_N^{-1}=32 \pi M_{\rm{Pl}}^2$.

\section{The Randall-Sundrum model}

\begin{figure}[t]
\begin{center}
\begin{picture}(300,150)(0,50)
\SetScale{1.3}
\SetOffset(-50,0)

\SetWidth{1}

\BCirc(150,100){50}
\DashLine(100,100)(200,100){3}

\CCirc(100,100){5}{Black}{Red}
\CCirc(200,100){5}{Black}{Green}

\Text(110,140)[]{$+$}
\Text(280,140)[]{$-$}
\Text(180,120)[]{$Z_2$}
\Text(250,140)[]{$L$}

\SetWidth{1.5}
\LongArrow(150,100)(150,115)
\LongArrow(150,100)(150,85)

\end{picture}
\end{center}
\caption{The Randall-Sundrum model with one $''+''$ and one $''-''$
brane at the fixed points of a $S^1/Z_2$ orbifold.}\label{+-}
\label{RSfig}
\end{figure}
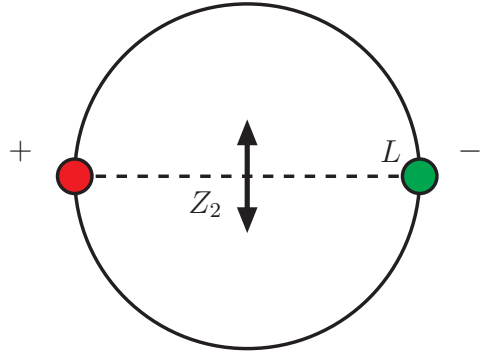

Since, throughout this thesis we will study warped models it is
instructive to briefly present the Randall-Sundrum model. This
construction consists of two  parallel three-branes sitting
on the fixed points of an $S^1/Z_2$ orbifold in five dimensional
spacetime which is essentially a slice of $AdS_5$ (see
Fig.\ref{RSfig}). The negative bulk cosmological constant and the
tensions of the branes are fine tuned in such a way that the
four-dimensional slices of the five-dimensional spacetime, and thus
the two branes, are flat. The metric that describes this configuration 
has the characteristic warped form:
\begin{equation}
ds^2=e^{-2k |y|}\eta_{\mu \nu} dx^{\mu}dx^{\nu}+dy^2
\end{equation}

If one now considers a scalar field, in particular the SM Higgs,
confined on one of the two branes, then the minimal coupling to the
metric will result to an action:
\begin{equation}
{\cal S}_H=\int d^4x\underbrace{~~\sqrt{-g_{ind}}~~}_{e^{-4k |y_i|}}\left\{\underbrace{~~~-~g_{ind}^{\mu \nu}~~~}_{-e^{2k |y_i|}\eta^{\mu \nu}}D_{\mu}H^\dagger D_{\nu}H-\lambda\left(H^2-v_0^2\right)^2\right\}
\end{equation}
where $g^{ind}_{\mu \nu}$ is the induced metric on the brane at
position $y_i$. After a field rescaling $H \to e^{k |y_i|}H$ which
brings the kinetic term to be canonical, the action reads:
\begin{equation}
{\cal S}_H=\int d^4x\left\{-\eta^{\mu \nu}D_{\mu}H^\dagger D_{\nu}H-\lambda\left(H^2-v_0^2e^{-2k |y_i|}\right)^2\right\}
\end{equation}
which clearly shows that the vacuum expectation value (vev) of the
Higgs is given as:
\begin{equation}
v=v_0e^{-k |y_i|}
\end{equation}

\begin{figure}[t]
\begin{center}
\begin{picture}(300,220)(0,50)

\Text(-10,250)[c]{$+$}
\Text(310,250)[c]{$-$}
\Text(310,140)[c]{$L$}

\SetWidth{2}

\SetColor{Red}
\Curve{(10,140)(50,150)(240,240)(290,220)}
\Text(235,220)[c]{\Red{1st-KK}}

\SetColor{Blue}
\Curve{(10,240)(50,192)(65,181)(80,173)(100,167)(160,159)(200,157)(290,155)}
\Text(235,170)[c]{\Blue{0-mode}}

\SetColor{Red}
\DashCurve{(10,140)(50,150)(140,240)(270,60)(290,80)}{4}
\Text(110,250)[c]{\Red{2nd-KK}}


\SetWidth{.5}
\SetColor{Black}
\Line(10,150)(290,150)
\Line(10,50)(10,250)
\Line(290,50)(290,250)

\Text(370,220)[c]{$m$}
\LongArrow(350,50)(350,220)
\SetColor{Red}
\Vertex(350,190){3}
\Vertex(350,155){3}
\Vertex(350,120){3}
\Vertex(350,85){3}
\SetColor{Blue}
\Vertex(350,50){3}

\end{picture}
\end{center}

\caption{The graviton (blue line), first (red line) and second
(red dashed line) KK states wavefunctions in the RS1 model. On the right the pattern of the
spectrum is sketched.}
\label{wfunctRS}
\end{figure}
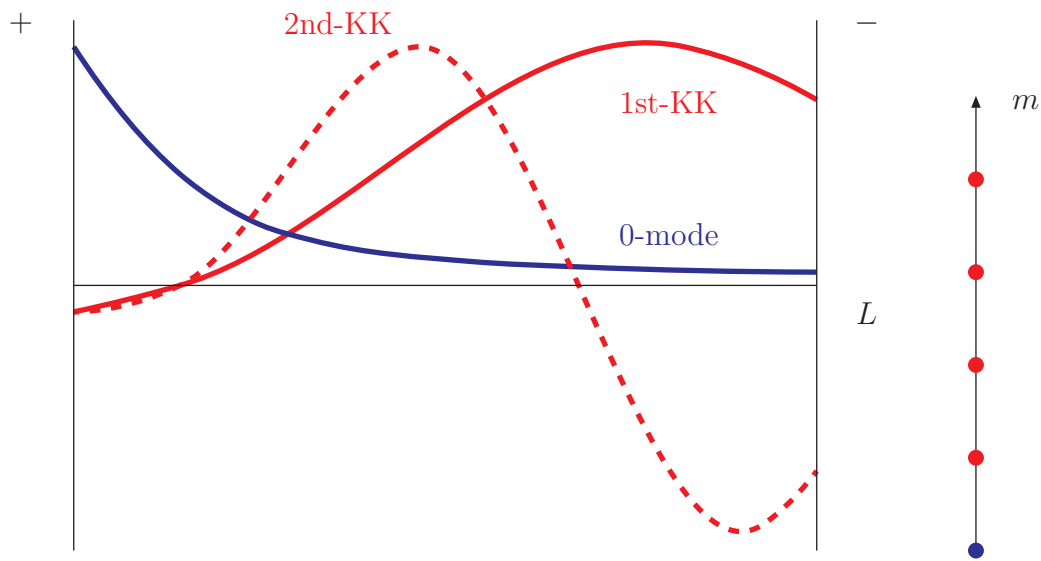

Assuming that all five dimensional parameters and thus $v_0$ are of
the order of the Planck scale, then a Higgs on the first brane will
have a vev of the order of the Planck scale, whereas a Higgs on the
second brane will have exponentially reduced vev. It was exactly this 
generation of exponentially different mass scales that was utilized
in the Randall-Sundrum scenario to resolve the Planck hierarchy
problem by placing the SM on the second brane (TeV brane). A
compactification radius of only some $35$ times larger than the Planck length was sufficient to
bring the Higgs vev to the TeV scale.

The phenomenology of  this model
has been extensively explored over the past years (see
\cite{Davoudiasl:2001wi} for a comprehensive account).  In the
TeV hierarchy scenario, the KK tower of the spin-2 graviton resonances
starts from the TeV scale with approximately TeV  spacing between
them giving rise to characteristic signals in high energy
colliders. The graviton and KK states wavefunctions (\textit{i.e.}
their profiles along the extra  dimension) are shown in
Fig.\ref{wfunctRS}. The massless zero mode wavefunction follows the
warp factor and is localized on the positive tension brane, while the
wavefunctions of the remaining tower are suppressed on the positive
tension brane and enhanced on the negative tension one. We will return 
to this picture later and compare it with the considerably different pattern that we get in
the multigravity models.
\chapter{Flat brane multigravity models in five dimensions}
\label{ch2}

In this chapter we will see how we can realize the multigravity scenario,
as it has been described in the introduction, with flat multi-brane
configurations in five dimensions. We will see that these models are infested with the presence
of freely moving negative tension branes. We will discuss the problems 
arising from the negative tension branes in the following
chapter. Here, we will concentrate on the gravitational potential
coming from the tensor fluctuations of the 
background metric and discuss the general characteristics of the
multigravity models. Although flat brane multigravity scenarios have difficulties, they are very easy to handle toy models  and can
be used to understand physics of more complicated but realistic multigravity models.

In the following we will discuss the $''+-+''$ bigravity modes, the
GRS model of quasi-localized gravity and the relation between them as well as the crystal multigravity model.

\sect{The $''+-+''$ bigravity model}
\label{sect+-+}

\begin{figure}[t]
\begin{center}
\begin{picture}(300,170)(0,50)
\SetScale{1.3}
\SetOffset(-50,0)

\SetWidth{1}

\BCirc(150,100){50}
\DashLine(100,100)(200,100){3}

\CCirc(100,100){5}{Black}{Red}
\CCirc(200,100){5}{Black}{Red}

\CCirc(150,150){5}{Black}{Green}
\CCirc(150,50){5}{Black}{Green}

\Text(196,210)[]{$-$}
\Text(196,50)[]{$-$}
\Text(196,175)[]{$L_1$}
\Text(196,85)[]{$-L_1$}

\Text(110,130)[]{$+$}
\Text(280,130)[]{$+$}
\Text(180,120)[]{$Z_2$}
\Text(245,140)[]{$L_2$}

\LongArrowArc(150,100)(58,6,86)
\LongArrowArcn(150,100)(58,84,4)
\Text(275,188)[]{$x=kl$}

\LongArrowArc(150,100)(58,96,176)
\LongArrowArcn(150,100)(58,174,94)
\Text(118,188)[]{$x=kl$}

\SetWidth{1.5}
\LongArrow(150,100)(150,115)
\LongArrow(150,100)(150,85)

\end{picture}
\end{center}
\caption{The $''+-+''$ bigravity model with two $''+''$ branes at the
fixed points of a $S^1/Z_2$ orbifold and
a $''-''$ brane in between.}
\label{+-+}
\end{figure}
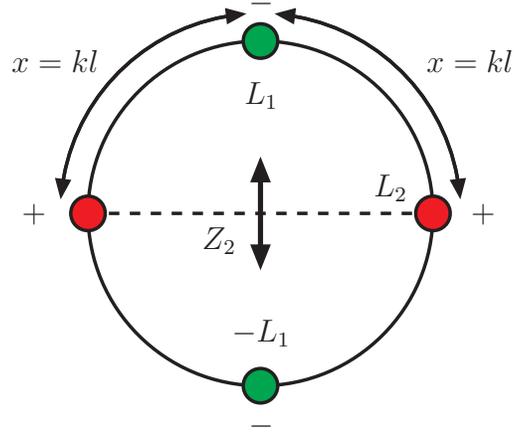

We begin our presentation of flat multigravity models in five
dimensions with the $''+-+''$ bigravity model
\cite{Kogan:2000wc}. This model is a straightforward extension of the
RS1 model with one more positive tension brane. In more detail, we have  three parallel three-branes in an $AdS_5$ space with
cosmological constant $\Lambda<0$. The fifth dimension has the geometry
of an $S^1/Z_2$ orbifold and the branes are located at
$L_0=0$, $L_1$ and $L_2$ where $L_0$ and $L_2$
are the orbifold  fixed points (see Fig.\ref{+-+}). This model has
three parameters, namely $L_1$, $L_2$ and $\Lambda$. For our present
purposes we will concentrate in the symmetric configuration where the intermediate
brane lies in the middle of the orbifold, \textit{i.e.}
$L_2=2L_1\equiv 2l$, leaving two parameters $l$, $\Lambda$. Firstly, we consider the branes having no  matter on them in
order to find a suitable vacuum solution. The action of this setup is:
\begin{equation}
{\cal S}_G=\int d^4 x \int_{-2l}^{2l} dy \sqrt{-G} 
\{  2 M^3
R - \Lambda \}-\sum_{i}\int_{y=L_i}d^4xV_i\sqrt{-\hat{G}^{(i)}}
\label{gravityaction}
\end{equation}
where $\hat{G}^{(i)}_{\mu\nu}$ is the induced metric on the branes,
$V_i$ their tensions and $M$ the five dimensional fundamental scale. The integration is performed throughout the orbifold, 
so there is no need to introduce Gibbons-Hawking boundary terms. The
Einstein equations that arise from this action are:
\begin{equation}
R_{MN}-\frac{1}{2}G_{MN}R=-\frac{1}{4M^3}
\left(\Lambda G_{MN}+
\sum_{i}V_i\frac{\sqrt{-\hat{G}^{(i)}}}{\sqrt{-G}}
\hat{G}^{(i)}_{\mu\nu}\delta_M^{\mu}\delta_N^{\nu}\delta(y-L_i)\right)
\end{equation}

At this point we demand that our metric respects four dimensional  Poincar\'{e}
invariance. The metric ansatz with this property is the following:
\begin{equation}
ds^2=e^{-2\sigma(y)}\eta_{\mu\nu}dx^\mu dx^\nu +dy^2
\label{warp-metric}
\end{equation}

Here the warp function $\sigma(y)$ is essentially a conformal
factor that rescales the four dimensional  component of the metric. A straightforward
calculation gives us the following differential equations for $\sigma(y)$:
\begin{eqnarray}
\left(\sigma '\right)^2&=&k^2\\
\sigma ''&=&\sum_{i}\frac{V_i}{12M^3}\delta(y-L_i)
\end{eqnarray}
where $k=\sqrt{\frac{-\Lambda}{24M^3}}$ is a measure of the curvature of the bulk. 

The solution of these equations consistent with the orbifold geometry is 
precisely:
\begin{equation}
\sigma(y)=k\left\{l-\left||y|-l\right|\right\}
\end{equation}
with the additional requirement that $V_0=V_2=-V_1 \equiv
V$. Furthermore, four dimensional Poincar\'{e} invariance like in the
RS1 model requires the 
fine tuning relation:
\begin{equation}
V=-\frac{\Lambda }{k}
\end{equation}

If we consider massless fluctuations of this vacuum metric  as in
\cite{Randall:1999ee} and then integrate over the fifth dimension, we find
the four dimensional fundamental scale\footnote{The factor $M_*$
multiplies the four dimensional Ricci scalar in the zero mode
truncated Lagrangian.} $M_*$, the five dimensional
fundamental scale $M$ and the curvature of the bulk $k$ are related as:
\begin{equation}
M_{*}^2=\frac{2M^3}{k}\left[1-e^{-2kl}\right]
\end{equation}

We do not identify the mass scale $M_*$ with the four dimensional
Planck scale for reasons which will be obvious in the following. The above formula tells us that for large enough $kl$ the
three mass scales $M_{*}$, $M$, $k$ can be taken to be of the same
order. Thus we take $k\sim {\mathcal {O}}(M)$ in order
not to introduce a hierarchy, with the additional restriction
$k<M$ so that the bulk curvature is small compared to the five dimensional Planck
scale and we can trust our solution. Furthermore, if we put matter on the  branes all the physical masses $m$ on the brane will be related to 
the fundamental mass parameters $m_0 \sim M$ of the five dimensional
theory by the conformal warp factor:
\begin{equation}
m=e^{-\sigma\left(L_i\right)}m_0
\end{equation}

Thus, in the symmetric configuration which we are examining, the mass 
scales on the first and third brane will be of the order of the
fundamental scale $M$ while on the intermediate negative tension brane they will be
exponentially suppressed. It is obvious that in an asymmetric
configuration we could obtain an electroweak scale hierarchy a la
Randall-Sundrum even on a positive tension brane. For this case we
would only need $2L_1-L_2\approx35k^{-1}$ in order to get
$e^{-\sigma\left(L_3\right)} \sim
{\mathcal{O}}\left(10^{-15}\right)$ (see \cite{Kogan:2000wc,Mouslopoulos:2000er}).

To determine the phenomenology of the model we need to know the KK
spectrum that follows from the dimensional reduction. This is
determined by considering the linear fluctuations of the metric of the
form:
\begin{equation}
ds^2=\left[e^{-2\sigma(y)}\eta_{\mu\nu} +h_{\mu\nu}(x,y)\right]dx^\mu
dx^\nu +dy^2
\label{fluct}
\end{equation}

Here we have ignored the dilaton and radion modes which will be the
subject of the following chapter. We expand the field $h_{\mu\nu}(x,y)$
in graviton and KK states plane waves according to:
\begin{equation}
h_{\mu\nu}(x,y)=\sum_{n=0}^{\infty}h_{\mu\nu}^{(n)}(x)\Psi^{(n)}(y)
\end{equation}
where $\left(\partial_\kappa\partial^\kappa-m_n^2\right)h_{\mu\nu}^{(n)}=0$
with
$\partial^{\alpha}h_{\alpha\beta}^{(n)}=h_{\phantom{-}\alpha}^{(n)\alpha}=0$.
The latter conditions for the zero mode correspond to a choice of
gauge fixing. The function $\Psi^{(n)}(y)$ will obey the following second order differential
equation:
\begin{equation}
\left\{-
\frac{1}{2}\partial_y^2+2k^2-2k\left[\delta(y)+\delta(y-2l)-\delta(y-l)-\delta(y+l)\right]\right\}\Psi^{(n)}(y)=\frac{m_n^2}{2}e^{2\sigma}\Psi^{(n)}(y)
\end{equation}

After a change of variables and a redefinition of the wavefunction the 
above equation reduces to an ordinary
Schr\"{o}dinger-type equation:
\begin{equation}
\left\{-
\frac{1}{2}\partial_z^2+V(z)\right\}\hat{\Psi}^{(n)}(z)=\frac{m_n^2}{2}\hat{\Psi
}^{(n)}(z)
\label{sch}
\end{equation}
with potential:
\begin{equation}
V(z)=\frac{15k^2}{8[g(z)]^2}-
\frac{3k}{2g(z)}\left[\delta(z)+\delta(z-2z_l)-\delta(z-z_l)-\delta(z+z_l)\right]
\end{equation}

The new variables and wavefunction in the above equation are defined as:
\begin{equation}
\renewcommand{\arraystretch}{1.5}
z\equiv {\rm sgn}(y)\left\{\begin{array}{cl}\frac{2e^{kl}-e^{2kl-k|y|}-
1}{k}&,l \leq |y|\leq 2l\\\frac{e^{k|y|}-1}{k}&,|y|\leq l\end{array}\right.
\
\end{equation}
\begin{equation}
\hat{\Psi}^{(n)}(z)\equiv \Psi^{(n)}(y)e^{\sigma/2}
\end{equation}
and the function $g(z)$ as $
g(z)\equiv k\left\{z_l-\left||z|-z_l\right|\right\}+1$, where
$z_l=z(l)$. The change of variables has been
chosen so that there are no first derivative terms in the
differential equation.  Furthermore, in this coordinate system it can be easily seen that
the vacuum metric takes the conformally flat form: 
\begin{equation}
ds^{2}=\frac{1}{g(z)^{2}}\left( \eta _{\mu \nu }dx^{\mu }dx^{\nu
}+dz^{2}\right)
\end{equation}

\begin{figure}
\begin{center}
\begin{picture}(300,250)(0,50)
\LongArrow(140,0)(140,280)
\LongArrow(-80,120)(370,120)

\SetWidth{2}
\SetColor{Red}
\Line(140,0)(148,250)
\Line(140,0)(132,250)
\SetColor{Black}
\Vertex(140,120){4}

\SetColor{Green}
\Line(220,270)(225,130)
\Line(220,270)(215,130)
\Line(60,270)(65,130)
\Line(60,270)(55,130)
\Text(220,100)[c]{$z_l$}
\Text(60,100)[c]{$-z_l$}
\SetColor{Black}
\Vertex(220,120){4}
\Vertex(60,120){4}

\SetColor{Red}
\Line(-12,250)(-20,0)
\Line(-28,250)(-20,0)
\Line(308,250)(300,0)
\Line(292,250)(300,0)
\Text(310,100)[l]{$2z_l$}
\Text(-30,100)[r]{$-2z_l$}
\SetColor{Black}
\Vertex(-20,120){4}
\Vertex(300,120){4}

\Text(150,270)[l]{$V(z)$}
\Text(370,130)[rb]{$z$}

\Text(-50,155)[c]{$+$}
\Text(330,155)[c]{$+$}
\Text(165,155)[c]{$+$}
\Text(241,155)[c]{$-$}
\Text(39,155)[c]{$-$}

\SetColor{Blue}
\Curve{(148,250)(153,249)(170,190)(200,138)(215,130)}
\Curve{(225,130)(240,138)(270,190)(287,249)(292,250)}

\Curve{(-12,250)(-7,249)(10,190)(40,138)(55,130)}
\Curve{(65,130)(80,138)(110,190)(127,249)(132,250)}

\end{picture}
\end{center}
\vskip15mm
\caption{The potential $V(z)$ in the $''+-+''$ bigravity model. The
$\delta$-function wells correspond to $''+''$ branes, while the
$\delta$-function barriers to $''-''$ branes.}
\label{+-+pot}
\end{figure}
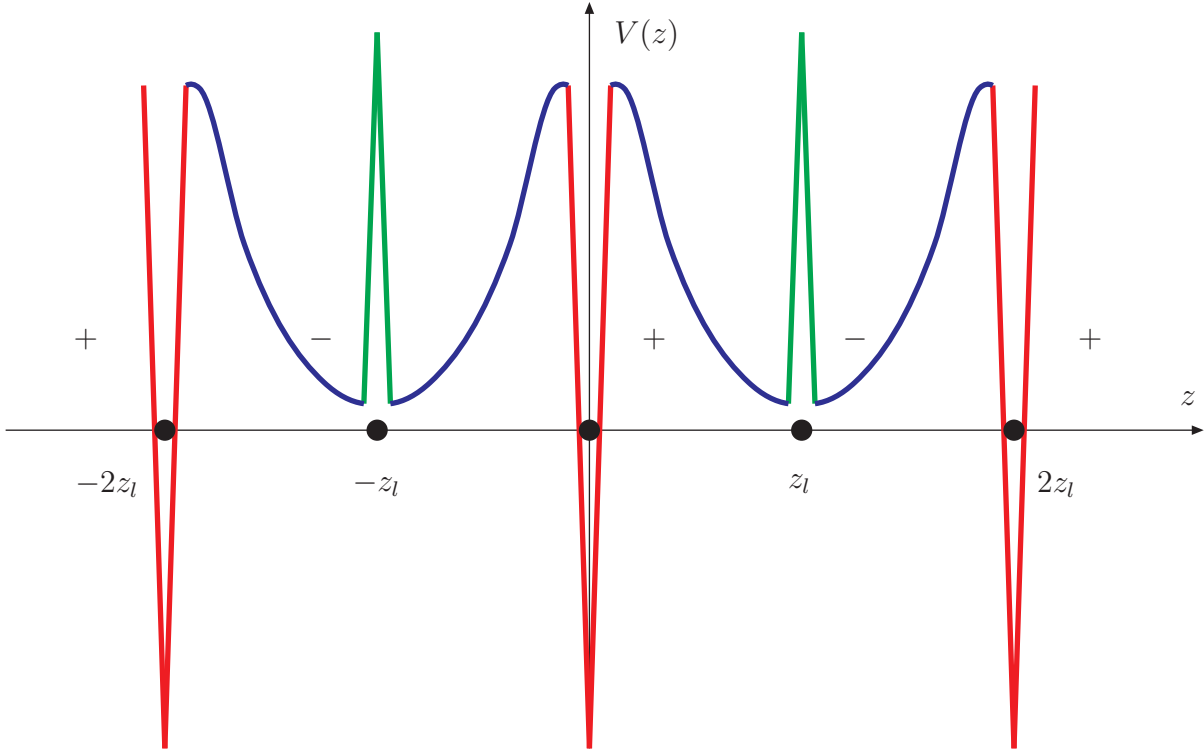

This is a quantum mechanical problem with $\delta$-function potentials and an extra $1/g^2$ smoothing term (due to the $AdS$ geometry) 
that gives the
potential a double ``volcano'' form (see Fig.\ref{+-+pot}). An
interesting characteristic of this Schr\"{o}dinger equation is that it has a 
``supersymmetric'' form which guarantees that it has no tachyonic
modes:
\begin{equation}
AA^{\dagger}\hat{\Psi}^{(n)}(z)=m_n^2\hat{\Psi}^{(n)}(z)~~~{\rm{with}}~~~A={d 
\over dz}-{3 \over 2}{\de_z g \over g }
\end{equation}

Since the system is compact, this differential equation always
gives rise to a massless zero mode which reflects the fact that
Poincar\'{e} invariance is preserved in the four dimensional spacetime. Its wavefunction is given by:
\begin{equation}
\hat{\Psi}^{(0)}=\frac{A}{[g(z)]^{3/2}}
\label{zero}
\end{equation}
The normalization factor $A$ is determined by the requirement 
$\displaystyle{\int_{-2z_l}^{2z_l}
dz\left[\hat{\Psi}^{(0)}(z)\right]^2=1}$, chosen so that we get the standard 
form of the Fierz-Pauli Lagrangian.

In the symmetric case that we are examining, the potential, and thus
the zero mode wavefunction, is symmetric with respect to the second
brane. This immediately implies that the next level will have a
wavefunction antisymmetric with respect to  the second brane. This is
exactly what happens.  The solution for  the graviton KK modes is
given in terms of Bessel functions. For $y$ lying in the regions ${\bf A}\equiv\left[0,l\right]$ and
${\bf B}\equiv\left[l,2l\right]$, we have:
\begin{equation}
\hat{\Psi}^{(n)}\left\{\begin{array}{cc}{\bf A}\\{\bf 
B}\end{array}\right\}=\sqrt{\frac{g(z)}{k}}\left[\left\{\begin{array}{cc}A_1\\B_
1\end{array}\right\}J_2\left(\frac{m_n}{k}g(z)\right)+\left\{\begin{array}{cc}A_
2\\B_2\end{array}\right\}Y_2\left(\frac{m_n}{k}g(z)\right)\right]
\end{equation}

The boundary conditions (one for the
continuity of the wavefunction at $z_1$ and three for the
discontinuity of its first derivative at $0$, $z_1$, $z_2$) result in a
$4\times4$ homogeneous linear system which, in order to have a
non-trivial solution, should have vanishing determinant:
\begin{equation}
\renewcommand{\arraystretch}{1.5}
\left|\begin{array}{cccc}J_1\left(\frac{m}{k}\right)&Y_1\left(\frac{m}{k}\right)
&\phantom{-}0&\phantom{-}0\\0&0&\phantom{-
}J_1\left(\frac{m}{k}\right)&\phantom{-
}Y_1\left(\frac{m}{k}\right)\\J_1\left(\frac{m}{k}g(z_l)\right)&Y_1\left(
\frac{m}{k}g(z_l)\right)&\phantom{-}J_1\left(\frac{m}{k}g(z_l)\right)&\phantom{-
}Y_1\left(\frac{m}{k}g(z_l)\right)\\J_2\left(\frac{m}{k}g(z_l)\right)&Y_2\left(
\frac{m}{k}g(z_l)\right)&-J_2\left(\frac{m}{k}g(z_l)\right)&-
Y_2\left(\frac{m}{k}g(z_l)\right)\end{array}\right|=0
\end{equation}
Here we have suppressed the subscript $n$ on the masses $m_n$. This is essentially the mass quantization condition which gives the
spectrum of the KK states. For each mass we can then determine the wavefunction 
with the same normalization as the zero mode. 

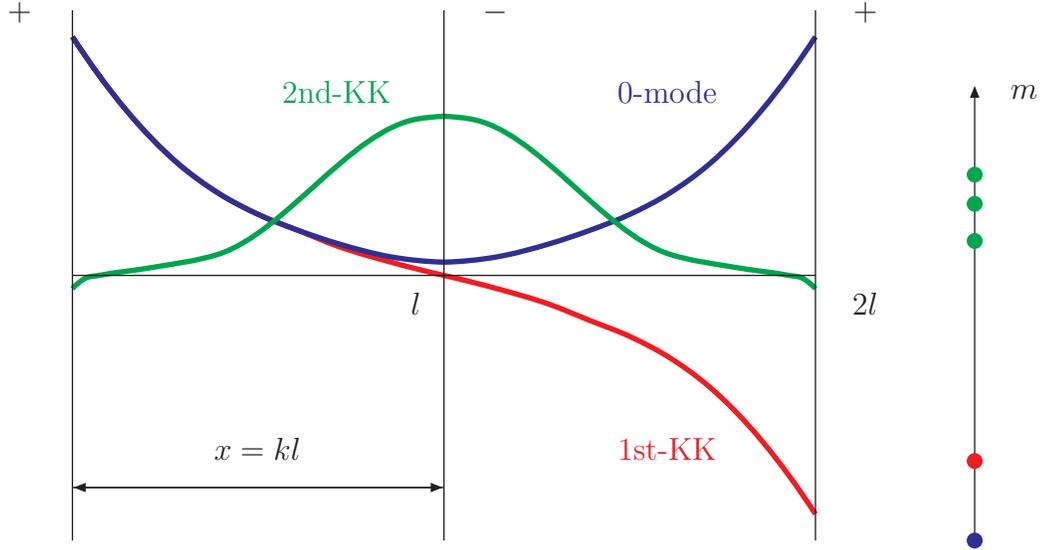
\begin{figure}[t]
\begin{center}
\begin{picture}(300,220)(0,50)

\Text(-10,250)[c]{$+$}
\Text(310,250)[c]{$+$}
\Text(170,250)[c]{$-$}
\Text(140,140)[c]{$l$}
\Text(310,140)[c]{$2l$}
\LongArrow(10,70)(148,70)
\LongArrow(150,70)(12,70)
\Text(80,85)[c]{$x=kl$}

\SetWidth{2}

\SetColor{Red}
\Curve{(10,240)(50,192)(65,181)(80,173)(95,167)(110,161)(130,155)(150,150)}
\Curve{(150,150)(170,145)(190,139)(205,133)(220,127)(235,119)(250,108)(290,60)}
\Text(235,85)[c]{\Red{1st-KK}}

\SetColor{Blue}
\Curve{(10,240)(50,192)(65,181)(80,173)(95,167)(110,162)(130,157)(150,155)}
\Curve{(150,155)(170,157)(190,162)(205,167)(220,173)(235,181)(250,192)(290,240)}
\Text(235,220)[c]{\Blue{0-mode}}

\SetColor{Green}
\Curve{(10,145)(15,149)(20,150)(40,153)(65,158)(140,209)(150,210)}
\Curve{(150,210)(160,209)(235,158)(260,153)(280,150)(285,149)(290,145)}
\Text(110,220)[c]{\Green{2nd-KK}}


\SetWidth{.5}
\SetColor{Black}
\Line(10,150)(290,150)
\Line(10,50)(10,250)
\Line(290,50)(290,250)
\Line(150,50)(150,250)

\Text(370,220)[c]{$m$}
\LongArrow(350,50)(350,220)
\SetColor{Green}
\Vertex(350,163){3}
\Vertex(350,177){3}
\Vertex(350,188){3}
\SetColor{Red}
\Vertex(350,80){3}
\SetColor{Blue}
\Vertex(350,50){3}

\end{picture}
\end{center}

\caption{The graviton (blue line), first (red line) and second
(green line) KK states wavefunctions in the symmetric $''+-+''$ model. The
wavefunctions of the zero mode and the second KK state have
discontinuities on the $''-''$ branes. On the right the pattern of the
spectrum is sketched.}
\label{wfunct}
\end{figure}

The exact calculation of the wavefunction of the first graviton KK
state verifies our previously stated intuition. Another way to
understand the form of the wavefunction is that since we have two
delta functions in the potential each of which supports a bound state, 
there should be a massive KK state that resembles the zero
mode. Indeed, if one calculates the wavefunctions of the higher levels 
it is evident that the first KK state has a special characteristic. It 
is the only one that is peaked in the positive tension branes, whereas 
all the higher levels are suppressed on the positive tension branes
and are spread in the bulk (see Fig.\ref{wfunct}). This is in contrast 
to the RS1 case where the wavefunctions of all the KK states are
spread in the bulk (see Fig.\ref{wfunctRS}). Moreover, since the
difference of the modulus of the
wavefunctions of the first special KK state and the zero mode comes from 
a very small region of the $y$-space around the position of the minimum of 
the warp factor, we expect its mass to be anomalously light in
comparison to the masses of the remaining of the KK tower.

This is a generic characteristic of potentials which have the shape of 
a bounce. In five dimensions with flat branes the only possibility of 
obtaining a bounce is by introducing negative tension branes. This is
because the warp factors generated by flat positive tension branes
always have the form of a falling exponential. The only way of pasting 
two warp factors which are falling exponentials is through a
discontinuity generated by a negative tension brane.

 If we define  $x=kl$, as the distance between the negative and
positive branes measured in units of $k$,  then for $x \gg 1$ we may obtain a reliable
approximation to its mass by using the first terms of the Bessel power
series. Then we find that the first KK mode has mass given by: 
\begin{equation}
m_1=2\sqrt{2}ke^{-2x}
\label{m1}
\end{equation}

 The masses of the other KK states  are found to depend in a different
way on the parameter $x$. Analytically we find that for $x \gg 1$ the
masses are:
\begin{equation}
m_{n+1}=\xi_nke^{-x}~~~,~~~n=1,2,\dots
\label{mn}
\end{equation}
where $\xi_{2i+1}$ is the $(i+1)$-th root of $J_1(x)$
$(i=0,1,2,\dots)$ and   $\xi_{2i}$ is the $i$-th root of $J_2(x)$
$(i=1,2,\dots)$. Using  the asymptotic values of the Bessel roots, we
find that the level splitting for high $n$ levels approaches a constant 
value:
\begin{equation}
\Delta m={\pi \over 2}ke^{-x}
\label{dmn}
\end{equation}

The coupling also of the first  KK state to matter on the branes is
 very different from the one of the remaining tower. To see this,
we consider a minimally coupled to gravity  matter  Lagrangian 
${\cal L}_m(\Phi_j,\hat{G}^{(i)})$, where we denote  with  
$\Phi_j$ the matter fields living on the branes. The total action of the
system is the sum of the gravity action (\ref{gravityaction}) plus the
matter action: 
\begin{equation}  
{\cal S}={\cal S}_G +\sum_i\int d^4xdy\sqrt{-\hat{G}^{(i)}}{\cal L}_m(\Phi_j,\hat{G}^{(i)})  
\end{equation} 
  
The effective four dimensional Lagrangian taking into account only the 
graviton perturbations, has the following form:  
\begin{equation}  
{\cal L}= {\cal L}_m(\Phi_j^c,\eta)  + 2M^3 \sum_n {\cal  
L}_{PF}(h^{(n)}(x))-\sum_i \sum_n\frac{e^{2\sigma(L_i)}\Psi^{(n)}(L_i)}{ 2}  h_{\mu\nu}^{(n)}(x) 
T^{\mu\nu}_{(i)}   
\end{equation} 
where ${\cal  L}_{PF}$ is the Pauli-Fierz Lagrangian for the $n$-level
graviton mode. The matter fields have been rescaled $\Phi_j \rightarrow \Phi_j^c$, to make  
them canonically normalized and the energy momentum tensor $T^{\mu\nu}_{(i)}$ for matter localized on the $i$-th brane  is  
defined with respect to the rescaled fields $\Phi_j^c$. By construction the  
induced background metric on the branes is the flat four dimensional
Minkowski metric ($\eta_{\mu\nu}$). Finally, defining the canonically normalized graviton fields:
\begin{equation} 
\bar{h}_{\mu\nu}^{(n)}(x)=\sqrt{2M^3}h_{\mu\nu}^{(n)}(x)
\label{canh}
\end{equation}  
the Lagrangian  reads:  
\begin{equation}  
{\cal L}= {\cal L}_m(\Phi_j^c,\eta)  +  \sum_n {\cal  
L}_{PF}(\bar{h}^{(n)}(x)) -\sum_i \sum_n
\frac{\hat{\Psi}^{(n)}\left(z_{i}\right)g\left(z_{i}\right)^{3/2}}{ 2\sqrt{2M^3}}  \, \bar{h}_{\mu\nu}^{(n)}(x) 
\, T^{\mu\nu}_{(i)}
\end{equation}   

Thus, the gravitons have the following couplings to matter on the positive
tension branes: 
\begin{equation} 
a_n=\frac{\hat{\Psi}^{(n)}\left(z_{+}\right)g\left(z_{+}\right)^{3/2}}{ 2\sqrt{2M^3}}
\label{coupling}
\end{equation}   

In more detail we find that the zero mode and the first ultralight KK
state couple with the same strength on the positive tension branes:
\begin{equation} 
a_0=a_1=\frac{1}{ 2\sqrt{2}M_*}
\end{equation}   
while the coupling of the remaining tower is exponentially suppressed:
\begin{equation}
a_{n+1}=\frac{1}{2\sqrt{2}M_{\ast }}~\frac{e^{-x}}{\sqrt{J_{1}^{2}\left( \xi_{n}\right) +J_{2}^{2}\left( \xi_{n}\right) }}
~~~~~~n=1,2,3,\ldots  
\label{bigcoupl}
\end{equation}

The reason for the strong coupling of the first KK state in comparison 
with the remaining tower is readily understood because, being
dominantly a bound  state of the volcano potential on the positive tension brane, it is largely localized on it. 

Let us now demonstrate how the bigravity scenario can be realized in
this model. We have already seen from  (\ref{m1}) and (\ref{mn}) that,
for large $x$, the lightest KK mode  splits off from the remaining
tower. This allows the possibility that the Newtonian gravity that we
experience at intermediate distances ($1{\rm mm}<r<10^{26}{\rm cm}$),
is due to {\textit{both}} the massless graviton {\textit{and}} the
ultralight KK state.  The gravitational potential is computed by the tree level exchange diagrams
of the massless graviton and KK states which in the Newtonian limit is: 
\begin{equation}
V(r)=-\sum_{n=0}^{N_{\Lambda}}{a_n^2 \over 4 \pi}\frac{e^{-m_{n}r}}{r}  \label{gravipot}
\end{equation}
where $a_n$ is the coupling (\ref{coupling}) and $n=0$ accounts for the
massless graviton. The summation stops at some very high level $N_{\Lambda}$
with mass of the order of the cutoff scale $\sim M$. In the bigravity scenario, at intermediate distances $m_2^{-1}\ll r\ll m_{1}^{-1}$, the first KK
state and the massless graviton contribute equally to the
gravitational potential:
\begin{equation}
V_{id}(r)= -\frac{1}{32 \pi M_{\ast }^{2}}\left( \frac{1}{r}+\frac{1}{r}\right) = -\frac{G_{N}}{r}
\label{bigri}
\end{equation}
where $G_{N}\equiv \frac{1}{16 \pi M_{\ast }^{2}}$ ($M_{\rm{Pl}}={M_* \over
\sqrt{2}}$). For long distances  $r\gtrsim
m_{1}^{-1}$ the Yukawa suppression effectively reduces gravity to half its
strength:
\begin{equation}
V_{ld}(r)= -\frac{1}{32 \pi M_{\ast }^{2}}\left( \frac{1}{r}+\frac{0}{r}\right) = -\frac{G_{N}}{2}\frac{1}{r}
\label{bigrl}
\end{equation}

\begin{figure}[t]
\begin{center}
\begin{picture}(320,80)(0,0)
\LongArrow(0,55)(300,55)
\C2Text(175,55){Black}{Red}{Excluded by Observational Data}{and by the
Cavendish Experiments}
\Text(300,65)[l]{$r$}
\Text(-2,65)[l]{$0$}
\Vertex(0,55){2}
\Vertex(45,55){2}
\Vertex(100,55){2}
\Text(25,67)[l]{$10 {\rm \mu m}$}
\Text(40,37)[l]{$m_2^{-1}$}
\Text(65,67)[l]{$0.2 {\rm mm}$}
\Vertex(250,55){2}
\Text(252,67)[l]{$10^{26} {\rm cm}$}
\Text(252,37)[l]{$m_1^{-1}$}
\LongArrow(45,-10)(45,20)
\LongArrow(252,-10)(252,20)
\Curve{(45,-10)(252,-10)}
\end{picture}
\end{center}
\caption{Exclusion regions for the bigravity case and correlation of
the first two KK states.}
\label{bigr-excl}
\end{figure}
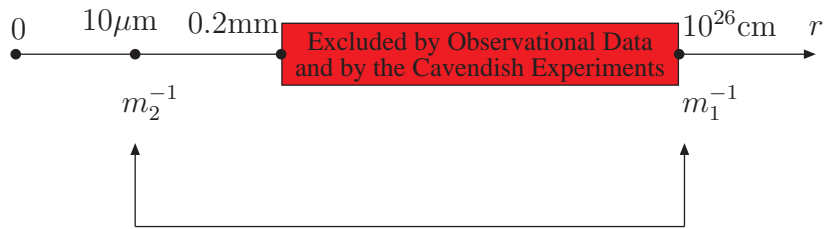

 Astronomical constraints and the requirement of the observability
of this effect demand that for $k\sim M_{\rm{Pl}}$ we should have $x$ in
the region $65-70$. The observational and experimental constrains in
the context of the graviton KK modes are  $m<10^{-31}{\rm eV}$ or
$m>10^{-4}{\rm eV}$ for modes with Planck suppressed coupling. If we choose $m_1\approx 10^{-31}{\rm
eV}$, from (\ref{m1}) and (\ref{mn}) we find  $m_2\approx 10^{-2}{\rm
eV}$ which comfortably satisfies the Cavendish bound (see Fig.\ref{bigr-excl}).

However, the Cavendish bound in this case is not even applicable
because the coupling of the remaining KK tower is exponentially
smaller than the usual Planck suppressed coupling. To see this let us
calculate the correction to the gravitational law  at distances $r\lesssim m_{2}^{-1}$. The Yukawa
interactions of the remaining KK states then turn on and will give rise
to a short distance correction. This can be evaluated by using the
asymptotic expression of the Bessel functions in (\ref{bigcoupl}) since we
are dealing with large $x$ and summing over a very dense spectrum, giving: 
\begin{equation}
\Delta V_{sd}(r)=-\frac{G_{N}}{k}\sum_{n=2}^{N_{\Lambda }}\frac{k\pi }{2e^{x}}~
\frac{m_{n}}{2k}~\frac{e^{-m_{n}r}}{r}  
\label{shortpotbigr}
\end{equation}

At this point we exploit the fact that the spectrum is nearly continuum
above $m_{2}$ and turn the sum to an integral with the first factor in (\ref
{shortpotbigr}) being the integration measure, {\textit{i.e.}} $\sum \frac{k\pi 
}{2e^{x}}=\sum \Delta m\rightarrow \int dm$ from (\ref
{dmn}). Moreover, we can
extend the integration to infinity because, due to the exponential
suppression of the integrand, the integral saturates very quickly and thus
the integration over the region of very large masses is irrelevant. The
resulting potential is now: 
\begin{equation}
\Delta V_{sd}(r)=-\frac{G_{N}}{k}\int_{m_{2}}^{\infty }dm~\frac{m}{2k}~\frac{%
e^{-m r}}{r}
\end{equation}

The integration is easily performed and gives: 
\begin{equation}
\Delta V_{sd}(r)= -\frac{G_{N}}{2r}~\frac{1+m_{2}r}{(kr)^{2}}~e^{-m_{2}r}
\end{equation}

We see these short distance corrections are significant only at Planck scale
distances $\sim k^{-1}$. Thus, the lower bound of the Cavendish
experiments is not important in the symmetric configuration simply
because although the low lying KK states have Compton wavelength at
the  millimeter scale, they are however very weakly coupled.

Finally, let us comment about the realization of the bigravity
scenario in asymmetric configurations where the positive tension
branes are not scale equivalent. In this case a whole rich of
bigravity scenarios emerge. Depending on which positive brane we live
in, we can have either a dominant massless graviton component or a
dominant massive graviton component. In the first case the correction
to Newtonian gravitational law at large scales will be negligible. In
the second case, where the massive KK state is the dominant source of 
Newtonian gravity, one would see gravity to essentially switch off at
large scales.

\section{The GRS model and its relation to the $''+-+''$ bigravity model}

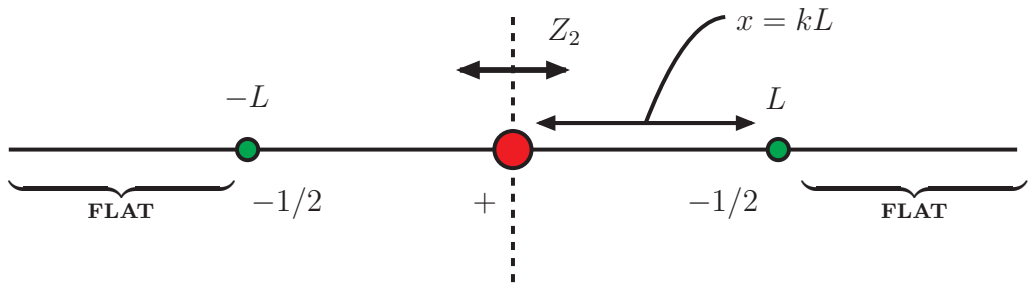
\begin{figure}
\begin{center}
\begin{picture}(300,120)(0,50)
\SetWidth{1.5}

\Line(-40,100)(340,100)
\DashLine(150,50)(150,150){3}
\CCirc(150,100){7}{Black}{Red}

\CCirc(50,100){4}{Black}{Green}
\CCirc(250,100){4}{Black}{Green}
\LongArrow(160,110)(240,110)
\LongArrow(240,110)(160,110)
\Curve{(200,110)(220,145)(230,150)}
\Text(235,150)[l]{$x=kL$}

\Text(65,80)[]{$-1/2$}
\Text(230,80)[]{$-1/2$}
\Text(140,80)[]{$+$}

\Text(250,120)[c]{$L$}
\Text(50,120)[c]{$-L$}

\Text(170,145)[]{$Z_2$}
\Text(-40,100)[lt]{$\underbrace{\phantom{abcdefghijklmn}}_{{\Large{\bf FLAT}}}$}
\Text(260,100)[lt]{$\underbrace{\phantom{abcdefghijklmn}}_{{\Large{\bf FLAT}}}$}

\SetWidth{2}
\LongArrow(130,130)(170,130)
\LongArrow(170,130)(130,130)

\end{picture}
\end{center}

\caption{The GRS model with one $''+''$ brane at the
fixed point of an $R^1/Z_2$ orbifold and a $''-1/2''$ brane at a distance
$x=kL$. The curvature of the bulk between the $''+''$ and $''-1/2''$
branes is $k$ and zero in the outside region.}
\label{GRS}
\end{figure}

An other interesting model which also has the feature of predicting
modifications of gravity at ultralarge scales is the one by Gregory,
Rubakov and  Sibiryakov (GRS) \cite{Gregory:2000jc}. In this section
we will briefly review the physics of the GRS model and demonstrate
how it is related to the $''+-+''$ bigravity model.  The GRS
model modifies the RS2 model by adding a
negative tension brane of half the tension of the positive tension
one at some distance $L$ in the bulk. The space between the positive and negative tension branes is a
slice  of $AdS_5$, while in the outside region it is flat (see
Fig.\ref{GRS}). The background metric for this setup has the four
dimensional Poincar\'{e} invariant form (\ref{warp-metric}) with the
warp factor being:
\begin{equation}
\sigma (y)=\left\{
\begin{array}{cl}k|y| & ,|y|<L \\kL & ,|y|>L \end{array} \right.
\end{equation}

This model does not have a normalizable massless graviton because the
volume of the system is infinite, but has a continuum of 
KK states that starts from zero. Despite the absence of a zero mode,
it generates four dimensional gravity at intermediate
distances due to the ``resonance''-like behaviour of the values of the wavefunctions of the KK
states continuum at the position of the positive tension brane. In that respect, gravity in this picture appears to
be quasi-localized.  The coupling for small masses of the KK
states has a Breit-Wigner form:
\begin{equation}
a(m)^2=\frac{{\mathcal A}}{m^2+(\Gamma/2)^2}
\end{equation}
where $\Gamma=4ke^{-3x}$ is the width of the resonance. This feature
of the coupling gives for intermediate distances  $r \ll \Gamma^{-1}$ normal four dimensional 
gravitational law:
\begin{equation}
V_{id}(r)= -G_N\frac{1}{r}
\end{equation}
where $G_N=\frac{k}{32 \pi M^3}$. For long
distances $r \gg  \Gamma^{-1}$, however, the gravitational law becomes 
five dimensional:
\begin{equation}
V_{ld}(r) = - \frac{4 G_N}{\pi \Gamma} \frac{1}{r^2}
\end{equation}

If we have $L \approx 45k^{-1}$, then $\Gamma^{-1} \sim 10^{26}$cm, so
this scenario cannot be excluded by current observations. The short
distance corrections, as in the $''+-+''$ model,  appear only for $r \ltsim
k^{-1}$ and read:
\begin{equation}
\Delta V_{sd}(r)=- \frac{G_N}{2r} \frac{1}{k^2r^2}
\end{equation}

Although the  GRS and the $''+-+''$ bigravity models look quite different, they share the key
element of freely moving negative tension branes. It was shown in \cite{Kogan:2000xc} that these two models are the limiting cases of a
more general $''+--+''$ multi-brane model that
interpolates between them. This model can be readily derived
from the bigravity model by cutting in half the intermediate negative
tension brane, {
\textit{i.e.}} instead of having one negative tension brane one can
take two negative tension branes of half
the tension of the original one, and considering flat space between
the two new negative tension branes. If the
second positive tension brane is taken to
infinity together with one of the negative tension branes, we shall get precisely the GRS picture.

\begin{figure}[t]
\begin{center}
\begin{picture}(300,180)(0,50)

\SetScale{0.9}
\SetOffset(20,40)
\SetWidth{1.5}

\BCirc(150,100){80}
\DashLine(70,100)(230,100){3}

\CCirc(70,100){7}{Black}{Red}
\CCirc(230,100){7}{Black}{Red}

\CCirc(183,172){4}{Black}{Green}
\CCirc(183,28){4}{Black}{Green}
\CCirc(117,172){4}{Black}{Green}
\CCirc(117,28){4}{Black}{Green}

\LongArrowArc(150,100)(90,298,358)
\LongArrowArcn(150,100)(90,358,298)
\LongArrowArc(150,100)(90,182,242)
\LongArrowArcn(150,100)(90,242,182)

\LongArrowArc(150,100)(90,70,110)
\LongArrowArcn(150,100)(90,110,70)

\Text(45,100)[]{$+$}
\Text(225,100)[]{$+$}
\Text(178,168)[l]{$-1/2$}
\Text(92,168)[r]{$-1/2$}
\Text(165,0)[c]{$-1/2$}
\Text(105,0)[c]{$-1/2$}

\Text(120,110)[]{$Z_2$}

\Text(215,45)[l]{$x=kl$}
\Text(55,45)[r]{$x=kl$}
\Text(135,185)[c]{$x_-=kl_-$}

\Text(100,140)[c]{$L_1$}
\Text(100,40)[c]{$-L_1$}
\Text(170,140)[c]{$L_2$}
\Text(170,40)[c]{$-L_2$}
\Text(195,105)[c]{$L_3$}

\Text(136,149)[c]{$\underbrace{\phantom{abcdefghi}}_{{\Large{\bf FLAT}}}$}
\Text(136,31)[c]{$\overbrace{\phantom{abcdefghi}}^{{\Large{\bf FLAT}}}$}

\SetWidth{2}
\LongArrow(150,100)(150,115)
\LongArrow(150,100)(150,85)

\end{picture}
\end{center}
\caption{The $''+--+''$ configuration with scale
equivalent $''+''$ branes. The distance between
the $''+''$ and $''-1/2'' $ branes is $l=L_1=L_3-L_2$ while the distance between the $''-1/2''$ branes is $l_-=L_2-L_1$. The curvature
of the bulk between the $''+''$ and $''-1/2''$ branes is $k$ and zero
between the $''-1/2''$ branes.}
\label{+--+}
\end{figure}
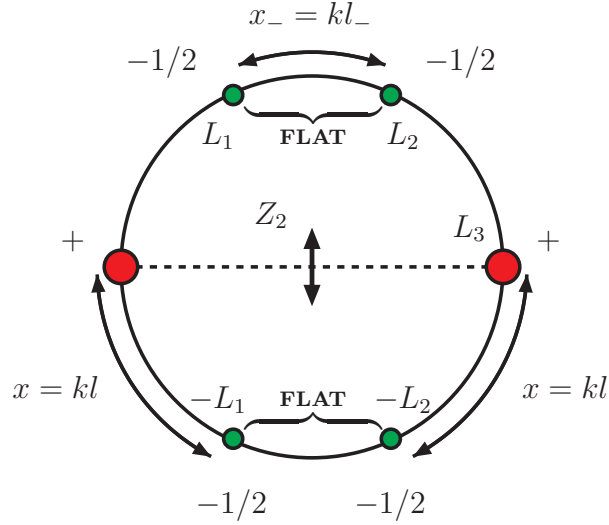

Let us discuss the four-brane $''+--+''$ model
in more detail. It consists of a five dimensional spacetime with orbifold topology with four
parallel three-branes located at $L_{0}=0$, $L_{1}$, $L_{2}$ and $L_{3}$, where $L_{0}$ and $L_{3}$ are the orbifold fixed points (see Fig.\ref{+--+}). The
bulk cosmological constant $\Lambda $ is negative  between the positive and negative tension  branes and zero between the two negative tension
branes. The model has four parameters, namely $L_{1}$, $L_{2}$ and $L_{3}$
and $\Lambda $. For our present purposes we consider the symmetric
configuration, leaving three parameters, $l$, $l_{-}$ and $\Lambda $ where
$l\equiv L_1=L_3-L_2$ and $l_{-}\equiv L_{2}-L_{1}$. In the absence of
matter the model is described by (\ref{gravityaction}) with 
\begin{equation}
\Lambda (y)=\left\{ 
\begin{array}{cl}
{0} & ,y\in \lbrack L_{1},L_{2}] \\ 
{\Lambda}  & ,y\in \lbrack 0,L_{1}]\bigcup [L_{2},L_{3}]
\end{array}
\right.
\end{equation}

By considering the ansatz (\ref{warp-metric}) the warp function $\sigma (y)$
must satisfy: 
\begin{eqnarray}
\sigma ^{\prime \prime }&=&\sum_{i}\frac{V_{i}}{12M^{3}}\delta (y-L_{i})\\( \sigma') ^2&=&\left\{ 
\begin{array}{cl}
{0} & ,y\in \lbrack L_{1},L_{2}] \\ 
k^{2} & ,y\in \lbrack 0,L_{1}] \bigcup [L_{2},L_{3}]
\end{array}
\right. \ 
\end{eqnarray}
where $k=\sqrt{\frac{-\Lambda }{24M^{3}}}$ is a measure of the bulk
curvature. The solution for $y>0$ is: 
\begin{equation}
\sigma (y)=\left\{ 
\begin{array}{cl}
{ky} & ,y\in \lbrack 0,L_{1}] \\ 
{kl} & ,y\in \lbrack L_{1},L_{2}] \\ 
{kl+k(l+l_--y)} & ,y\in \lbrack L_{2},L_{3}]
\end{array}
\right. \ 
\end{equation}
with the additional requirement that  $V_{0}=V_{3}=-2V_{1}=-2V_{2} \equiv
V$. Furthermore, four dimensional Poincar\'{e} invariance requires the fine tuning relation: 
\begin{equation}
V=-\frac{\Lambda }{k}
\end{equation}

In order to determine the mass spectrum and the couplings of the KK modes we
consider linear fluctuations around the background metric as in (\ref{fluct}). Following the same procedure we find that the function $\hat{\Psi}^{(n)}(z)$
obeys a Schr\"{o}dinger-like equation (\ref{sch}) with potential $V(z)$ of the form: 
\begin{eqnarray}
\hspace*{0.5cm}V(z) &=&\frac{15k^{2}}{8[g(z)]^{2}}\left[ \theta (z)-\theta
(z-z_{1})+\theta (z-z_{2})-\theta (z-z_{3})\right]  \nonumber \\
&&-\frac{3k}{2g(z)}\left[ \delta (z)-\frac{1}{2}\delta (z-z_{1})-\frac{1}{2}
\delta (z-z_{2})+\delta (z-z_{3})\right]  
\label{pot2}
\end{eqnarray}

The conformal coordinates now are given by: 
\begin{equation}
\renewcommand{\arraystretch}{1.5}z\equiv \left\{ 
\begin{array}{cl}
\frac{e^{ky}-1}{k} & ,y\in \lbrack 0,L_{1}] \\ 
\ (y-l)e^{kl}+\frac{e^{kl}-1}{k} & ,y\in \lbrack L_{1},L_{2}] \\ 
\ -\frac{1}{k}e^{2kl+kl_{-}}e^{-ky}+l_{-}e^{kl}+\frac{2}{k}e^{kl}-\frac{1}{k}
& ,y\in \lbrack L_{2},L_{3}]
\end{array}
\right. \ 
\end{equation}
with
\begin{equation}
g(z)=\left\{ 
\begin{array}{cl}
{kz+1} & ,z\in \lbrack 0,z_{1}] \\ 
{kz_{1}+1} & ,z\in \lbrack z_{1},z_{2}] \\ 
{k(z_{2}-z)+kz_{1}+1} & ,z\in \lbrack z_{2},z_{3}]
\end{array}
\right. \ 
\end{equation}
where $z_{1}=z(L_{1})$, $z_{2}=z(L_{2})$ and $z_{3}=z(L_{3})$.

The potential (\ref{pot2}) again gives rise to a massless graviton zero mode
whose wavefunction is given by (\ref{zero}) with the same normalization
convention. Note, however, that in the limit $l_- \rightarrow \infty$ this
mode becomes non-normalizable (GRS case). The solution of the
Schr\"{o}dinger equation for the massive KK modes is: 
\begin{equation}
\hat{\Psi}^{(n)}\left\{ 
\begin{array}{c}
\mathbf{A} \\ 
\mathbf{B} \\ 
\mathbf{C}
\end{array}
\right\}=N_n \renewcommand{\arraystretch}{1.7} \left\{ 
\begin{array}{c}
\sqrt{\frac{g(z)}{k}}\left[\phantom{A_1}Y_{2}\left(\frac{m_n}{k}%
g(z)\right)+A_{\phantom{2}}J_{2}\left(\frac{m_n}{k}g(z)\right)\right] \\ 
~~~~~~~~B_1\cos(m_{n}z)+B_{2}\sin(m_{n}z) \\ 
\sqrt{\frac{g(z)}{k}}\left[C_{1}Y_{2}\left(\frac{m_n}{k}g(z)%
\right)+C_{2}J_{2}\left(\frac{m_n}{k}g(z)\right)\right]
\end{array}
\right\} \   \label{wavemulti}
\end{equation}
where $\mathbf{A}=[0,z_{1}]$, $\mathbf{B}=[z_{1},z_{2}]$, and $\mathbf{C}%
=[z_{2},z_{3}]$. We observe that the solution in the first and third
interval has the same form as in the $^{\prime\prime}+ - +^{\prime\prime}$
model. The new feature is the second region (flat spacetime). The
coefficients that appear in the solution are determined by imposing the
boundary conditions and normalizing the wavefunction.

The boundary conditions (two for the continuity of the wavefunction at $%
z_{1} $, $z_{2}$ and four for the discontinuity of its first derivative at $%
0 $, $z_{1}$, $z_{2}$ and $z_{3}$) result in a $6\times 6$ homogeneous
linear system which, in order to have a non-trivial solution, should have
vanishing determinant. It is readily reduced to a $4\times 4$ set of
equations leading to the quantization condition: 
\begin{equation}
\renewcommand{\arraystretch}{2}{\footnotesize {\left| 
\begin{array}{cccc}
Y_{2}\left( \frac{m}{k}g\right) -\frac{Y_{1}\left( \frac{m}{k}\right) }{
J_{1}\left( \frac{m}{k}\right) }J_{2}\left( \frac{m}{k}g\right) & -\cos
(mz_{1}) & -\sin (mz_{1}) & 0 \\ 
Y_{1}\left( \frac{m}{k}g\right) -\frac{Y_{1}\left( \frac{m}{k}\right) }{
J_{1}\left( \frac{m}{k}\right) }J_{1}\left( \frac{m}{k}g\right) & 
\phantom{-}\sin (mz_{1}) & -\cos (mz_{1}) & 0 \\ 
0 & -\sin (mz_{2}) & \phantom{-}\cos (mz_{2}) & \phantom{-}Y_{1}\left( 
\frac{m}{k}g\right) -\frac{Y_{1}\left(\frac{m}{k}\right) }{J_{1}\left(
\frac{m}{k}\right) }J_{1}\left( \frac{m}{k}g\right) \\ 
0 & \phantom{-}\cos (mz_{2}) & \phantom{-}\sin (mz_{2}) & -Y_{2}\left( 
\frac{m}{k}g\right) +\frac{Y_{1}\left( \frac{m}{k}\right) }{J_{1}\left(
\frac{m}{k}\right) }J_{2}\left( \frac{m}{k}g\right)
\end{array}
\right| =0}}  \label{det1}
\end{equation}
with $g=g(z_{1})=g(z_{2})$. Here we have
suppressed the subscript $n$ on the masses $m_{n}$.

\subsection{Spectrum Interpolation}

The above quantization condition provides the mass spectrum of the model. It
is convenient to introduce two dimensionless parameters, $x=kl$ and $x_{-}=kl_{-}$ (see Fig.\ref{+--+}) and we work from now on with the set
of parameters $x$, $x_{-}$ and $k$. The mass spectrum depends crucially on the
distance $x_{-}$. We must recover the $''+-+''$ bigravity model spectrum in the limit $x_{-}\rightarrow 0$, and the GRS spectrum in the limit $x_{-}\rightarrow
\infty .$ From the quantization condition (\ref{det1}) it is easy to
verify these features and show how the $^{\prime \prime }+--+^{\prime \prime
}$ spectrum smoothly interpolates between the $''+-+''$ bigravity model and the GRS one.
It turns out that the structure of the spectrum has simple $x_{-}$ and $x$
dependence in three separate regions of the parameter space:

\subsubsection{The \textbf{{$^{\prime \prime }+-+^{\prime \prime
}$ Bigravity Region}}}

For $x_{-}\lower3pt\hbox{$\, \buildrel < \over \sim \, $}1$ we find that the
mass spectrum is effectively $x_{-}$-independent given by the approximate
form: 
\begin{eqnarray}
m_{1} &=&2\sqrt{2}ke^{-2x} \\
m_{n+1} &=&\xi _{n}ke^{-x}~~~~~~n=1,2,3,\ldots  \label{bimass}
\end{eqnarray}
where $\xi _{2i+1}$ is the $(i+1)$-th root of $J_{1}(x)$ ($i=0,1,2,\ldots $)
and $\xi _{2i}$ is the $i$-th root of $J_{2}(x)$ ($i=1,2,3,\ldots $). As
expected the mass spectrum is identical to the one of the $''+-+''$
bigravity model for scale equivalent positive tension branes. The first mass is manifestly singled out from
the rest of the KK tower and for large $x$ leads to the possibility of
bigravity (see Fig.\ref{+--+masses}).

\subsubsection{The \textbf{Saturation Region}}

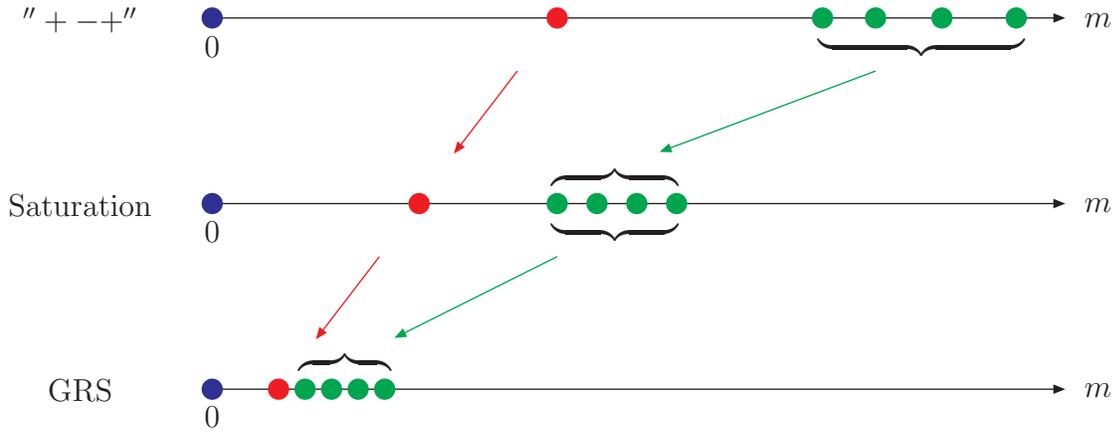
\begin{figure}[t]
\begin{center}
\begin{picture}(300,200)(0,50)

\LongArrow(0,220)(320,220)
\LongArrow(0,150)(320,150)
\LongArrow(0,80)(320,80)

\SetColor{Blue}
\Vertex(0,220){4}
\Vertex(0,150){4}
\Vertex(0,80){4}

\SetColor{Red}
\Vertex(130,220){4}
\SetColor{Green}
\Vertex(230,220){4}
\Vertex(250,220){4}
\Vertex(275,220){4}
\Vertex(303,220){4}

\SetColor{Red}
\Vertex(78,150){4}
\SetColor{Green}
\Vertex(130,150){4}
\Vertex(145,150){4}
\Vertex(160,150){4}
\Vertex(175,150){4}

\SetColor{Red}
\Vertex(25,80){4}
\SetColor{Green}
\Vertex(35,80){4}
\Vertex(45,80){4}
\Vertex(55,80){4}
\Vertex(65,80){4}

\Text(335,220)[]{$m$}
\Text(335,150)[]{$m$}
\Text(335,80)[]{$m$}

\Text(0,210)[]{$0$}
\Text(0,140)[]{$0$}
\Text(0,70)[]{$0$}

\Text(50,85)[c]{$\overbrace{\phantom{abcdef}}$}
\Text(152,145)[c]{$\underbrace{\phantom{abcdefgh}}$}
\Text(152,155)[c]{$ \overbrace{\phantom{abcdefgh}}$}
\Text(268,215)[c]{$ \underbrace{\phantom{abcdefghijklm}}$}

\SetColor{Green}
\LongArrow(250,200)(170,170)
\LongArrow(130,130)(70,100)

\SetColor{Red}
\LongArrow(115,200)(92,170)
\LongArrow(63,130)(40,100)

\Text(-50,220)[c]{$''+-+''$}
\Text(-50,150)[c]{Saturation}
\Text(-50,80)[c]{GRS}

\end{picture}
\end{center}
\caption{The behaviour of the mass of the first five KK states in the three
regions of simple $x$, $x_{-}$ dependence. The first dot at zero stands for
the graviton.}
\label{+--+masses}
\end{figure}

For $1\ll x_{-}\ll e^{2x}$ we find a simple dependence on $x_{-}$ given by
the approximate analytic form: 
\begin{eqnarray}
m_{1} &=&2k\frac{e^{-2x}}{\sqrt{x_{-}}} \\
m_{n+1} &=&n\pi k\frac{e^{-x}}{x_{-}}~~~~~~n=1,2,3,\ldots
\end{eqnarray}
As $x_{-}$ increases the first mass decreases less rapidly than the other
levels (see Fig.\ref{+--+masses}).

\subsubsection{The \textbf{GRS Region}}

For $x_{-}\gg e^{2x}$ the first mass is no longer special and scales with
respect to {\textit{both}} $x$ and $x_{-}$ in the same way as the remaining
tower: 
\begin{equation}
m_{n}=n\pi k\frac{e^{-x}}{x_{-}}~~~~~~n=1,2,3,\ldots  \label{longmass}
\end{equation}

The mass splittings $\Delta m$ tend to zero as $x_{-}\rightarrow \infty $
and we obtain the GRS continuum of states (see Fig.\ref{+--+masses}).

\subsection{Multigravity}

Armed with the details how the spectrum smoothly changes between the $''+-+''$ bigravity 
model ($x_{-}=0)$ and the GRS model ($x_{-}\rightarrow \infty )$, we can now
discuss the possibilities for modifying gravity at large distances. The
couplings of the KK states with matter on one of the scale equivalent
positive tension branes are readily calculated by (\ref{coupling}): 
\begin{equation}
a_{n}={1 \over 2 \sqrt{2}}\left[ \frac{g(0)}{M}\right] ^{3/2}\hat{\Psi}^{(n)}(0)
\label{couplmulti}
\end{equation}

\subsubsection{\textbf{Bigravity Region}}

As  $x_{-}\rightarrow 0$, we are in the bigravity region which has
been extensively explored in the previous section. 

\subsubsection{The \textbf{GRS Region}}

In the GRS limit, $x_{-}\gg e^{2x},$ we should reproduce the
``resonance''-like behaviour of the coupling in the GRS model. In the
following we shall see that indeed this is the case and we will calculate
the first order correction to the GRS potential for the case that $x_{-}$ is
large but finite.

For the rest of the section we split the wavefunction (\ref{wavemulti}) in
two parts, namely the normalization $N_{n}$ and the unnormalized
wavefunction $\psi^{(n)}(z)$, {\textit{i.e.}} $\hat{\Psi}^{(n)}(z)=N_{n}\psi^{(n)}(z)$. The former is as usual chosen so that
we get a canonically normalized Pauli-Fierz Lagrangian for the four
dimensional  KK modes $\bar{h}_{\mu \nu }^{(n)}$ (see (\ref{canh})) and is given by: 
\begin{equation}
N_{n}^{2}=\frac{1/2}{2\displaystyle{\int_{\phantom{.}0}^{z_{1}}dz\left[ 
\psi^{(n)}(z)\right] ^{2}}+\displaystyle{\int_{\phantom{.}%
z_{1}}^{z_{2}}dz\left[\psi^{(n)}(z)\right] ^{2}}}
\end{equation}

The value of $\psi^{(n)}(z)$ on the left positive tension brane is, for $m_{n}\ll k$: 
\begin{equation}
\psi_{(n)}^{2}(0)= \frac{16k^{3}}{\pi ^{2}m_{n}^{4}}  \label{un}
\end{equation}

It is convenient to split the gravitational potential given by the relation (
\ref{gravipot}) into two parts: 
\begin{equation}
V(r)=-\frac{1}{8M^{3}}\sum_{n=1}^{N_{x_{-}}-1}\frac{e^{-m_{n}r}}{r}N_{n}^{2}
\psi_{(n)}^{2}(0)-\frac{1}{8M^{3}}\sum_{n=N_{x_{-}}}^{N_{\Lambda }}
\frac{e^{-m_{n}r}}{r}N_{n}^{2}\psi_{(n)}^{2}(0)  \label{gravpot}
\end{equation}
where we denote by $N_{x_{-}}$ the level after which the coupling of
the KK states becomes linear to their mass. It corresponds to a mass scale
about $m_{s}=ke^{-x}$. As we shall see this separation is useful because the first $N_{x_{-}}$
states give rise to the long distance gravitational potential $V_{ld}$ while
the remaining ones will only contribute to the short distance
corrections $\Delta V_{sd}$.

\begin{itemize}
\item  \textbf{\ Short Distance Corrections}
\end{itemize}

We first consider the second term. The normalization constant in this region
is computed by considering the asymptotic expansions of the Bessel functions
with argument $\frac{g(z_{1})m_{n}}{k}$. It is calculated to be: 
\begin{equation}
N_{n}^{2}=\frac{\pi ^{3}m_{n}^{5}}{32k^{3}g(z_{1})x_{-}}~\left[ \frac{1}{1+%
\frac{2}{x_{-}}}\right]
\end{equation}

If we combine the above normalization constant with the unnormalized wavefunction (\ref{un}), we find:
\begin{equation}
\Delta V_{sd}(r)= -\frac{1}{32 \pi M^{3}}\sum_{n=N_{x_{-}}}^{N_{\Lambda }}\frac{k\pi 
}{x_{-}e^{x}}~\frac{m_{n}}{2k}~\frac{e^{-m_{n}r}}{r}~\left[ \frac{1}{1+\frac{%
2}{x_{-}}}\right]  \label{shortpot}
\end{equation}

Since we are taking $x_{-}\gg e^{2x}$, the spectrum tends to continuum, {%
\textit{i.e.}} $N_{n}\rightarrow N(m)$, $\psi_{(n)}(0)\rightarrow 
\psi(m)$, and the sum turns to an integral where the first factor in
(\ref{shortpot}) is the integration measure, {\textit{i.e.}} $\sum \frac{%
k\pi }{x_{-}e^{x}}=\sum \Delta m\rightarrow \int dm$ from (\ref
{longmass}). Moreover, we can extend the integration to
infinity because the integral saturates rather quickly. Finally, we expand the fraction involving $x_{-}$ keeping the
first term in the power series to obtain the potential: 
\begin{equation}
\Delta V_{sd}(r)= -\frac{1}{32 \pi M^{3}}\int_{m_{s}}^{\infty }dm~\frac{e^{-mr}}{r}~%
\frac{m}{2k}\left( 1-\frac{2}{x_{-}}\right)
\end{equation}
 The integral is easily calculated and the potential
reads: 
\begin{equation}
\Delta V_{sd}(r)= -\frac{G_{N}}{2r}~\frac{1+m_{s}r}{(kr)^{2}}~(1-\frac{2}{x_{-}})~e^{-m_{s}r}
\end{equation}
where we identified $G_{N}\equiv \frac{k}{32 \pi M^{3}}$ for reasons to be seen
later. The second part of the above potential is the first correction coming
from the fact that $x_{-}$ is finite. Obviously this correction vanishes
when $x_{-}\rightarrow \infty $. Note that the above potential gives
corrections to the Newton's law only at distances comparable to the Planck
length scale.

\begin{itemize}
\item  \textbf{\ Multigravity: 4D and 5D gravity}
\end{itemize}

We turn now to the more interesting first summation in (\ref{gravpot}) in
order to show that the coupling indeed has the ``resonance''-like behaviour
for $\Delta m\rightarrow 0$ responsible for four dimensional Newtonian gravity at
intermediate distances and the five dimensional gravitational law for cosmological
distances. This summation includes the KK states from the graviton zero mode
up to the $N_{x_{-}}$-th level. The normalization constant in this region is
computed by considering the series expansion of all the Bessel functions
involved. It is calculated to be: 
\begin{equation}
N_{n}^{2}= \frac{\pi ^{2}m_{n}^{4}}{4g(z_{1})^{4}x_{-}}~\left[ \frac{1}{%
m_{n}^{2}+\frac{\Gamma ^{2}}{4}+\frac{8k^{2}}{g(z_{1})^{4}x_{-}}}\right]
\end{equation}
where $\Gamma =4ke^{-3x}$. If we combine the above normalization
constant with the
unnormalized wavefunction (\ref{un}), we find that the long distance
gravitational potential is: 
\begin{equation}
V_{ld}(r)=-\frac{1}{32 \pi M^{3}}\sum_{n=0}^{N_{x_{-}}}\frac{\pi k}{x_{-}e^{x}}~%
\frac{4k^{2}}{\pi g(z_{1})^{3}}~\frac{e^{-m_{n}r}}{r}~\left[ \frac{1}{%
m_{n}^{2}+\frac{\Gamma ^{2}}{4}+\frac{8k^{2}}{g(z_{1})^{4}x_{-}}}\right]
\end{equation}

Again, since we are taking $x_{-}\gg e^{2x},$ the above sum will turn to an
integral with $\sum \Delta m\rightarrow \int dm$. Moreover, we can safely
extend the integration to infinity since the integral saturates very fast
for $m\lesssim \Gamma /4\equiv r_{c}^{-1}\ll ke^{x}$. If we also expand the
fraction in brackets keeping the first term in the power series, we find the
potential: 
\begin{eqnarray}
V_{ld}(r)= &-&\frac{1}{32 \pi M^{3}}\int_{0}^{\infty }dm~\frac{4k^{2}}{\pi
g(z_{1})^{3}}~\frac{e^{-mr}}{r}~\frac{1}{m^{2}+\frac{\Gamma ^{2}}{4}} 
\nonumber \\
&+&\frac{1}{32 \pi M^{3}}\int_{0}^{\infty }dm~\frac{32k^{4}}{x_{-}\pi g(z_{1})^{7}}~%
\frac{e^{-mr}}{r}~\frac{1}{(m^{2}+\frac{\Gamma ^{2}}{4})^{2}}
\end{eqnarray}

The first part is the same as in the GRS model potential, whereas the second
one is the first correction that comes from the fact that $x_{-}$ is still
finite though very large. Note that the width of the ``resonance'' scales
like $e^{-3x}$, something that is compatible with the scaling law of the
masses ($m_{n}=n\pi k\frac{e^{-x}}{x_{-}}$), since we are working at the
region where $x_{-}\gg e^{2x}$, {\textit{i.e.}} $m_{n}\ll n\pi ke^{-3x}$.
The above integrals can be easily calculated in two interesting limits:

For $k^{-1}\ll r\ll r_{c}$ the potential is given approximately by: 
\begin{equation}
V_{ld}(r\ll r_{c})= -\frac{G_{N}}{r}~(1-\frac{e^{2x}}{x_{-}})
\end{equation}
where we have identified $G_{N}\equiv \frac{k}{32 \pi M^{3}}$ to obtain the normal
four dimensional Newtonian potential. Note that since $x_{-}\gg e^{2x}$, the $1/x_{-}$
term is indeed a small correction.

In the other limit, $r\gg r_{c}$, the integrand is only significant for
values of $\ m$ for which the $m^{2}$ term in the denominator of the
Breit-Wigner can be dropped and the potential becomes five dimensional: 
\begin{equation}
V_{ld}(r\gg r_{c})= -\frac{G_{N}r_{c}}{\pi r^{2}}~(1-\frac{2e^{2x}}{%
x_{-}})
\end{equation}

The fact that Newtonian gravity has been tested close to the present horizon
size require that for $k\sim M_{\mathrm{Pl}}$ we should have $x\gtrsim $
45-50. Finally, we note that if we take the $x_{-}\rightarrow \infty $ we recover
the GRS result:
\begin{equation}
\lim_{x_{-}\rightarrow \infty }V_{+--+}(r,x_{-})=V_{GRS}(r)
\end{equation}

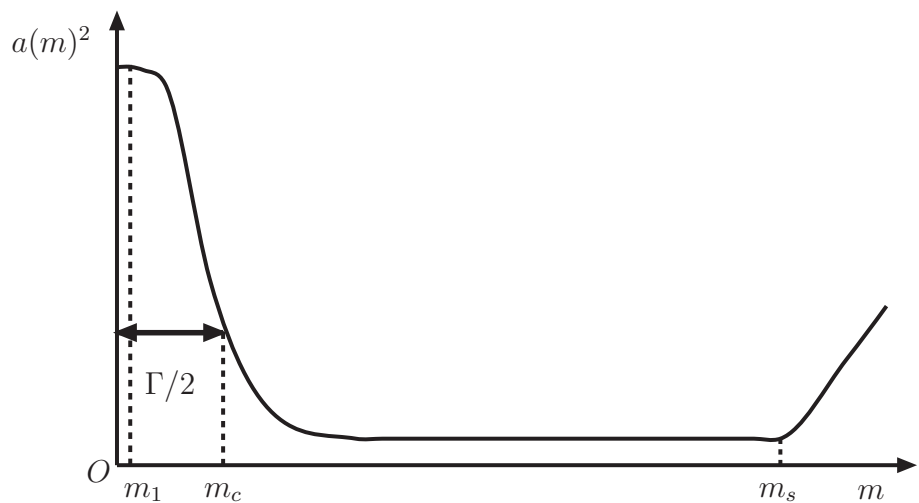
\begin{figure}[tbp]
\begin{center}
\begin{picture}(400,200)(-50,50)
\SetOffset(50,0)
\SetWidth{1.5}

\LongArrow(-50,80)(250,80)
\LongArrow(-50,80)(-50,250)
\Curve{(-50,230)(-40,229)(-32,224)(-15,150)(40,90)(50,90)(75,90)(100,90)(125,90)(150,90)(175,90)(190,90)(200,90)(225,120)(240,140)}
\DashLine(-45,80)(-45,231){2.5}
\DashLine(-10,80)(-10,130){2.5}
\DashLine(200,80)(200,90){2.5}

\Text(-90,240)[l]{$a(m)^2$}
\Text(-40,70)[c]{$m_{1}$}
\Text(-30,110)[c]{$\Gamma/2$}
\Text(230,70)[l]{$m$}
\Text(200,70)[c]{$m_{s}$}
\Text(-10,70)[c]{$m_{c}$}
\Text(-57,78)[c]{$O$}

\SetWidth{2}
\LongArrow(-50,130)(-10,130)
\LongArrow(-10,130)(-50,130)

\end{picture}
\end{center}
\caption{The behaviour of the coupling, $a(m),$ in the limit of $x_{-}\gg
e^{2x}$. Three regions of interest are indicated. The region $m>m_{s}$ gives
rise to short distance corrections. The $m_{1}\ll m\ll m_{c}$ region gives
rise to 4D gravity at intermediate distances and 5D gravity at ultra large
distances. For distances $r\gg m_{1}^{-1}$, the zero mode gives the dominant
contribution and thus we return to 4D gravity.}
\label{+--+coupling}
\end{figure}

\begin{itemize}
\item  \textbf{\ Back to 4D gravity}
\end{itemize}

As we have just seen, probing larger distances than $r_{c},$ the four dimensional
gravitational potential changes to a five dimensional one. This is the most significant
characteristic of the GRS model. In the case that $x_{-}$ is large compared
to $e^{2x}$ but still finite, there is another distinct region of interest,
namely $r\gg m_{1}^{-1}$. This follows from the fact that, in this limit,
the spectrum is still discrete. For distances larger than of the order of
the corresponding wavelength of the first KK mode, the contribution to
gravity from the KK tower is suppressed and thus the zero mode gives the
dominant contribution, leading to the four dimensional Newtonian potential again. In this
case the strength of the gravitational interaction is a small fraction of
the strength of the intermediate scale four dimensional gravity. More precisely, the
contribution of the massless graviton is $1/x_{-}$ suppressed and
thus vanishes when $x_{-}\rightarrow \infty $, something that is expected
since in this limit there is no nomalizable zero mode. The gravitational
potential in this case is: 
\begin{equation}
V_{4D}(r)=-\frac{1}{32 \pi M^{3}}~\frac{1}{r}~N_{0}^{2}\psi_{(0)}^{2}(0)=-
\frac{G_{N}}{r}~\frac{e^{2x}}{x_{-}}
\end{equation}

Obviously this four dimensional region disappears at the limit $x_{-}\rightarrow \infty $
since the spectrum becomes continuum and thus the five dimensional gravity ``window''
extents to infinity. We should note finally that for the values of $x$ that
we consider here this final modification of gravity occurs at distances well
above the present horizon.

\section{The Crystal Universe Model}

The last model we consider is the crystal universe introduced in \cite{Kaloper:2000et} (see also \cite{Nam:2000tz,Nam:2000ke}). It consists of an
infinite array of parallel three-branes in five dimensional $AdS$ space with cosmological
constant $\Lambda $. For simplicity we assume that the branes are
equidistant with distance, $l$, between two successive branes. Needless to
say, in this case all the positive tension branes are scale equivalent. The metric
ansatz that respects four dimensional Poincar\'{e} invariance is again given by (\ref{warp-metric}) where the $\sigma (y)$ function is constrained by the the Einstein
equations to have the sawtooth form: 
\begin{equation}
\sigma (y)=k\sum_{j=-\infty }^{+\infty }(y-2jl)\left[ 2\theta (y-2jl)-\theta
(y-(2j-1)l)-\theta (y-(2j+1)l)\right]
\end{equation}
The tensions of successive branes are required to be opposite and equal to $%
\pm \Lambda /k$, where $k=\sqrt{\frac{-\Lambda }{24M^{3}}}$ is a measure of
the curvature of the bulk and $M$ the five dimensional fundamental scale.

\begin{figure}[t]
\begin{center}
\begin{picture}(300,80)(0,50)
\SetOffset(0,-20)
\SetWidth{1.5}

\Line(-40,100)(340,100)
\CCirc(150,100){7}{Black}{Red}
\CCirc(10,100){7}{Black}{Red}
\CCirc(290,100){7}{Black}{Red}

\CCirc(80,100){7}{Black}{Green}
\CCirc(220,100){7}{Black}{Green}

\LongArrow(160,110)(210,110)
\LongArrow(210,110)(160,110)
\Curve{(185,110)(180,130)(170,140)}
\Text(168,140)[r]{$x=kl$}

\Text(80,75)[c]{$-$}
\Text(220,75)[c]{$-$}
\Text(150,75)[c]{$+$}
\Text(10,75)[c]{$+$}
\Text(290,75)[c]{$+$}

\Text(80,125)[c]{$-l$}
\Text(220,125)[c]{$l$}
\Text(10,125)[c]{$-2l$}
\Text(290,125)[c]{$2l$}

\Text(340,125)[c]{$\cdots$}
\Text(-40,125)[c]{$\cdots$}

\end{picture}
\end{center}
\caption{The Crystal Universe made up of an infinite array of $^{\prime
\prime }+^{\prime \prime }$ and $^{\prime \prime }-^{\prime \prime }$ branes
with lattice \mbox{
spacing $l$} and bulk curvature $k$.}
\label{crystal}
\end{figure}

We consider the general fluctuations around the  vacuum ansatz of
the form (\ref{fluct}) and decompose as usual the perturbation tensor
$h_{\mu \nu }(x,y)$ as: 
\begin{equation}
h_{\mu \nu }(x,y)=\int dm\,h_{\mu \nu }(m,x)\Psi (m,y)
\end{equation}
where $\Psi (m,y)$ is a complex function that arises from the linear
combination of the independent graviton polarizations. This leads to the
Schr\"{o}dinger equation (\ref{sch}) for the wavefunction $\hat{\Psi}(m,z)\equiv \Psi (m,y)e^{\sigma /2}$ with potential: 
\begin{equation}
V(z)=\frac{15k^{2}}{8[g(z)]^{2}}-\frac{3k}{2g(z)}\sum_{j=-\infty }^{+\infty
}(-)^{j}\delta (z-z_{i})
\end{equation}

The $z$ coordinates have been defined as usual to be the ones that make the
background metric conformally flat. In these coordinates the branes sit at
the points \mbox{$z_j=j\frac{e^{kl}-1}{k}\equiv jz_l$} and the function $g(z) $ is: 
\begin{equation}
g(z)=1+k\sum_{j=-\infty }^{+\infty }(z-z_{2j})\left[ 2\theta
(z-z_{2j})-\theta (z-z_{2j-1})-\theta (z-z_{2j+1})\right]
\end{equation}

The solution of the Schr\"{o}dinger equation for the wavefunctions for two
adjacent cells is given \cite{Kaloper:2000et} in terms of Hankel functions: 
\begin{equation}
\hat{\Psi}(m,z)=N_{m}\sqrt{\frac{g(z)}{k}}\left[ \left\{ 
\begin{array}{c}
1 \\ 
B \\ 
e^{2iqz_{l}} \\ 
e^{2iqz_{l}}B
\end{array}
\right\} H_{2}^{+}\left( \frac{m}{k}g(z)\right) +\left\{ 
\begin{array}{c}
A \\ 
C \\ 
e^{2iqz_{l}}A \\ 
e^{2iqz_{l}}C
\end{array}
\right\} H_{2}^{-}\left( \frac{m}{k}g(z)\right) \right]  \label{sol}
\end{equation}
in the regions $z\in \lbrack 0,z_{l}]$, $z\in \lbrack z_{l},2z_{l}]$, $z\in
\lbrack 2z_{l},3z_{l}]$ and $z\in \lbrack 3z_{l},4z_{l}]$ respectively. In the above expression $N_{m}$ is an
overall normalization constant, $q$ is the Bloch wave quasi-momentum and the
constants $A$,$B$,$C$ are given by: 
\begin{eqnarray}
A &=&\frac{(h_{1}^{+}h_{2}^{-}+h_{1}^{-}h_{2}^{+})\hat{h}%
_{2}^{+}-(h_{1}^{-}h_{2}^{+}-h_{1}^{+}h_{2}^{-})e^{2iqz_{l}}\hat{h}%
_{2}^{+}-2h_{1}^{+}h_{2}^{+}\hat{h}_{2}^{-}}{%
(h_{1}^{+}h_{2}^{-}+h_{1}^{-}h_{2}^{+})\hat{h}%
_{2}^{-}+(h_{1}^{-}h_{2}^{+}-h_{1}^{+}h_{2}^{-})e^{2iqz_{l}}\hat{h}%
_{2}^{-}-2h_{1}^{-}h_{2}^{-}\hat{h}_{2}^{+}} \\
B &=&\frac{h_{1}^{+}h_{2}^{-}+h_{1}^{-}h_{2}^{+}+2h_{1}^{-}h_{2}^{-}A}{%
h_{1}^{-}h_{2}^{+}-h_{1}^{+}h_{2}^{-}} \\
C &=&\frac{2h_{1}^{+}h_{2}^{+}+(h_{1}^{+}h_{2}^{-}+h_{1}^{-}h_{2}^{+})A}{%
h_{1}^{+}h_{2}^{-}-h_{1}^{-}h_{2}^{+}}
\end{eqnarray}
where $h_{n}^{\pm }\equiv H_{n}^{\pm }\left( \frac{m}{k}g(z_{l})\right) $
and $\hat{h}_{n}^{\pm }\equiv H_{n}^{\pm }\left( \frac{m}{k}\right) $. The
above coefficients of the Hankel functions were determined by the boundary
conditions, {\textit{i.e.}} continuity of the wavefunction at $z_{l}$, $2z_{l}$, $%
3z_{l} $, discontinuity of its first derivative at $z_{l}$, $3z_{l}$ and the
Bloch wave conditions relating the wavefunction at the edges of each
cell\footnote{The Bloch wave condition reads $\hat{\Psi}(z_i+2z_l)=e^{2iqz_l}\hat{\Psi}(z_i)$.}.
The last remaining boundary condition of the discontinuity of the first
derivative of the wavefunction at $2z_{l}$ gives us the band equation
connecting $q$ and $m$. We disagree with the relation given in \cite{Kaloper:2000et}
and instead we find: 
\begin{equation}
\cos(2qz_{l})=\frac{(j_{2}y_{1}+j_{1}y_{2})(\hat{\jmath}_{2}\hat{y}_{1}+\hat{%
\jmath}_{1}\hat{y}_{2})-2\hat{\jmath}_{1}\hat{\jmath}%
_{2}y_{1}y_{2}-2j_{1}j_{2}\hat{y}_{1}\hat{y}_{2}}{(j_{2}y_{1}-j_{1}y_{2})(%
\hat{\jmath}_{2}\hat{y}_{1}-\hat{\jmath}_{1}\hat{y}_{2})}\equiv f(m)
\label{band}
\end{equation}
where again $j_{n}\equiv J_{n}\left( \frac{m}{k}g(z_{l})\right) $, $%
y_{n}\equiv Y_{n}\left( \frac{m}{k}g(z_{l})\right) $ and $\hat{\jmath}%
_{n}\equiv J_{n}\left( \frac{m}{k}\right) $, $\hat{y}_{n}\equiv Y_{n}\left( 
\frac{m}{k}\right) $. This dispersion relation {\textit{always}} gives a
band at zero as is to be expected intuitively. Defining the parameter $%
x\equiv kl$, we see that for $x\lower3pt%
\hbox{$\, \buildrel > \over \sim \,
$}10$ we may reliably approximate the width of the zero mode band by: 
\begin{equation}
\Gamma _{0}=2\sqrt{2}ke^{-2x}
\end{equation}
while the separation of the zeroth and the first band is 
\begin{equation}
\Delta \Gamma _{1}=\xi _{1}ke^{-x}
\end{equation}
with $\xi _{1}$ the 1st root of $J_{1}(x)$. This characteristic behaviour of
the above widths with respect to $x$ implies that we can have a viable
multigravity scenario if $\Gamma _{0}^{-1}\lower3pt%
\hbox{$\, \buildrel >
\over \sim \, $}10^{26}\mathrm{cm}$ and at the same time $\Delta \Gamma
_{1}^{-1}\lower3pt\hbox{$\, \buildrel < \over \sim \, $}0.2\mathrm{mm}$. This
can happen for $k\sim M_{\mathrm{Pl}}$ and $x\approx 68$.

The band structure has the following form. The first bands have very narrow
widths $\Gamma _{i}$ with spacing $\Delta \Gamma _{i}$ between them ({\textit{i.e.}} between the $i$-th and the $(i-1)$-th band), which are
approximately: 
\begin{eqnarray}
\Gamma _{i} &= &\epsilon _{i}\xi _{i}^{2}ke^{-3x} \\
\Delta \Gamma _{i} &= &(\xi _{i}-\xi _{i-1})ke^{-x}
\end{eqnarray}
where $\xi _{2n+1}$ is the $(n+1)$-th root of $J_{1}(x)$, $\xi _{2n}$ is the
{n}-th root of $J_{2}(x)$, and 
\mbox {
$\epsilon_{2n+1}=\frac{\pi}{4}
\frac{Y_1(\xi_{2n+1})}{J_2(\xi_{2n+1})}$}, $\epsilon _{2n}=\frac{\pi }{4}%
\frac{Y_{2}(\xi _{2n})}{J_{1}(\xi _{2n})}$. As we move on to higher bands
their width is increasing and the spacing between them is decreasing. In the
limit $m\rightarrow \infty $, the spacing disappears, {\textit{i.e.}} $\Delta \Gamma
_{i}\rightarrow 0$, while their width tends to $\Gamma _{i}\rightarrow \frac{%
\pi }{2}ke^{-x}$. In this limit the function $f(m)$ in (\ref{band}) tends to 
$f(m)=\cos(2z_{l}m)$ and we have no forbidden zones.

In order to find the gravitational potential we should at first find how the
KK modes couple to matter and determine their spectral density. Instead of
working with wavefunctions in an infinite extra dimension with continuous
normalization, it is more convenient effectively to compactify the system to
a circle of length $2{\mathcal {N}}l$, by assuming that we have a finite crystal of ${\mathcal {N}}$
cells $[2jl,2(j+1)l]$. We shall find the gravitational potential is
independent of ${\mathcal {N}}$, so we will be able to take the limit ${\mathcal {N}}\rightarrow
\infty $ and decompactify the geometry.

In this regularization scheme we can readily see that the ever-present zero
mode, corresponding to the four dimensional graviton, has wavefunction: 
\begin{equation}
\hat{\Psi}(0,z)=\frac{\sqrt{k/{\mathcal {N}}}}{[g(z)]^{3/2}}  \label{grav}
\end{equation}
and its coupling to matter on the central positive brane is: 
\begin{equation}
a(0)=\frac{1}{\sqrt{\mathcal
{N}}}\frac{\sqrt{k}}{2\sqrt{2M^{3}}}\equiv \frac{1}{\sqrt{\mathcal
{N}}}\frac{1}{2 \sqrt{2}M_{\ast }}
\end{equation}

The wavefunction (\ref{grav}) is a smooth limit of (\ref{sol}) for $m\rightarrow 0$. Indeed, for large $x$ we find the normalization constant in
(\ref{sol}) in the region of the zero mode band is approximately: 
\begin{equation}
N_{m}=\frac{\sqrt{2}}{z_{l}^{2}m}\frac{1}{\sqrt{\mathcal {N}}}
\end{equation}

The approximation of the Hankel functions with their first term of their
power series proves our previous claim. Furthermore, we can see that the KK
states in the whole zero mode band for large $x$ will couple to matter in
the central positive brane as: 
\begin{equation}
a(m)=\frac{1}{\sqrt{\mathcal {N}}}\frac{1}{2 \sqrt{2}M_{\ast }}e^{i\phi (m)}
\end{equation}
with $\phi (m)$ an unimportant $m$-dependent phase that does not appear in
observable quantities.

In the limit $m\rightarrow \infty $ the Hankel functions can be approximated
by their asymptotic form. The wavefunction is then almost constant in
z-space and the KK states couple to matter on the central positive brane as: 
\begin{equation}
a(m)=\frac{1}{\sqrt{\mathcal {N}}}\frac{1}{4e^{x/2}M_{\ast }}e^{i\phi (m)}
\end{equation}
where again $\phi (m)$ is an unimportant $m$-dependent phase. The couplings in
the low lying bands interpolate between the above limiting cases.

The standard procedure to calculate the spectral density is at first to find
the number of states with masses smaller than $m$ and then differentiate
with respect to $m$. In our regularization prescription it is
straightforward to count the states in the quasi-momentum space with masses
lighter that $m$. The latter is simply: 
\begin{equation}
N_{tot}(m)=\frac{{\mathcal {N}}z_l}{\pi}q
\end{equation}
so the spectral density is its derivative: 
\begin{equation}
\rho(m)=\frac{dN_{tot}(m)}{dm}=\frac{{\mathcal {N}}z_l}{\pi} \frac{dq}{dm}= (\mp) \frac{{\mathcal {N}}}{2 \pi} \frac{1}{\sqrt{1-f(m)^2}} \frac{df(m)}{dm}
\end{equation}
where the plus and minus signs will succeed each other for neighbouring
bands, starting with minus for the zero mode band. For the zero band we can
have a reliable approximation for large $x$ of the above function: 
\begin{equation}
\rho(m)={\mathcal {N}} \frac{1}{\pi \Gamma_0} \frac{1}{\sqrt{1 -
\left(m/\Gamma_0\right)^2}}
\end{equation}

\begin{figure}[t]
\begin{center}
\begin{picture}(300,160)(0,40)
\SetScale{0.7}
\SetOffset(70,20)
\SetWidth{1.5}

\LongArrow(-50,30)(300,30)
\LongArrow(-50,30)(-50,250)
\DashLine(30,30)(30,250){3}

\Text(-45,170)[r]{\Large {$\frac{\rho(m)}{\mathcal {N}}$}}
\Text(-45,90)[r]{\Large {$\frac{1}{\pi \Gamma_0}$}}
\Text(210,30)[l]{$m$}
\Text(200,60)[l]{$\cdots$}

\Curve{(-50,130)(-35,131)(-15,134)(-8,136)(0,141)(5,146)(15,175)(25,250)}

\Text(-5,10)[c]{$\Gamma_0$}
\Text(117,10)[c]{$\Gamma_1$}
\Text(174,10)[c]{$\Gamma_2$}

\SetOffset(90,20)
\DashLine(120,30)(120,250){3}
\DashLine(150,30)(150,250){3}
\DashLine(200,30)(200,250){3}
\DashLine(230,30)(230,250){3}
\Curve{(125,250)(126,130)(126.5,110)(127,104)(128,102)(130,101)(135,100)(140,101)(142,102)(143,104)(143.5,110)(144,130)(145,250)}
\Curve{(205,250)(206,110)(206.5,90)(207,84)(208,82)(210,81)(215,80)(220,81)(222,82)(223,84)(223.5,90)(224,110)(225,250)}

\end{picture}
\end{center}
\caption{Spectral density for the first three bands in the crystalline 
model.}
\label{density}
\end{figure}
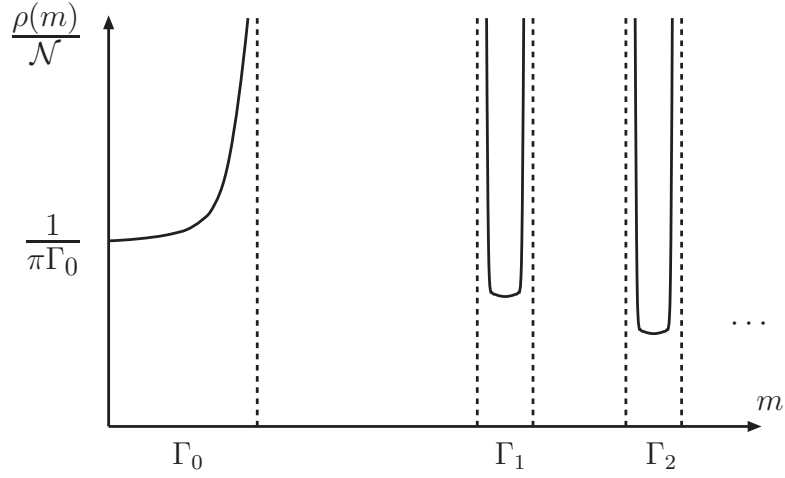

Thus, the spectral density diverges as $m \rightarrow \Gamma_0$ but the
divergence is integrable and does not cause any problem to the following
calculations. Actually, the spectral density diverges at the edges of every
band since at these points $\sqrt{1-f(m)^2} \rightarrow 0$ while $\frac{df(m)%
}{dm} \neq 0$. The only point that this does not happen is at $m=0$ where $%
\frac{df(m)}{dm} \rightarrow 0$ in such a way that the result is finite.

In the limit $m\rightarrow \infty $ the bands disappear and the spectrum has
constant spectral density: 
\begin{equation}
\rho (m)={\mathcal {N}}\frac{e^{x}}{k\pi }
\end{equation}

The gravitational potential, taking into account the Yukawa suppressions of
the KK states, is simply: 
\begin{equation}
V(r)=-\int_{0}^{\infty }dm{|a(m)|^{2} \over 4\pi}\rho (m)\frac{e^{-mr}}{r}
\end{equation}

For distances $r\gg \Delta \Gamma _{1}^{-1}$ we have the effective
potential: 
\begin{eqnarray}
V(r) &=&-\frac{1}{32 \pi M_{* }^{2}}\frac{1}{r}\int_{0}^{\Gamma _{0}}dm\frac{
\rho (m)}{{\mathcal {N}}}e^{-mr}=-\frac{1}{32 \pi^2 M_{\ast }^{2}}\frac{1}{r}\int_{0}^{1}d\xi 
\frac{e^{-(\Gamma _{0}r)\xi }}{\sqrt{1-\xi ^{2}}}  \nonumber \\
&=&-\frac{1}{64 \pi M_{\ast }^{2}}\frac{1}{r}[I_{0}(\Gamma _{0}r)-L_{0}(\Gamma
_{0}r)]
\end{eqnarray}
where $I_{0}$ is the $0$-th modified Bessel function and $L_{0}$ is the $0$-th modified Struve function. The gravitational potential is ${\mathcal {N} }$-independent as expected. For distances $\Delta \Gamma _{1}^{-1}\ll r\ll
\Gamma _{0}^{-1}$ the above functions tend to $I_{0}(\Gamma
_{0}r)\rightarrow 1$ and $L_{0}(\Gamma _{0}r)\rightarrow 0$, so we recover
the four dimensional Newton law with $G_{N}=\frac{1}{64 \pi M_{\ast }^{2}}$: 
\begin{equation}
V(r\ll \Gamma _{0}^{-1})=-\frac{G_{N}}{r}
\end{equation}

On the other hand, for distances $r \gg \Gamma_0^{-1}$ the asymptotic
expansion of the difference of the modified Bessel and Struve functions
gives a five dimensional Newton's law: 
\begin{equation}
V(r \gg \Gamma_0^{-1})= -\frac{2G_N}{\pi \Gamma_0} \frac{1}{r^2}
\end{equation}

In case we take the crystal to be finite, there will appear a region for $%
r\gg m_{1}^{-1}$ where gravity will turn again to four dimensional. However, as explained
in the previous section, this region is well above the universe horizon and
thus of no phenomenological interest.

Finally, for distances $r\lower3pt\hbox{$\, \buildrel < \over \sim \, $}%
\Delta \Gamma _{1}^{-1}$ we will start to feel the short distance Yukawa
type modifications to gravity due to the presence of the bands above the
zeroth one. As in the zero hierarchy bigravity and $^{\prime \prime
}+--+^{\prime \prime }$ multigravity model we expect these corrections
to be important at scales of the order of the Planck length. In summary, the
physics of the crystalline model is very similar to the GRS model as
far as modifications of gravity due to  tensor modes is concerned. In
both cases we have a crossover from four dimensional to five
dimensional gravity at cosmological scales.

\section{Propagator structure}
\label{sectFLprop}

We have shown that in models in which negative tension branes are sandwiched
between positive tension branes there are anomalously light KK states that
can lead to modifications of gravity. So far we have implicitly assumed that
the gravitational force generated by a massive graviton is identical to that
generated by a massless one but this is not, in general, the case because
the massive graviton has additional degrees of freedom which do not decouple
in the massless limit. As a result the interaction generated by a massive
graviton violates the normal four dimensional relation between the gravitational
interactions of matter and light in Einsteinian gravity \cite{vanDam:1970vg,Zakharov}  . This may be seen explicitly from the form
of the massive and massless graviton propagators.

In  more detail, the form of the massless graviton propagator in flat
spacetime (in momentum space) has the form: 
\begin{equation}
G^{\mu\nu;\alpha\beta}=\frac{1}{2}~ \frac{\left(\eta^{\mu\alpha}\eta^{\nu\beta} +
\eta^{\nu\alpha}\eta^{\mu\beta}\right) - \eta^{\mu\nu}\eta^{\alpha\beta} 
  }{p^2 }+ \cdots
\end{equation}
where we have omitted terms that do not contribute when contracted
with a conserved energy momentum tensor $T_{\mu \nu}$.

On the other hand, the 
four-dimensional  massive graviton propagator (in momentum space) has the form: 
\begin{equation}
G^{\mu\nu;\alpha\beta}=\frac{1}{2} ~ \frac{\left(\eta^{\mu\alpha}\eta^{\nu\beta} +
\eta^{\nu\alpha}\eta^{\mu\beta}\right) - \frac{2}{3}~ \eta^{\mu\nu}\eta^{\alpha\beta} 
 }{p^2 +m^2}+\cdots
\end{equation}

The difference between the two propagators 
\begin{equation}
\delta G^{\mu \nu ,\alpha \beta }=\frac{1}{6}~\frac{g^{\mu \nu }g^{\alpha \beta }}{p^{2}+m^{2} } +\cdots
\end{equation}
is due to the additional helicity components needed for a massive graviton.
There are three additional components needed. Two, corresponding to the
graviphotons, decouple at low energies as they are derivatively coupled to
the conserved energy-momentum tensor. The third, corresponding to a scalar
component, does not decouple and is responsible for $\delta G.$ The
contribution of $\delta G$ to the one-graviton exchange amplitude for any
two four dimensional sources $T_{\mu \nu }^{1}$ and $T_{\alpha \beta }^{2}$
vanishes for light for which $T=T_{\mu }^{\mu }=0$ but not for matter
because $T=T_{0}^{0}\neq 0$, hence the normal four dimensional gravity relation between
the interactions of matter and light is violated as was first pointed
out by \cite{Dvali:2000rv}.  For example, if
gravity is due to the exchange of a massive spin 2 particle and one
insists in the validity of Newton's law for static sources,
then the deflection angle of light would be $25\%$ smaller than if it
corresponds to the  exchange of the massless graviton.  
The fact that the  bending of the light by the sun agrees with 
the prediction of  Einstein's theory  to $0.02\%$  accuracy,  rules out
the possibility that gravity is due to massive graviton exchange
irrespective  of
how small its mass is. Of course there remains the possibility 
that a small
fraction of the gravitational interactions are associated with a
massive graviton component in the presence of a dominant massless
graviton component. 
 This can be realized by having an ultralight spin-2 particle with a
very small coupling compared to graviton's. Such an example could be
the massless - massive graviton system in the asymmetric
bigravity model case.

However, in our discussion so far we have omitted an important
gravitational mode, the moduli (radions) corresponding to moving
negative tension branes. These radions as we shall see in the following 
chapter cancel the troublesome additional polarization states of the
massive gravitons \cite{Gregory:2000iu,Csaki:2000ei,Pilo:2000et,Kogan:2001qx} . This restores the phenomenologically favoured
propagator structure of the above models at intermediate
distances. However, these fields must have unusual properties
because it is known that a normal scalar contributes with the same
sign as $\delta G$ and therefore cannot cancel it. An explicit
computation presented in the following chapter reveals
that it  has a negative kinetic energy term leading to a contribution
which cancels the troublesome $\delta G$ contribution. As a result the
theory indeed gives rise to normal four dimensional gravity at
intermediate distances, but at large distances the ghost radions
dominate giving in all cases scalar antigravity \cite{Gregory:2000iu}. The
presence of the ghost radion field is worrying for the consistency of the
theory. We will elaborate on this issue in the following chapter.

\chapter{Radions and dilaton}

In this chapter we are interested in describing the dynamics of the  
scalar gravitational perturbations for a system with branes in a warped
background in five dimensions. These excitations describe  
the effect of the fluctuation of the size of the extra dimension and/or  
of the relative positions of the branes. We will distinguish   
these two kinds of modes by calling the former dilaton 
\cite{Charmousis:2000rg} and the  
latter radions \cite{Charmousis:2000rg,Pilo:2000et,Kogan:2001qx}.  
An orbifold symmetry $Z_2$ acting on the extra dimensional coordinate 
as $y \to -y$ is also imposed. When the topology of the extra dimension 
is $S^1$, the compact case, the $Z_2$ action has two fixed points
$y=0, \, L$. If branes are sitting on the orbifold fixed points, the
corresponding radions are frozen. For example in a system of three
three-branes where two of them are sitting on fixed points at  $y=0,
\, L$,  we are left with just 
one radion field corresponding to the fluctuation of the position of the 
third brane and the dilaton corresponding to the fluctuation of the 
size of the orbifold.   
  
For the sake of simplicity we will only consider a three flat  
three-branes system embedded in $AdS_5$ with in general different
cosmological constants between the branes. This will be sufficient to
clarify the role of the radion and dilaton in the multibrane
models. We will calculate the full gravitational effective action for
this general model, compute the kinetic energy of the moduli and study
the nature of gravity on the branes. As we have already noted the radion excitations play an important role in the context of   
multigravity models. They restore the
correct propagator structure for the massive graviton excitations. However, the radions corresponding to negative tension branes
are physical ghosts and therefore disastrous for the theory. 

Afterwards, we will concentrate on the dilaton of a five dimensional
model with identical four geometries and calculate its mass
for  bulk dynamics involving
arbitrary minimally coupled scalar fields. A useful tool
for this calculation is the sum rules derived by Gibbons, Kallosh and
Linde  \cite{Gibbons:2001tf} which we briefly review.  We find a generic formula
relating the dilaton mass with the sum of the tensions of the
branes and the curvature of the four-geometry. Demanding that this
mass is not tachyonic we can derive the necessary condition for
stabilizing the overall size of the system.

\section{The general three three-brane system}   
 
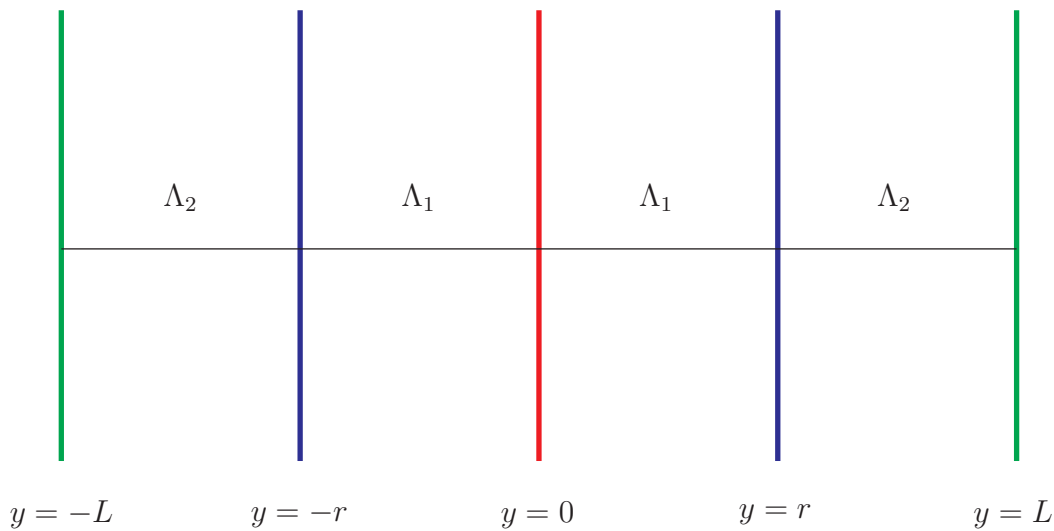
\begin{figure}[t]  
\begin{center}  
\begin{picture}(300,200)(0,50)  
\SetOffset(0,-20)  
  
\SetWidth{2}  
\SetColor{Red}  
\Line(150,80)(150,250)  
  
\SetColor{Blue}  
\Line(240,80)(240,250)  
\Line(60,80)(60,250)  
  
\SetColor{Green}  
\Line(-30,80)(-30,250)  
\Line(330,80)(330,250)  
  
\SetWidth{0.5}  
\SetColor{Black}  
\Line(-30,160)(330,160)

\Text(150,60)[c]{$y=0$}  
\Text(60,60)[c]{$y=-r$}  
\Text(240,60)[c]{$y=r$}  
\Text(-30,60)[c]{$y=-L$}  
\Text(330,60)[c]{$y=L$}  
\Text(195,180)[c]{$\Lambda_1$}  
\Text(285,180)[c]{$\Lambda_2$}  
\Text(105,180)[c]{$\Lambda_1$}  
\Text(15,180)[c]{$\Lambda_2$}

\end{picture}  
\end{center}  
  
\caption{General three three-brane model on an orbifold with different 
cosmological constants between the branes.}  
\label{model}  
\end{figure}

We will consider a three three-brane model on an $S^1/Z_2$ 
orbifold. Two of the branes sit on the orbifold fixed points $y=y_0=0$, 
$y=y_2=L$ respectively. A  third brane is sandwiched in between at 
position $y=y_1=r$ as in Fig.\ref{model}. In each region between the 
branes the space is $AdS_5$ and  in general the various $AdS_5$ regions  
have  different cosmological constants $\Lambda_1$, $\Lambda_2$. The 
action describing the above system is:   
\begin{equation} 
{\cal S}_G = \int d^4xdy \sqrt{-G} \left[ 2 M^3 \, R \,- \, \Lambda(y)  
\, - \, \sum_i V_i \, \delta(y - y_i)  \,   
\frac{\sqrt{-\hat{G}^{(i)}}}{ \sqrt{-G}} \right] 
\label{act}  
\end{equation} 
where $M$ is the five dimensional Planck mass, $V_i$ the tensions of 
the gravitating branes and $\hat{G}^{(i)}_{\mu\nu}$ the induced metric
on the branes. The orbifold symmetry $y \to -y $ is imposed.

We seek a background static solution of Einstein equations for the 
following four dimensional Poincar\'{e} invariant metric ansatz: 
\begin{equation} 
ds^2 = a^2(y) \, \eta_{\mu \nu} dx^\mu dx^\nu \, + \, dy^2    
\label{back}  
\end{equation} 
 
The solution for the warp factor has the usual exponential form:   
\begin{equation} 
a(y) = \begin{cases} e^{-k_1 y } &  ,0 < y < r  \\[0.3cm]  
e^{-k_2 y + r(k_2 - k_1)} & ,r < y < L   
\end{cases} 
\label{solu}  
\end{equation} 
where $k_1$ and $k_2$ are the curvatures of the bulk in the two 
regions and are related to the bulk cosmological constants as:  
\begin{equation}  
k_1^2 = -\frac{\Lambda_1}{24 M^3}  \, , \qquad  
k_2^2= -\frac{\Lambda_2}{24 M^3} 
\end{equation} 
 
Moreover, the Einstein equations impose the following fine tuning 
between the brane tensions and the bulk cosmological constants: 
\begin{equation} 
V_0=24 M^3 \, k_1 \, , \qquad V_1=24 M^3 \frac{(k_2-k_1)}{ 2} \, , 
\qquad V_2=-24 M^3 \, k_2 
\end{equation}  
 
It is straightforward to recover some special models 
that have been considered in the literature. The RS1 model is obtained 
for $k_1=k_2$ where the intermediate brane is absent (zero tension).  
For $k_1>0$ and $k_2=-k_1$ we get the $''+-+''$ bigravity model 
\cite{Kogan:2000wc,Mouslopoulos:2000er}. For $k_1>0$ and $k_2>k_1$ we 
get the $''++-''$ brane model considered in \cite{Kogan:2000xc}. In 
the decompactification limit where $L \rightarrow \infty$ we get also 
two interesting models: For  $k_1>0$ and  $k_2=0$ we obtain the 
GRS model \cite{Gregory:2000jc} and 
for  $k_1>0$ and $k_2=0>k_1$ the non-zero tension version 
\cite{Pilo:2000et} of the model considered in \cite{Lykken:2000nb}.

\section{Effective action}

Our purpose is to study fluctuations of the background (\ref{back}).  
The first important observation is that there exists a generalization of   
Gaussian normal coordinates such that  in the perturbed geometry the $g_{\mu y}$ components of the metric are zero and at the same time the embedding  
of branes is still  described by $y=0, \, y=r$ and $y=L$  (see for instance  
the  Appendix of \cite{Pilo:2000et}). In that case the metric takes the form:
\begin{equation}  
ds^2 = g_{\mu \nu}(x,y) \,  dx^\mu dx^\nu  + g_{yy}(x,y) \, dy^2 \quad .  
\end{equation}  
  
When analyzed from a four dimensional point of view, in each region,  perturbations are of  
of three types.  
\begin{itemize}  
\item  
{\bf Spin two}:  
 
\no 
Tensor-like perturbation $\tilde{h}_{\mu\nu}(x,y)$ corresponding to massive   
(and massless) four dimensional gravitons\footnote{We denote
$\tilde{h}_{\mu\nu}$ the conformal perturbation in contrast to the previous
chapter graviton perturbation $h_{\mu\nu}$. Their relation is $h_{\mu\nu}=a^2\tilde{h}_{\mu\nu}$.}:   
\begin{equation}  
ds^2 = a^2(y) \left[\eta_{\mu \nu} + \tilde{h}_{\mu \nu}(x,y) \right] \, dx^\mu dx^\nu  
+ dy^2 
\end{equation}  
\item  
{\bf Spin zero: Dilaton}   
 
\no 
Scalar perturbation $f_1(x)$ corresponding to the fluctuation of the size of the orbifold \cite{Charmousis:2000rg}: 
\begin{equation}  
ds^2 = a^2(y) \left[1+ Q(y) f_1(x) \right] \eta_{\mu \nu} \, dx^\mu dx^\nu   
+ \left[1+ q(y) f_1(x) \right] dy^2 
\end{equation}   
\item  
{\bf Spin zero: Radion}  
 
\no 
Scalar perturbation $f_2(x)$ corresponding to a fluctuating distance   
of the intermediate  brane \cite{Pilo:2000et}:  
\begin{equation}  
ds^2 = a^2(y) \left[ \left(1+ B(y) f_2(x) \right) \eta_{\mu \nu} +    
2 \epsilon(y) \, \de_\mu \de_\nu f_2(x) \right]  
dx^\mu dx^\nu   
+ \left[1+  A(y) \,  f_2(x) \right] dy^2 
\end{equation}   
\end{itemize}   

In the above decomposition we should note that we have kept only the zero modes of the radion and the dilaton. Their massive KK-states are ``eaten'' by the graviton's corresponding modes in order to give them mass, in a geometrical realization of the Higgs phenomenon. The function $\epsilon(y)$ is needed in order that the metric satisfies the Israel junction
conditions in the presence of moving branes. As it will be shown later, the values of 
$\de_y \epsilon$ at $y=0, \, r, L$ are gauge invariant. When $k_1 \neq k_2$ a non-trivial $\epsilon$ is required.

The generic perturbation can be written as:   
\begin{equation}  
\begin{split}  
ds^2 =& a^2(y) \left\{ \left[1+ \varphi_1(x,y) \right] \eta_{\mu \nu} + 2   
\epsilon(y) \, \de_\mu \de_\nu f_2(x) +   
\tilde{h}_{\mu \nu}(x,y) \right \} dx^\mu dx^\nu \,   \\  
& +\left[1+  \varphi_2(x,y) \right] dy^2 
\end{split}  
\label{pert}  
\end{equation}  
where   
\begin{equation}  
\begin{gathered}  
\varphi_1(x,y)  =  Q(y) f_1(x) + B(y) f_2(x)\\  
\varphi_2(x,y)  = q(y) f_1(x) + A(y) f_2(x)     
\end{gathered} 
\label{ans}  
\end{equation}  

Given the expression (\ref{solu}) for $a$, the Israel junctions conditions (continuity of metric components and jump conditions) at $y=0, r, L$, simply require that $A, \, B, \,  \de_y \epsilon, \, Q, \, q$ are  
continuous there \cite{Pilo:2000et}.   
The four dimensional effective action ${\cal S}_{eff}$ for the various modes is obtained inserting   
the ansatz (\ref{ans}) in the action (\ref{act}) and integrating out $y$. So far the functions $A, \, B, \, \epsilon, \, Q, \,   
q$ have not been specified, however imposing that ${\cal S}_{eff}$ contains no   
mixing terms among $\tilde{h}$ and  $f_i$ one determines $Q, \, q$ and finds
$A$ in terms of $B$ (see Appendix A). The latter satisfies:  
\begin{gather}  
\frac{d}{dy} \left(B a^2 \right) + 2 a^{-1} \, \frac{da}{dy} \, \frac{d}{dy}   
\left(a^4 \de_y \epsilon \right) = 0  \label{bdiff} \\  
 \int_{- L}^{L} dy \; a \left(\frac{da}{dy} \right)^{-1} \,   
\frac{dB}{dy} = 0 \label{nomix}  
\end{gather}  

The condition that  $\tilde{h}$ and  $f_i$ do not mix between them allows us to give a clear particle interpretation for the quanta of  $\tilde{h}$ and  $f_i$. However, despite the fact that we have chosen that gauge, the results that we obtain are bound to be the same as in the general case where there would be quadratic mixing between the modes, because of the five dimensional gauge invariance of the theory. As a consequence of the no-mixing conditions the linearized Einstein equations  
for (\ref{ans}) will consist in a set independent equations for the graviton  
and the scalars.  
  
The effective Lagrangian reads (see Appendix A):  
\begin{equation}  
\begin{split}  
{\cal S}_{eff} &= \int d^4 x  \, {\cal L}_{eff} =  \int d^4 x  \, \left(  
{\cal L}_{Grav} + {\cal L}_{Scal} \right) \\  
{\cal L}_{eff} &=  2 M^3 \int_{- L}^{L} dy  \, \Big \{a^2    
{\cal L}^{(0)}_{PF}(\tilde{h}) +   
\frac{a^4}{4} \left[(\de_y \tilde{h})^2 - \de_y \tilde{h}_{\mu \nu} \, \de_y \tilde{h}^{\mu \nu}   
\right] + {\cal L}_{Scal}\Big \} 
\end{split}  
\label{eff1}  
\end{equation}  
with  
\begin{equation}  
\begin{split}  
{\cal L}_{Scal} &= {\cal K}_1 \, f_1 \Box f_1 +  {\cal K}_2 \, f_2 \Box f_2 \\  
 {\cal K}_1 =  2 M^3 \frac{3}{2} c^2\int_{- L}^{L} a^{-2}   
\, dy \, ,& \quad   
{\cal K}_2 = - 2 M^3 \, \frac{3}{4} \int_{- L}^{L} a   
\left(\frac{da}{dy} \right)^{-1} \, \frac{d}{dy} \left( B^2 a^2 \right) \, dy
\end{split}  
\label{kt}  
\end{equation}  
where $c$ is an integration constant.

In (\ref{eff1}), the spin-2 part,  as expected,  contains the four dimensional   
Pauli-Fierz Lagrangian ${\cal L}^{(0)}_{PF}(\tilde{h})$ for the
gravitons without the mass term (see Appendix A), plus a mass term  
coming from the dimensional reduction. In the scalar part ${\cal L}_{Scal}$  
the mass terms are zero since $f_i$ are moduli fields. The indices in
the above expressions are raised and lowered with $\eta_{\mu\nu}$. Notice that after the  
no-mix conditions are enforced, the effective Lagrangian  
contains the undetermined function $\epsilon (y)$.  
 
The metric ansatz $G_{{}_{MN}}$ in (\ref{ans}) is related to a special  
coordinate choice, nevertheless, a residual  
gauge (coordinate) invariance is still present. Consider the class of  
infinitesimal coordinate transformations  $x^M \to {x^\prime}^M= x^M +  
\xi^M(x)$ such that  
the transformed metric ${G_{{}_{MN}}}^\prime = G_{{}_{MN}} + \delta  
G_{{}_{MN}}$ retains the original form (\ref{pert}) up to a  
redefinition of the functions $q,Q,A,B, \epsilon$ and the dilaton 
and the radion field. Consistency with the orbifold geometry and the  
requirement 
that the brane in $y=r$ is kept fixed by the diffeomorphism lead to $ 
\xi^5(x,0) = \xi^5(x,r) = \xi^5(x,L) =0 $. From  
\begin{equation} 
\delta G_{{}_{MN}} = - \xi^A \, \de_A  G_{{}_{MN}} - \de_{{}_M} \xi^A \,  
 G_{{}_{AN}} - \de_{{}_N} \xi^A \, G_{{}_{MA}}
\end{equation} 
one can show that $\xi^M$ has to be of the form: 
\begin{equation} 
\begin{gathered} 
\xi^\mu(x,y) = \hat{\xi}^\mu(x)  - W(y) \, \eta^{\mu \nu} \,  
\de_\nu f_2(x) 
\, , \qquad \xi^5(x,y) = a^2 W^\prime(y) \, f_2(x)  \\ 
\text{with } ~ W^\prime(0) =  W^\prime(r) =  W^\prime(L) = 0  
\end{gathered} 
\end{equation} 

The case $W = 0$ corresponds the familiar pure four dimensional diffeomorphisms, under which $\tilde{h}_{\mu \nu}$  
transforms as spin two field, $f_i$ as scalars and $q,Q,A,B, \epsilon$ are 
left unchanged. On the contrary the case $\hat{\xi}^\mu = 0, \; W \neq 0 $ is 
relic of five dimensional diffeomorphisms and one can check that $q,Q,A,B, \epsilon$ are not 
invariant and in particular $\delta \epsilon = W$.   
As a result, the values of $\de_y \epsilon$ in  
$0$, $r$ and $L$ are gauge independent and this renders   
${\cal L}_{eff}$ free from any gauge ambiguities.

\section{Scalars Kinetic Energy}  
  
\subsection{The compact case}  
In this section we will focus on the part of the effective Lagrangian 
involving the scalars and concentrate on the dilaton and radion kinetic 
term coefficients  ${\cal K}_1$ and ${\cal K}_2$. In particular we are  
interested in the cases when the radion becomes a ghost field, 
\textit{i.e.} ${\cal K}_2<0$. Firstly, it is trivial to obtain the
dilaton kinetic term ${\cal K}_1$ by carrying out the first integral of (\ref{kt}):  
\begin{equation}  
{\cal 
K}_1=3c^2M^3\left[\frac{a^{-2}(r)-1}{k_1}+\frac{a^{-2}(L)-a^{-2}(r)}{k_2}\right]  
\label{K1} 
\end{equation}

It turns out that for any possible values of $k_1$, $k_2$ and $r$, $L$  
the above quantity is positive definite. The radion kinetic term on 
the other hand is more involved.  
Integrating (\ref{bdiff}) we  get the  radion wavefunction for the  
two regions ($y>0$):  
\begin{equation}  
B(y)=   
\begin{cases}    
c_1 \, a^{-2} + 2 k_1 \, \de_y \epsilon \, a^2  &  ,0 < y < r \\[0.3cm]   
c_2 \, a^{-2} + 2 k_2 \, \de_y \epsilon \, a^2  &  ,r < y < L   
\end{cases} 
\label{B}   
\end{equation}  
where $c_1$ and $c_2$ are integration constants. The orbifold boundary   
conditions demand that $\de_y \epsilon(0) =\de_y \epsilon(L)= 0$ since  
$\epsilon$ is an even function of $y$. From the non-mixing conditions  
for radion and dilaton (\ref{nomix}) and the  
continuity of $B$ we are able to determine $c_2$ and $\de_y  
\epsilon(r)$ as the following:  
\begin{gather}   
c_2=c_1 \frac{k_2}{k_1}~\frac{a^2(r)-1}{\left(\frac{a(r)}{a(L)}\right)^2-1}\\   
\epsilon^\prime(r)=\frac{c_1k_2}{2k_1(k_2-k_1)a^4(r)}\left[\frac{k_1}{k_2}- 
\frac{a^2(r)-1}{\left(\frac{a(r)}{ a(L)}\right)^2-1}\right]  
\end{gather}   

Therefore, the values of the radion  wavefunction $B$ at the brane positions 
are given by the following expressions:  
\begin{gather}  
B(0)=c_1 \\  
B(r)=\frac{c_1k_2}{(k_2-k_1)} \frac{1-a^2(L)}{a^2(L)\left[\left( 
\frac{a(r)}{ a(L)}\right)^2-1\right]} \\  
B(L)=\frac{c_1k_2}{k_1}\frac{a^2(r)-1}{a^2(L)\left[\left(\frac{a(r)}{  
a(L)}\right)^2-1\right]} \; 
\end{gather}  

Thus, we can carry out the second integral on   
(\ref{kt}) to find  the radion kinetic term coefficient:  
\begin{align}   
{\cal K}_2&= 3 M^3  \left[\left(\frac{1}{  k_1} - \frac{1}{ k_2} \right)  
 B^2(r) \, a^2(r) + \frac{1}{   
k_2} \, B^2(L) \, a^2(L) - \frac{1}{ k_1} \, B^2(0)\right] \nonumber \\  
&= \frac{3 M^3c_1^2}{ k_1}  \left\{ \frac{k_2}{  
(k_2-k_1)}\frac{a^2(r)(a^2(L)-1)^2}{a^4(L)\left[\left(\frac{a(r)}{ 
a(L)}\right)^2-1\right]^2}+\frac{k_2}{  
k_1}\frac{(a^2(r)-1)^2}{a^2(L)\left[\left(\frac{a(r)}{  
a(L)}\right)^2-1\right]^2}-1 \right\}   
\label{K2} 
\end{align}  

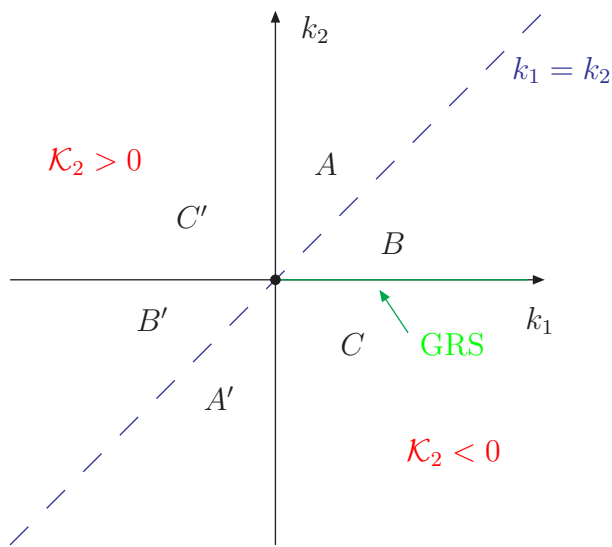
\begin{figure}[t]  
\begin{center}  
\begin{picture}(200,200)(0,0)  
\LongArrow(0,100)(200,100)   
\LongArrow(100,0)(100,200)  
\SetColor{Blue}  
\Text(110,190)[lb]{$k_2$}  
\Text(195,80)[lb]{$k_1$}  
\DashLine(0,0)(200,200){10}  
\Text(190,185)[lt]{\Blue{$k_1=k_2$}}  
\Text(50,140)[rb]{\color{red}${\cal K}_2 >0$}  
\Text(150,40)[lt]{\color{red}${\cal K}_2 <0$}  
\SetColor{Red}  
\Text(155,80)[lt]{\color{green} GRS}  
\SetColor{Green} 
\Line(100,100)(195,100) 
\LongArrow(150,80)(140,95)  
\SetColor{Black}  
\Vertex(100,100){2} 
\Text(115,140)[lb]{$A$}  
\Text(85,60)[rt]{$A'$} 
\Text(140,110)[lb]{$B$} 
\Text(60,90)[rt]{$B'$} 
\Text(125,80)[lt]{$C$} 
\Text(75,120)[rb]{$C'$} 
\end{picture}  
\end{center}  
\caption{Sign of the radion kinetic term in the $(k_1,k_2)$ plane. In 
the regions $A$, $B'$, $C'$ the moving brane has positive tension 
and the radion positive kinetic energy. In the regions $B$, $C$, $A'$ 
the moving brane has negative tension and the radion  negative kinetic 
energy. We show the GRS line for the non-compact case. The dashed line  
corresponds to $k_1=k_2$, \textit{i.e.} a \textit{tensionless} moving brane.}  
\label{phase}  
\end{figure}

The above quantity is not positive definite. In particular, it turns out 
that it is positive whenever the intermediate brane has positive 
tension and  negative whenever the intermediate brane has negative 
tension. This result is graphically represented in Fig.\ref{phase} 
where the $(k_1,k_2)$ plane is divided in two regions.

\subsection{The non-compact limit} 

It is instructive to discuss the decompactification limit in which the third brane is  sent  to infinity, \textit{i.e.}  
$L \to + \infty$. To examine this limit we distinguish two cases:  
\vskip 0.5truecm  
\centerline{\bf The case $\mathbf{k_2 > 0}$}  
\vskip 0.5truecm  
\noindent  
In this case we have $\displaystyle{a(\infty) \propto 
\lim_{L\rightarrow \infty} e^{-k_2L} =0}$. The dilaton 
kinetic term is trivial since from (\ref{K1}) we obtain ${\cal K}_1 \to \infty$. In other words 
the dilaton decouples  from the four dimensional effective theory and the condition 
of absence of kinetic mixing between the scalars (\ref{nomix}) plays 
no role. The radion kinetic term coefficient can be read off from (\ref{K2}):   \begin{equation}  
{\cal K}_2= \frac{3 M^3  c_1^2}{k_1} \left[e^{2k_1r} \frac{k_2}  
{(k_2-k_1)}-1  
\right]  
\end{equation}  

This result agrees with the computation of \cite{Pilo:2000et}. It is 
easy to see that the radion has positive kinetic energy when the moving 
brane has positive tension and negative kinetic energy when the 
tension is negative. Indeed, for $0 < k_1 < k_2$, or for $k_2 > 0$ and 
$k_1 < 0$ we have a 
positive brane and positive kinetic energy. On the other hand, for $0 < 
k_2 < k_1$  we have a negative tension brane 
and negative kinetic energy. In the limit $k_2 \rightarrow 0$ we get the GRS  
model with  negative kinetic energy as in \cite{Pilo:2000et}. 
\vskip 0.5truecm  
\centerline{\bf The case $\mathbf{k_2 < 0}$}  
\vskip 0.5truecm  
\noindent  
In this case $\displaystyle{a(\infty) \propto 
\lim_{L\rightarrow \infty} e^{-k_2L} \rightarrow \infty}$. This time 
the dilaton plays in the game since ${\cal K}_1$ is finite as seen 
from (\ref{K1}) and has the value: 
\begin{equation}  
{\cal K}_1 = \frac{3 M^3 \, c^2 }{2 k_1 |k_2|}  \,   
\left[ e^{2k_1r} (|k_2|+k_1)  - |k_2| \right]  
 \quad   
\end{equation}  
which is manifestly positive definite. The presence of dilaton mode  
is somewhat surprising since it describes the fluctuations of the overall size of  
the system which is infinite. Something similar happens in the  
Karch-Randall (KR) model \cite{Karch:2001ct} that has a dilaton mode although it is 
non-compact. The dilaton is a remnant of the decompactification 
process of the $''++''$ model \cite{Kogan:2001vb} which we will
examine in the following chapter and enters in the game 
because the inverse of the warp factor is normalizable. 
 
The radion kinetic term 
coefficient can be obtained from  taking the limit $L \to  
+ \infty$ in (\ref{K2}). We get 
\begin{equation}  
{\cal K}_2 = \frac{3 M^3 \,  c^2_1}{k_1} \left[e^{-2 k_1 r} \, \frac{k_2}  
{(k_2-k_1)} - 1 \right]  
\end{equation}  

The same considerations for the compact case applies here. When $k_1 < k_2 < 0$ we have a 
positive tension brane and a positive kinetic energy. On the other hand when  
$k_2 < k_1 < 0$, or for $k_2 < 0$ and $k_1 > 0$ we have a negative tension  
brane and a negative kinetic energy. In the GRS limit $k_2 \rightarrow 0$,  
the radion has negative kinetic energy.

\section{Gravity on the branes}   
  
In this section we will study how the moduli and graviton(s) couple to   
matter confined on the branes. For simplicity we will study the 
non-compact case. For this purpose, we consider a matter Lagrangian 
${\cal L}_m(\Phi_j,\hat{G}^{(i)})$, where we denote generically with  
$\Phi_j$ matter fields living on the branes;  $\hat{G}^{(i)}$ is 
the induced metric on the $i$-th brane. Our starting point is the following action:  
\begin{equation}
\begin{split}  
{\cal S}=&\int d^4xdy \, \sqrt{-G} \, [2M^3R-\Lambda(y)]-\sum_i\int d^4x\sqrt{-\hat{G}^{(i)}}V_i\\ &+\sum_i\int d^4xdy\sqrt{-\hat{G}^{(i)}}{\cal L}_m(\Phi_j,\hat{G}^{(i)})  
\end{split}
\end{equation}   

We have already calculated the effective action for the gravity sector   
for the perturbation (\ref{pert}) in the previous chapter. It is useful to decompose the perturbation  
$\tilde{h}(x,z)$ in terms of a complete set of  eigenfunctions\footnote{We denote by $\tilde{\Psi}(z)$
the gravitons wavefunctions in the conformal frame in contrast to the 
 previous chapter graviton wavefunction $\Psi(z)$. Their relation is
$\Psi(z)=a^2\tilde{\Psi}(z)$.} $\tilde{\Psi}^{(n)}(z)$ of the  
graviton kinetic operator (a suitable four dimensional gauge fixing like de Donder is  
understood):  
\begin{equation} 
\tilde{h}(x,z)=\sum_{n} \tilde{\Psi}^{(n)}(z) \, h_{\mu\nu}^{(n)}(x) \, + \,  
\int dm \, \tilde{\Psi}(y,m) \, h^{(m)}_{\mu\nu}(x) \;    
\end{equation} 
having taken into account both the discrete and the continuum part of 
the spectrum. The effective four dimensional Lagrangian has the following form:
\begin{equation}  
\begin{split} 
{\cal L}&= {\cal L}_m(\Phi_j^c,\eta)  + 2M^3 \sum_n^{~~~~*} {\cal  
L}_{PF}(h^{(n)}(x)) +{\cal K}_1 \, f_1 \Box f_1 +  {\cal K}_2 \, f_2 \Box  
f_2\\ 
&- \sum_i \sum_n^{~~~~*}\frac{\tilde{\Psi}^{(n)}(y_{i})}{ 2}  h_{\mu\nu}^{(n)}(x) 
T^{\mu\nu}_{(i)}- \sum_i \frac{Q(y_{i})}{ 2} f_1 T_{(i)} - \sum_i \frac{B(y_{i})}{ 2}f_2 T_{(i)}    
\end{split} 
\end{equation}
 
The matter fields have been rescaled $\Phi_j \rightarrow \Phi_j^c$, to make  
them canonically normalized and the energy momentum tensor $T^{\mu\nu}_{(i)}$ for matter localized on the $i$-th brane is  
defined with respect to the rescaled fields $\Phi_j^c$. By construction the  
induced background metric on the branes is the flat four dimensional Minkowski metric.  
The asterix denotes that the sum has to be converted into an  
integral for the continuum part of the spectrum. Finally, defining the   
canonically normalized fields:
\begin{equation} 
\bar{h}_{\mu\nu}^{(n)}(x)=\sqrt{2M^3}h_{\mu\nu}^{(n)}(x) \, , \quad   
\bar{f}_1=\sqrt{2|{\cal K}_1|}f_1   \, , \quad   \bar{f}_2=\sqrt{2|{\cal K}_2|}f_2
\end{equation}  
the Lagrangian  reads:  
\begin{equation}  
\begin{split} 
{\cal L}&= {\cal L}_m(\Phi_j^c,\eta)  + \sum_n^{~~~~*} {\cal  
L}_{PF}(\bar{h}^{(n)}(x)) +\frac{1}{2} \, \bar{f}_1 \Box \bar{f}_1 +  
\frac{1}{ 2} \, \text{sgn}({\cal K}_2) \, \bar{f}_2 \Box \bar{f}_2 \\ 
&-\sum_i \sum_n^{~~~~*} 
\frac{\tilde{\Psi}^{(n)}(y_{i})}{ 2\sqrt{2M^3}}  \, \bar{h}_{\mu\nu}^{(n)}(x) 
\, T^{\mu\nu}_{(i)}  - \sum_i \frac{Q(y_{i})}{ 2\sqrt{2{\cal K}_1}} \, \bar{f}_1 \, T_{(i)} 
- sum_i \frac{B(y_{i})}{  2\sqrt{2|{\cal K}_2|}} \, \bar{f}_2 \, T_{(i)}  
\end{split} 
\end{equation}   

Thus, the dilaton, the radion and the graviton (whenever it is 
normalizable) have the following couplings respectively: 
\begin{equation} 
{\cal C}_D={Q(y_{br}) \over 2\sqrt{2{\cal K}_1}} \, , \qquad {\cal C}_R={B(y_{br}) \over 2\sqrt{2|{\cal K}_2|}} \, , 
\qquad {\cal C}_G={\tilde{\Psi}^{(0)}(y_{br}) \over 2\sqrt{2M^3}}
\end{equation}  
For convenience we will define ${\cal C}_N \equiv {\sqrt{|k_1|} \over 
2 \sqrt{2 M^3}}$ and write the above quantities as: 
\begin{equation} 
{\cal C}_D^{(i)}={\cal C}_N{g_D^{(i)} \over \sqrt{3}} \, , \qquad {\cal C}_R^{(i)}={\cal C}_N{g_R^{(i)} \over \sqrt{3}} \, , 
\qquad {\cal C}_G={\cal C}_N {\tilde{\Psi}^{(0)}(y_{br}) \over \sqrt{|k_1|}} 
\end{equation}  
where the dimensionless radion couplings are: 
\begin{eqnarray} 
g_D^{(0)}=\frac{\theta(-k_2)}{\sqrt{\left|e^{2k_1r}{k_1+|k_2| \over 
|k_2|}-1\right|}} \, &,& \qquad g_D^{(r)}=e^{2k_1r}g_D^{(0)}\\ 
g_R^{(0)}=\frac{1}{\sqrt{\left|e^{2k_1r{\rm sgn}(k_2)}{k_2 \over 
k_2-k_1}-1\right|}} \, &,& \qquad g_R^{(r)}=e^{2k_1}{k_2 \over k_2 
-k_1}g_R^{(0)} 
\label{cc} 
\end{eqnarray} 
 
We should note at this point that there are certain cases as we will 
see in the following that multigravity is realized and four 
dimensional gravity in 
intermediate distances is due to more than one mode. In that case we 
will denote by ${\cal C}_G$ the coupling of the ``effective zero 
mode'' even though there might not be a genuine zero mode at all.

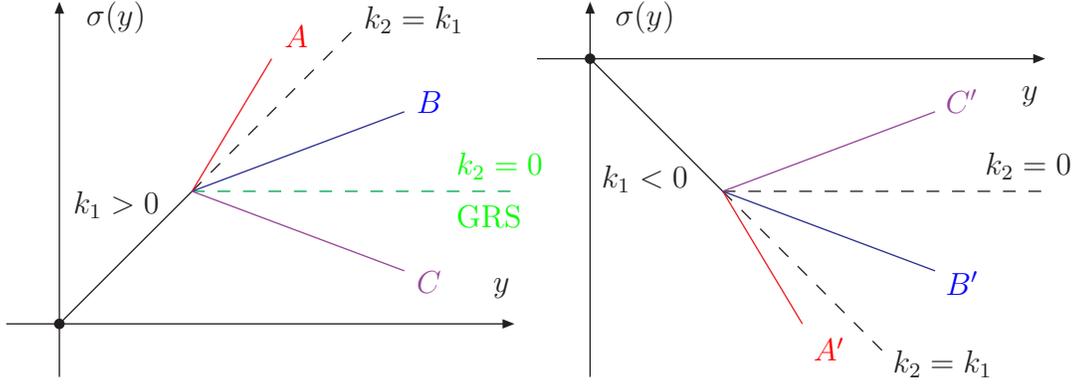
\begin{figure}[t]  
\begin{center}  
\begin{picture}(200,150)(0,0)

\Text(-70,130)[lb]{$\sigma(y)$} 
\Text(90,30)[rb]{$y$} 
\LongArrow(-80,0)(-80,140)  
\LongArrow(-100,20)(90,20) 
\Vertex(-80,20){2} 
\Line(-80,20)(-30,70) 
\Text(-75,60)[lb]{$k_1>0$} 
\SetColor{Green} 
\DashLine(-30,70)(90,70){5} 
\Text(70,75)[lb]{\color{green}$k_2=0$} 
\Text(70,65)[lt]{\color{green}GRS} 
\SetColor{Black} 
\DashLine(-30,70)(30,130){5} 
\Text(35,130)[lb]{$k_2=k_1$} 
\SetColor{Red} 
\Line(-30,70)(0,120) 
\Text(5,125)[lb]{\color{red}$A$}  
\SetColor{Blue} 
\Line(-30,70)(50,100) 
\Text(55,100)[lb]{\color{blue}$B$}  
\SetColor{Purple} 
\Line(-30,70)(50,40) 
\Text(55,40)[lt]{\Purple{$C$}} 
 
 
\Text(130,130)[lb]{$\sigma(y)$} 
\Text(290,110)[rt]{$y$} 
\SetColor{Black} 
\LongArrow(120,0)(120,140)  
\LongArrow(100,120)(290,120) 
\Vertex(120,120){2} 
\Line(120,120)(170,70) 
\Text(125,80)[lt]{$k_1<0$} 
\DashLine(170,70)(290,70){5} 
\Text(270,75)[lb]{$k_2=0$} 
\DashLine(170,70)(230,10){5} 
\Text(235,10)[lt]{$k_2=k_1$} 
\SetColor{Red} 
\Line(170,70)(200,20) 
\Text(205,15)[lt]{\color{red}$A'$} 
\SetColor{Blue} 
\Line(170,70)(250,40) 
\Text(255,40)[lt]{\color{blue}$B'$}  
\SetColor{Purple} 
\Line(170,70)(250,100) 
\Text(255,100)[lb]{\Purple{$C'$}} 
 
\end{picture}  
\end{center}  
\caption{The function $\sigma(y)=-\log [a(y)]$ for all possible 
combinations of $k_1$, $k_2$. The regions are named in accordance 
with the Fig.\ref{phase} phase diagram.}  
\label{warp}  
\end{figure}

We will now discuss how these couplings behave is the six distinct 
combinations of $k_1$ and $k_2$ shown in Fig.\ref{warp}.

\vskip 1truecm  
\centerline{\bf Region $A$}  
\vskip 0.5truecm  
\noindent  
 
In this case, we have the non-compact system of two positive tension branes discussed in \cite{Lykken:2000nb,Pilo:2000et,Kogan:2000xc}. The volume of the extra dimension is 
finite and therefore we have a normalizable graviton zero mode. The KK 
tower will be continuum and its coupling to matter on either 
branes will be exponentially suppressed for reasonably large $r$ as 
can be easily seen by the analysis of \cite{Kogan:2000xc}. The graviton 
wavefunction in the conformal gauge is constant, \textit{i.e.} 
$\tilde{\Psi}^{(0)}(y)=A$. Thus the universal coupling of the zero 
mode will be ${\cal C}_G={\cal C}_N$.  
 
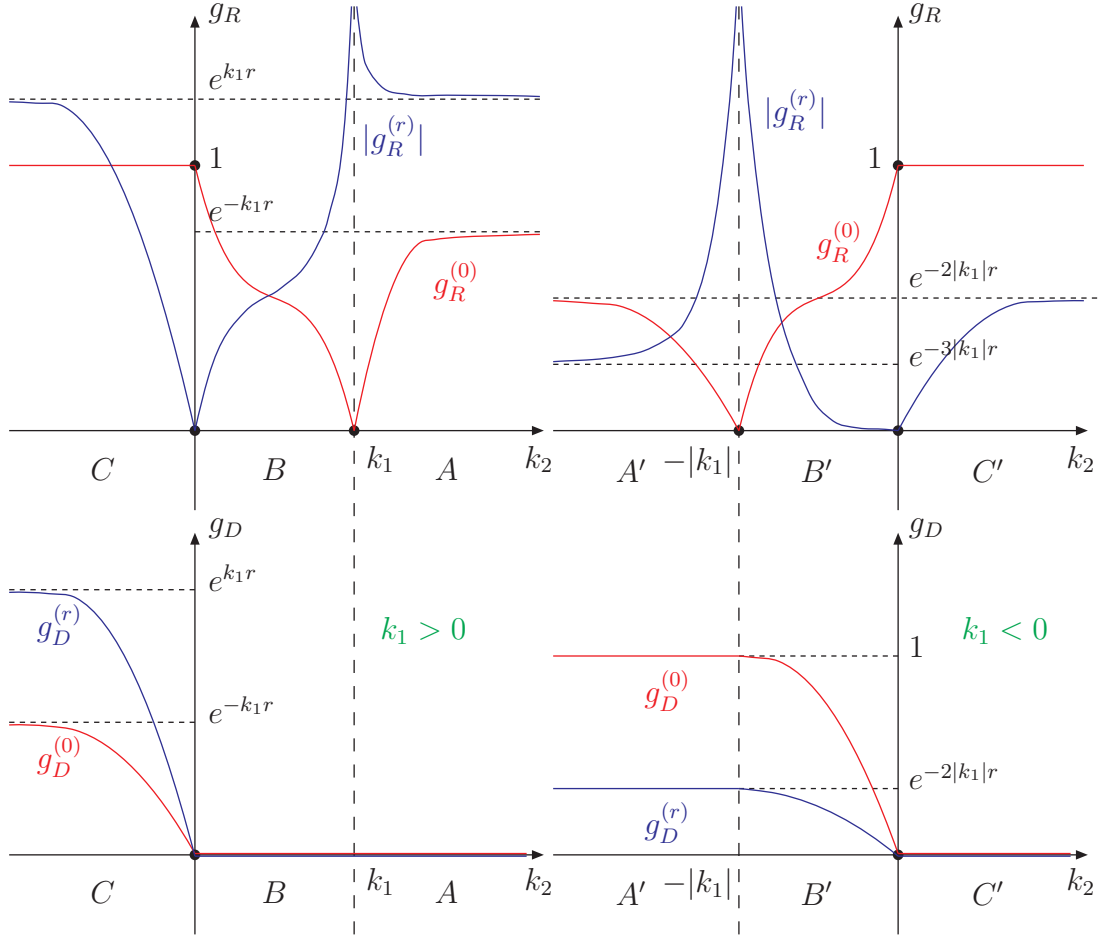
\begin{figure}[t]  
\begin{center}  
\begin{picture}(200,350)(0,0)

\DashLine(95,215)(225,215){2} 
\DashLine(95,240)(300,240){2} 
\DashLine(-40,265)(90,265){2} 
\DashLine(-110,315)(90,315){2} 
 
\DashLine(165,55)(225,55){2} 
\DashLine(-110,80)(-40,80){2} 
\DashLine(165,105)(225,105){2} 
\DashLine(-110,130)(-40,130){2} 
 
\DashLine(165,0)(165,350){5} 
\DashLine(20,0)(20,350){5}

\LongArrow(-40,160)(-40,345)  
\LongArrow(-110,190)(90,190) 
\Vertex(-40,190){2} 
\Vertex(-40,290){2} 
\Vertex(20,190){2} 
\SetColor{Red} 
\Line(-110,290)(-40,290) 
\Curve{(-40,290)(-39,286)(19,194)(20,190)} 
\Curve{(20,190)(45,261)(50,262)(60,263)(70,263.4)(80,263.7)(85,263.9)(90,264)} 
\SetColor{Blue} 
\Curve{(-110,314)(-100,313.5)(-90,312.5)(-40,190)} 
\Curve{(-40,190)(-35,210)(-21,235)(-14,240)(-10,242)(-6,245)(.7,251)(6.2,260)(10.2,270)(18,330)(19,350)} 
\Curve{(21,350)(23.2,333)(24,329)(24.4,328)(26,325)(30,320)(40,316.8)(55,316.5)(70,316.4)(80,316.3)(90,316)} 
\SetColor{Black} 
\Text(-35,345)[lb]{$g_R$} 
\Text(90,185)[tb]{$k_2$} 
\Text(30,185)[tb]{$k_1$} 
\Text(-35,320)[lb]{$e^{k_1r}$} 
\Text(-35,290)[lb]{$1$} 
\Text(-35,270)[lb]{$e^{-k_1r}$} 
\Text(-75,180)[tb]{$C$} 
\Text(-10,180)[tb]{$B$} 
\Text(55,180)[tb]{$A$} 
\Text(35,310)[tb]{$\Blue{|g_R^{(r)}|}$} 
\Text(50,255)[lt]{$\Red{g_R^{(0)}}$}

\LongArrow(-40,0)(-40,150)  
\LongArrow(-110,30)(90,30) 
\Vertex(-40,30){2} 
\SetColor{Red} 
\Line(-40,30.5)(85,30.5) 
\Curve{(-110,79)(-95,78.5)(-85,77)(-40,30.5)} 
\SetColor{Blue} 
\Line(-40,29.5)(85,29.5) 
\Curve{(-110,129)(-95,128.5)(-85,127)(-40,29.5)} 
\SetColor{Black} 
\Text(-35,150)[lb]{$g_D$} 
\Text(90,25)[tb]{$k_2$} 
\Text(30,25)[tb]{$k_1$} 
\Text(-35,130)[lb]{$e^{k_1r}$} 
\Text(-35,80)[lb]{$e^{-k_1r}$} 
\Text(-75,20)[tb]{$C$} 
\Text(-10,20)[tb]{$B$} 
\Text(55,20)[tb]{$A$} 
\Text(-100,125)[lt]{$\Blue{g_D^{(r)}}$} 
\Text(-100,75)[lt]{$\Red{g_D^{(0)}}$} 
\Text(30,120)[lt]{$\Green{k_1>0}$}

\LongArrow(225,160)(225,345)  
\LongArrow(95,190)(295,190) 
\Vertex(225,190){2} 
\Vertex(225,290){2} 
\Vertex(165,190){2} 
\SetColor{Red} 
\Line(225,290)(295,290) 
\Curve{(165,190)(166,194)(224,286)(225,290)} 
\Curve{(95,239)(110,238)(120,237)(165,190)} 
\SetColor{Blue} 
\Curve{(95,216)(100,216.3)(105,216.5)(110,216.8)(120,218)(129,220)(137,224)(140,226)(144,230)(160,300)(162,320)(164,350)} 
\Curve{(166,350)(168,320)(170,300)(191,205)(200,194)(210,191)(220,190.5)(225,190)} 
\Curve{(225,190)(270,238)(280,238.7)(295,239)} 
\SetColor{Black} 
\Text(230,345)[lb]{$g_R$} 
\Text(295,185)[tb]{$k_2$} 
\Text(150,185)[tb]{$-|k_1|$} 
\Text(220,290)[rb]{$1$} 
\Text(230,245)[lb]{$e^{-2|k_1|r}$} 
\Text(230,215)[lb]{$e^{-3|k_1|r}$} 
\Text(125,180)[tb]{$A'$} 
\Text(195,180)[tb]{$B'$} 
\Text(260,180)[tb]{$C'$} 
\Text(175,320)[lt]{$\Blue{|g_R^{(r)}|}$} 
\Text(195,270)[lt]{$\Red{g_R^{(0)}}$}

\LongArrow(225,0)(225,150)  
\LongArrow(95,30)(295,30) 
\Vertex(225,30){2} 
\SetColor{Red} 
\Line(225,30.5)(290,30.5) 
\Line(95,105)(165,105) 
\Curve{(165,105)(170,104.5)(180,103)(225,30.5)} 
\SetColor{Blue} 
\Line(225,29.5)(290,29.5) 
\Line(95,55)(165,55) 
\Curve{(165,55)(170,54.5)(180,53)(225,29.5)} 
\SetColor{Black} 
\Text(230,150)[lb]{$g_D$} 
\Text(295,25)[tb]{$k_2$} 
\Text(150,25)[tb]{$-|k_1|$} 
\Text(230,105)[lb]{$1$} 
\Text(230,55)[lb]{$e^{-2|k_1|r}$} 
\Text(125,20)[tb]{$A'$} 
\Text(195,20)[tb]{$B'$} 
\Text(260,20)[tb]{$C'$} 
\Text(130,100)[lt]{$\Red{g_D^{(0)}}$} 
\Text(130,50)[lt]{$\Blue{g_D^{(r)}}$} 
\Text(250,120)[lt]{$\Green{k_1<0}$}

\end{picture}  
\end{center}  
\caption{The dimensionless couplings of the radion (upper) and the 
dilaton (lower) to matter on the $y=0$ or the $y=r$ branes. The left 
diagrams correspond to $k_1>0$ while the right ones for $k_1<0$. The 
regions $A,~B, \dots$ are in accordance with the ones of the phase 
diagram in   Fig.\ref{phase}. The diagrams are not in scale.}  
\label{couplings}  
\end{figure}

The dilaton in this region will simply be absent, or equivalently it 
will have zero coupling ${\cal C}_D=0$ to the branes because ${\cal K}_1 \to 
\infty$. The radion in this region has positive kinetic term and from 
(\ref{cc}) we find that its coupling to matter on the $y=0$ 
and $y=r$ branes is bounded as following (see Fig.\ref{couplings}): 
\begin{equation} 
0 \leq g_R^{(0)} \lesssim e^{-k_1r} \, , \qquad g_R^{(r)} \gtrsim e^{k_1r} 
\end{equation} 
 
The radion on the central brane is always weakly coupled and decouples 
in the tensionless moving brane limit. On the other hand the radion is  
always strongly coupled on the moving brane and diverges in the 
tensionless limit where $k_2 \to k_1$.

\vskip 0.5truecm  
\centerline{\bf Region $B$}  
\vskip 0.5truecm  
\noindent

The volume of the extra dimension is still finite and therefore we have a 
normalizable graviton zero mode.  
When $k_1 = k_2$ the brane is tensionless and  as soon as $k_2 < k_1$ 
the KK continuum starts to develop  a ``resonance''. This ``resonance'' is initially rather 
broad as $k_2 \to k_1$ and its width decreases until it coincides with 
the GRS width $\Gamma \sim k_1 \exp(-3 k_1 r)$ 
as $k_2 \to 0$. In this region four dimensional 
gravity at intermediate distances is the net effect of the massless 
graviton and the lower part of the KK continuum that contributes more 
and more as $k_2 \to 0$. 
 
The dilaton is again absent, ${\cal C}_D=0$, since still we have 
${\cal K}_1 \to \infty$. The radion in this region is a ghost field, to  
compensate for the presence of the extra polarization states of the 
contributing massive gravitons,  
and its coupling to the  branes is bounded as (see Fig.\ref{couplings}): 
\begin{equation} 
0 \leq g_R^{(0)} \leq 1 \, , \qquad g_R^{(r)} \leq 0 
\end{equation} 
 
The radion on the central brane interpolates between the decoupling 
limit of the tensionless moving brane to the GRS limit where it 
couples with strength equal to the one of the graviton. On the other 
hand, the radion coupling on the moving brane interpolates between the 
infinitely strong limit of the tensionless brane (strictly speaking there is no brane 
in $y=r$ and we are left with RS2) to the GRS limit where  
it decouples.

\vskip 0.5truecm  
\centerline{\bf Region $C$}  
\vskip 0.5truecm  
\noindent  
 
In this region we have a infinite analogue of the $''+-+''$ model with  
no normalizable zero mode present. The  
graviton spectrum for $ k_2 \to 0$ is approximately equally 
spaced  and there is a resonance which coincides with 
the GRS width and gets more and more broad as $k_2$ gets more 
negative. Soon enough a special light state is 
singled out as in the $''+-+''$ model, a behaviour that persists for 
all values of $k_2$ in this region. From the above behaviour we 
deduce that  four dimensional gravity at intermediate distances is 
generated by the lower part of the discrete spectrum as  $k_2 \to 0$, 
whereas only by the special state for all other values of $k_2$. 
 
The dilaton in this region is present and its coupling to the branes 
is bounded as following (see Fig.\ref{couplings}): 
\begin{equation} 
0 \leq g_D^{(0)} \lesssim e^{-k_1r}  \, , \qquad 0 
\leq g_D^{(r)}  \lesssim e^{k_1r} 
\end{equation} 
 
It is always weakly coupled to matter on the central brane and becomes 
strongly coupled but saturated on the moving brane. 
 
The radion in this region is again a ghost field to cancel the 
unwanted extra massive graviton polarization states and its coupling to 
the branes is bounded as (see Fig.\ref{couplings}): 
\begin{equation} 
g_R^{(0)} \approx 1 \, , \qquad 0 \leq g_R^{(r)} \lesssim e^{k_1r} 
\end{equation} 
 
On the central brane it couples always with strength equal to the one 
of the ``effective graviton'' and on the moving brane it interpolates 
between the decoupling limit of the GRS case to a strongly coupled 
region with saturated coupling as the tension of the second brane gets  
infinite.

\vskip 1truecm  
\centerline{\bf Region $A'$}  
\vskip 0.5truecm  
\noindent

In this region we have no normalizable zero mode and the two negative 
tension brane system resembles an inverted version of the RS2 
model. The KK spectrum  is discrete and all excitations lie above the 
characteristic curvature scale $k_1$. Thus the low energy effective theory does not have four 
dimensional gravity at all. 
 
The dilaton in this system is present and the coupling to the branes 
is approximately constant (see Fig.\ref{couplings}): 
\begin{equation} 
g_D^{(0)} \approx 1  \, , \qquad g_D^{(r)}  \approx 1 
\end{equation} 
 
The radion field is a ghost and its coupling to matter on the branes 
is bounded as (see Fig.\ref{couplings}): 
\begin{equation} 
g_R^{(0)} \gtrsim e^{-3|k_1|r} \, , \qquad 0 \leq g_R^{(r)} \lesssim e^{-2|k_1|r} 
\end{equation} 
 
The radion on the central negative tension brane is always weakly 
coupled and vanishes in the limit of $k_2 \to k_1=-|k_1|$ of the 
tensionless moving brane. On the other hand the coupling on the moving brane 
is bounded from below in the limit of infinite negative tension brane 
and diverges at the limit of the tensionless brane. 
 
\vskip 0.5truecm  

\vskip 0.5truecm  
\centerline{\bf Region $B'$}  
\vskip 0.5truecm  
\noindent  
  
In this region there is again no normalizable zero mode and the system  
of the negative and positive tension branes still resembles the 
inverted RS2 model. The KK spectrum is almost identical with the 
one of the previous region except for the limit $k_2 \to 0$ when the 
spectrum drops below the curvature scale $k_1$ and a continuum develops. 
  
The dilaton coupling to the branes decreases from the constant value 
of the previous region, to zero as $k_2 \to 0$ (see Fig.\ref{couplings}): 
\begin{equation} 
0 \leq g_D^{(0)} \lesssim 1  \, , \qquad 0 
\leq g_D^{(r)}  \lesssim e^{-2|k_1|r} 
\end{equation} 
 
The radion has positive kinetic energy and its coupling to the branes 
is bounded as following (see Fig.\ref{couplings}): 
\begin{equation} 
0 \leq g_R^{(0)} \leq 1 \, , \qquad g_R^{(r)} \leq  0 
\end{equation} 
 
It is weakly coupled in the central brane and the coupling interpolates between 
zero from the previous region to one in the inverted GRS case where
$k_2 \to 0$. On the second brane  
it is divergent as $k_2 \to k_1=-|k_1|$ and vanishes as $k_2 \to 0$. 
 
\vskip 0.5truecm  
\centerline{\bf Region $C'$}  
\vskip 0.5truecm  
\noindent 
 
In this region we again have a system of a negative and a positive 
tension brane, but gravity  can be localized on the moving positive 
tension brane. There is a normalizable zero mode that mediates four 
dimensional gravity and a KK continuum with suppressed couplings on 
the branes. 
 
The dilaton in this region will be absent, or equivalently it 
will have zero coupling ${\cal C}_D=0$ to the branes because ${\cal K}_1 \to 
\infty$. 
 
The radion will have again positive kinetic term and its coupling will  
be bounded as following (see Fig.\ref{couplings}): 
\begin{equation} 
0 \leq g_R^{(0)} \lesssim e^{-2|k_1|r} \, , \qquad g_R^{(r)} \approx  1 
\end{equation} 
 
It is approximately constant on the central brane with strength equal 
to the one of the graviton  and is weakly coupled to the moving brane 
with coupling interpolating between zero in the 
inverted GRS case with $k_2 \to 0$ to a certain value as the tension of 
the moving brane becomes infinitely large.

\section{Review of sum rules}

In this section we will review the sum rules presented in
\cite{Gibbons:2001tf} in order to discuss the dilaton stabilization in the
presence of bulk dynamics of arbitrary minimally
coupled bulk scalar fields. We will concentrate  as in \cite{Gibbons:2001tf} in the case  where 
the background metric of the five
dimensional spacetime can be written in the form:
\begin{equation}
ds^2=a(y)^2g_{\mu\nu}(x)dx^{\mu}dx^{\nu}+dy^2
\label{background}
\end{equation}
with $g_{\mu\nu}(x)$ a general background four dimensional metric
and $a(y)$ a generic warp factor. We should stress here that this is
not the most general choice of metric in five dimensions as we have
explicitly assumed that the all four dimensional sections have the
same conformal geometry. The particular choice $g_{\mu\nu}=\eta_{\mu\nu}$ is
valid for the models which we have discussed so far.

We can now consider an arbitrary number of minimally coupled scalar bulk fields
$\Phi^I(x,y)$ with internal metric ${\mathcal{G}}_{IJ}$ and arbitrary
bulk potential $V(\Phi)$ (which includes bulk cosmological constant), 
coupled to an again arbitrary number of branes with brane potential
$\lambda_{i}(\Phi)$ (which again includes the brane tensions). The
action describing the above system is the following:
\begin{equation}
{\cal S} = \int d^{4}x dy \sqrt{-G}\left(2M^{3}R
-{1 \over
2}{\mathcal{G}}_{IJ}\partial_M\Phi^I\partial^M\Phi^J-V(\Phi)-\sum_{i}\lambda_{i}(\Phi)\delta(y-y_i){\sqrt{-\hat{G}^{(i)}} 
\over\sqrt{-G}} \right)
\label{action}
\end{equation}
where $\hat{G}^{(i)}_{\mu\nu}$ is the induced metric on the brane and
$M$ the fundamental five dimensional scale. The
Einstein equations arising from the above metric can be written in
the form:
\begin{eqnarray}
4M^3R_{\mu}^{\mu}&=&-{1 \over 3}T_{\mu}^{\mu}-{4 \over 3}T_5^5\label{e1}\\
4M^3R_5^5&=&-{1 \over 3}T_{\mu}^{\mu}+{2 \over 3}T_5^5\label{e2}
\end{eqnarray}
where the energy-momentum tensor components are:
\begin{eqnarray}
T_{\mu}^{\mu}&=&-\partial_{\mu}\Phi\cdot\partial^{\mu}\Phi-2\Phi'\cdot\Phi'-4V(\Phi)-4\sum_{i}\lambda_{i}(\Phi)\delta(y-y_i)\label{T}\\
T_5^5&=&-{1 \over 2}\partial_{\mu}\Phi\cdot\partial^{\mu}\Phi+{1 \over 2}\Phi'\cdot\Phi'-V(\Phi)\label{T5}
\end{eqnarray}
with the indices in the above formulas raised and lowered by
$G_{\mu\nu}=a(y)^2g_{\mu\nu}(x)$ and where dot product denotes
contraction with the internal metric ${\mathcal{G}}_{IJ}$. Since we are interested in a
background configuration, the $\partial_{\mu}\Phi\cdot\partial^{\mu}\Phi$ 
terms can be dropped. The Ricci tensor is easily calculated to be:
\begin{eqnarray}
R_{\mu}^{\mu}&=&a^{-2}R_g -12(a')^2a^{-2}-4a''a^{-1}\label{r}\\
R_5^5&=&-4a''a^{-1}\label{r5}
\end{eqnarray}

If we now consider the function $(a^n)''$ with $n$ an arbitrary real
number, its integral around the compact extra dimension is zero. Using
(\ref{e1}), (\ref{e2}), (\ref{r}), (\ref{r5}) we arrive at an infinite 
number of constraints \cite{Gibbons:2001tf}:
\begin{equation}
\oint dy~a^n(T_{\mu}^{\mu}+(2n-4)T_5^5)=4M^3(1-n)R_g\oint dy~a^{n-2}
\end{equation}

As it is obvious, these constraints are a natural consequence of the
equations of motion and are totally equivalent to them. It is straightforward to see for example  that they are
satisfied in the RS1 model \cite{Randall:1999ee} and the
bigravity/multigravity models \cite{Kogan:2000wc,Kogan:2000xc,Kogan:2001vb}.  We will
single out three constraints which we will be important for the
subsequent discussion, namely the ones for $n=0,1,2$:
\begin{eqnarray}
\oint dy~(T_{\mu}^{\mu}-4T_5^5)&=&4M^3R_g\oint dy~ a^{-2}\label{0gib}\\
\oint dy~a(T_{\mu}^{\mu}-2T_5^5)&=&0\label{gib}\\
\oint dy~a^2T_{\mu}^{\mu}&=&-4M^3R_g\oint dy\label{newgib}
\end{eqnarray}

One could also use the above constraints for non-compact models, but 
should be careful that the above derivation makes sense. For the
RS2 model \cite{Randall:1999vf}, all constraints are valid for $n\geq 0$. For the GRS model \cite{Gregory:2000jc}  only the $n=0$ constraint is
valid and for the KR model \cite{Karch:2001ct} (see next chapter) all constraints
are valid for $n \leq 0$.

\section{Dilaton in warped backgrounds}

We now consider the perturbation related to the overall size of the
compact system, namely the dilaton. As previously noted in the chapter, we do not consider the KK-tower of the dilaton since these states can be gauged away by five dimensional diffeomorphisms. The general form of the metric for 
the physical radion perturbations that do not mix with the graviton(s)
is given in \cite{Pilo:2000et,Kogan:2001qx}. For the dilaton, the ansatz is rather simple and following   \cite{Charmousis:2000rg}  we  write it in the form\footnote{We are interested only in the quadratic in $\gamma(x)$ Lagrangian derived from the above metric, so only the part of the metric up to linear order  in   $\gamma(x)$ is important. However, we prefer to  keep the same notation as  \cite{Charmousis:2000rg} and do the calculation with (\ref{ruth}). The result, of course, for the quadratic Lagrangian would be the same if we used the  metric up to linear order  in   $\gamma(x)$.}:
\begin{equation}
ds^2=e^{-a(y)^{-2}\gamma(x)}a(y)^2g_{\mu\nu}(x)dx^{\mu}dx^{\nu}+\left(1+a(y)^{-2}\gamma(x)\right)^2dy^2\label{ruth}
\end{equation}
 
Substituting the above metric in the action (\ref{action}) (see
Appendix B for analytic formulas),
integrating out total derivatives, throwing out $\gamma$-independent parts  and keeping terms up to quadratic order, we get:
\begin{equation}
{\cal S}= \int d^4xdy\sqrt{-g}\left\{-{1 \over
2}(6M^3a^{-2})g^{\mu\nu}\gamma_{,\mu}\gamma_{,\nu}+{\mathcal{L}}_1\gamma-{1 
\over 2}{\mathcal{L}}_2\gamma^2\right\}
\end{equation}
with
\begin{eqnarray}
{\mathcal{L}}_1&=&2M^3(4(a')^2+16a''a)+{3 \over
2}a^2\Phi'\cdot\Phi'+a^2V(\Phi)+2a^2\sum_{i}\lambda_{i}(\Phi)\delta(y-y_i)~~~~~~~~~~~\\
{\mathcal{L}}_2&=&2M^3(a^{-2}R_g-32(a')^2a^{-2}+32a''a^{-1})+5\Phi'\cdot\Phi'+4\sum_{i}\lambda_{i}(\Phi)\delta(y-y_i)
\end{eqnarray}


At this point, let us work out the integral over the extra dimension
of the tadpole term ${\mathcal{L}}_1$  of the Lagrangian. This gives:
\begin{equation}
\oint dy{\mathcal{L}}_1=-{1 \over 2}\oint dy~a^2\left[T_{\mu}^{\mu}-2T_5^5-16M^3\left({(a')^2 \over a^2}+4{a'' \over a}\right)\right]\label{improved}
\end{equation}
where we used the energy-momentum tensor components found in the previous 
section with respect to the unperturbed background metric (\ref{T}),
(\ref{T5}). We can further simplify this quantity if we use the
equations (\ref{e1}), (\ref{e2}), (\ref{r}), (\ref{r5}) which hold for 
the background metric. The resulting expression is:
\begin{equation}
\oint dy{\mathcal{L}}_1={1 \over 6}\oint dy~a^2\left[T_{\mu}^{\mu}+4M^3a^{-2}R_g\right]
\end{equation}
which is exactly zero because of the $n=2$ constraint
(\ref{newgib}). This result should have been expected since the
perturbation (\ref{ruth}) is bound to extremize the effective
potential when one evaluates the action using the
background equations of motion. It is worth
mentioning here that in the case that the warp factor is effectively
constant ($a \approx 1$), as it was assumed in \cite{Kanti:2000nz}, the condition that the expression
(\ref{improved}) vanishes, is identical with the   $n=1$ constraint (\ref{gib}).

Our next task is to read off the mass of the dilaton from the action
functional. For this reason we define the canonically normalized
dilaton field  with mass dimension one $\bar{\gamma}^2=\left(6M^3\oint
dya^{-2}\right)\gamma^2\equiv A \gamma^2$. Then the
mass of the canonical dilaton $\bar{\gamma}$ is:
\begin{equation}
m^2={1 \over A}\oint dy{\mathcal{L}}_2
\end{equation}

After a lot of simplifications using the relations
(\ref{e1})-(\ref{r5}) we obtain:
\begin{equation}
m^2=-{1 \over 3A}\oint dy(10M^3a^{-2}R_g +\Phi'\cdot\Phi'+4\sum_{i}\lambda_{i}(\Phi)\delta(y-y_i))
\end{equation}

We can further simplify the expression using the $n=0$ constraint
(\ref{0gib}) and get a more suggestive result:
\begin{equation}
m^2={1 \over A}\oint dy(\Phi'\cdot\Phi'-2M^3a^{-2}R_g)\label{fi}
\end{equation}
or equivalently,
\begin{equation}
m^2=-{1 \over A}\left\{\sum_{i}\lambda_{i}(\Phi)+3M^3R_g\oint dya^{-2}\right\}\label{V}
\end{equation}

From the second expression it is clear that we cannot have a massive
dilaton if the sum of the effective tensions of the branes
$\lambda_{i}(\Phi)$ is exactly zero and at the same time they are
kept flat. This is the same conclusion that appeared in \cite{Gibbons:2001tf}
regarding the Goldberger-Wise mechanism \cite{Goldberger:1999uk} in the RS1
scenario. Regarding the latter mechanism this is not a surprise, since 
promoting the Goldberger-Wise to an exact solution means that either
the  scalar field
energy will generate a four dimensional cosmological constant, or if
one insists on having flat branes, this will introduce a new and different
fine tuning of the tensions which  gives  a mismatch of brane
tensions. Moreover, the absence of a tachyonic mass will guarantee the
stabilization of the overall size of any system with the above
characteristics. By (\ref{fi}), (\ref{V}) we get two equivalent conditions:
\begin{eqnarray}
\oint dy(\Phi'\cdot\Phi'-2M^3a^{-2}R_g) &>& 0\\
\sum_{i}\lambda_{i}(\Phi)+3M^3R_g\oint dya^{-2} &<& 0
\end{eqnarray}

If one wishes to have flat branes, then  the sum of brane
tensions should be negative or equivalently one should have a non constant (in $y$) scalar field
configuration. In the case that the above expressions (and thus the mass)
vanish, one should look for the higher orders of the effective
potential to examine the stability of the system.

Finally, we should note that the formulas (\ref{fi}), (\ref{V}) are valid even
for non-compact models whenever the dilaton mode is normalizable. This 
happens for example in the KR model \cite{Karch:2001ct} and one can find from the above
expressions the mass of the dilaton, in agreement with \cite{Chacko:2001em}. In
that case, the dilaton mode cannot be attributed to the fluctuation of
the overall size of the system, but can be understood to be a remnant
mode if we start with the compact $''++''$ system \cite{Kogan:2001vb} and send
one of the branes to infinity (\textit{i.e.} after decompactifying the
system). We will see this possibility in the next chapter.

\chapter{$AdS_4$ brane multigravity models in five dimensions}

As we have seen clearly in the previous two chapters, the flat brane
multigravity models in five dimensions necessarily have moving negative tension
branes. These in turn have corresponding  radions with negative
kinetic energy which are disastrous for the theory.  It is clearly
desirable to have a model with only positive tension branes. In this
chapter we will demonstrate how one can obtain such a model in five
dimensions. The price that one has to pay is that the branes are no
longer flat, but instead $AdS_4$. A two brane model with only
positive tension branes was already known in
\cite{Kaloper:1999sm,Nihei:1999mt,Kim:2000ja,DeWolfe:2000cp,Kanti:2000nz,Kanti:2000rd,Kanti:2000cz} 
and exhibited a characteristic
bounce of the warp factor. We  know from chapter \ref{ch2} that a
model that has a bounce in the warp factor is  bound to have
bigravity, something that we will explicitly show for this particular model in
the present chapter. Furthermore, we can construct a crystal
model with an infinite array of positive tension branes, but we cannot 
reproduce the quasi-localized model because the presence of the
cosmological constant on the branes prevents the warp factor from
being asymptotically constant.

Since there is no ghost modulus in these configurations one may worry
about the structure of the graviton propagator at intermediate
distances. As we will show explicitly, the $AdS_4$ brane configurations 
can evade this problem because of an unusual property of the graviton
propagator in $AdS$ background. We will show in the second part of
this chapter that the van Dam-Veltman-Zakharov discontinuity can in
certain cases be  circumvented and a particular example is the the
$''++''$ model. In more detail, the troublesome additional polarizations are very
weakly coupled and do not disturb the normal structure of the graviton
propagator.

Nevertheless, the construction that we obtain faces phenomenological
difficulties because of the remnant negative cosmological constant on the
brane. This sets an horizon scale and unfortunately the Compton
wavelength of the  light state of  the system lies exponentially far
from this horizon. This does not change even if we consider a highly asymmetric
version of this model. As a result, this bigravity model makes
no predictions for observable deviations from Newtonian gravity at
ultra-large distances. Furthermore, the very fact that the
remnant cosmological constant is negative is observationally disfavoured.

\sect{The $''++''$ bigravity model}

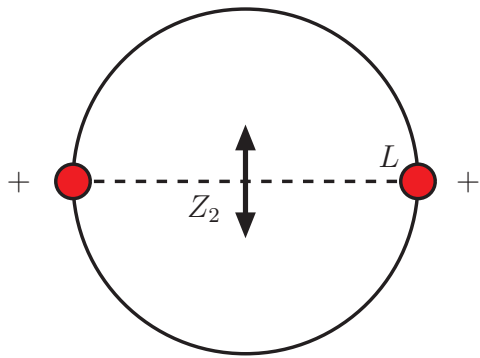
\begin{figure}[t]
\begin{center}
\begin{picture}(300,170)(0,50)
\SetScale{1.3}
\SetOffset(-50,0)

\SetWidth{1}

\BCirc(150,100){50}
\DashLine(100,100)(200,100){3}

\CCirc(100,100){5}{Black}{Red}
\CCirc(200,100){5}{Black}{Red}

\Text(110,130)[]{$+$}
\Text(280,130)[]{$+$}
\Text(180,120)[]{$Z_2$}
\Text(250,140)[]{$L$}

\SetWidth{1.5}
\LongArrow(150,100)(150,115)
\LongArrow(150,100)(150,85)

\end{picture}
\end{center}
\caption{The $''++''$ bigravity model with two $AdS_4$ $''+''$ branes at the
fixed points of a $S^1/Z_2$.}
\label{++}
\end{figure}

The model consists of two  three-branes with tensions $V_{1}$ and $V_{2}$ respectively, in an $AdS_5$ space with
five dimensional cosmological constant $\Lambda<0$. The fifth dimension has the geometry
of an orbifold and the branes are located at its fixed points,
\textit{i.e.}  $L_0=0$ and $L_{1}=L$ (see Fig.\ref{++}). Due to orbifolding, we can
restrict ourselves to the region $0\leq{y}\leq{L}$. Firstly, we  find a suitable vacuum solution. The action of this setup is:
\begin{equation}
{\cal S}_G =\int d^4 x \int_{-L}^{L} dy \sqrt{-G} 
[ 2 M^3 R - \Lambda]-\sum_{i}\int_{y=L_i}d^4xV_i\sqrt{-\hat{G}^{(i)}}
\end{equation}
where $\hat{G}^{(i)}_{\mu\nu}$ is the induced metric on the branes. The Einstein equations that arise from this
action are:
\begin{equation}
R_{MN}-\frac{1}{2}G_{MN}R=-\frac{1}{4M^3}
\left(\Lambda G_{MN}+
\sum_{i}V_i\frac{\sqrt{-\hat{G}^{(i)}}}{\sqrt{-G}}
\hat{G}^{(i)}_{\mu\nu}\delta_M^{\mu}\delta_N^{\nu}\delta(y-L_i)\right)
\end{equation}

In order to find a specific form of the equations of motion we need to
write a metric ansatz which will take in account the spacetime symmetries of the
three-brane. Since we would like not to restrict our model to flat solutions on the
branes, we should make a choice which will let us interpolate between
the maximally symmetric space-times in four dimensions, \textit{i.e.} de-Sitter,
Minkowski and Anti-de-Sitter. The metric ansatz that accomplishes this is the following:
\begin{equation}
ds^2=\frac{a^2(y)}{(1\mp \frac{H^{2}x^{2}}{4})^2}\eta_{\mu\nu}dx^{\mu}dx^{\nu} +dy^2
\end{equation}
where $x^2=\eta_{\mu\nu}x^{\mu}x^{\nu}$. It can be shown that the
Ricci scalar for this metric is $R=\mp 12H^2$. From now on, the upper
sign corresponds to $AdS_4$ space while the lower to $dS_4$
space and always $H^2>0$. Thus,  this metric represents all maximally symmetric spaces. A straightforward
calculation of the Einstein's equations gives us the following differential equations for $a(y)$:
\begin{eqnarray}
(a'(y))^2&=&\pm H^2-\frac{\Lambda}{24M^3}a^2(y)\label{ad2}\\
a
''(y)&=&-\sum_{i}\frac{V_i}{12M^3}a(y)\delta(y-L_i)-\frac{\Lambda}{24M^3}a(y)
\label{add}
\end{eqnarray}

By solving the above equations we find that the solution can be written
in the form:
\begin{equation}
a(y)=\cosh(k|y|)+\frac{V_{1}k}{\Lambda}\sinh(k|y|)
\end{equation}
with
\begin{equation}
\renewcommand{\arraystretch}{1.5}
H^2=\left\{\begin{array}{cl}\frac{k^2}{\Lambda^2}(V_{1}^{2}k^2-\Lambda^2)&,\frac{|\Lambda|}{k}<V_{1}
~~{\rm for}~dS_{4}~{\rm branes}\\
0&,\frac{|\Lambda|}{k}=V_{1}~~{\rm for~flat~branes}\\
\frac{k^2}{\Lambda^2}(\Lambda^2-V_{1}^{2}k^{2})&,\frac{|\Lambda|}{k}>V_{1}~~{\rm for}~AdS_{4}~{\rm branes}\end{array}\right.
\
\end{equation}
where we have normalized $a(0)=1$ and assumed $V_{1}>0$. Also we have
defined $k\equiv{\sqrt{\frac{-\Lambda}{24 M^3}}}$.

Additionally, in order to have this solution, the brane tensions
$V_{1}$, $V_{2}$, the bulk cosmological
constant $|\Lambda|$ and the position of the second brane $L$ must be
related through the equation: 
\begin{equation}
\tanh(kL)=k|\Lambda|\frac{V_{1}+V_{2}}{|\Lambda|^{2}+k^{2}V_{1}V_{2}}
\label{ten}
\end{equation}

Let us now restrict ourselves to the case of $AdS_{4}$ spacetime  on the two
branes which will turn out to be the most interesting. In this case
the condition $\frac{|\Lambda|}{k}>V_{1}$ must hold. Hence, we can
define $\tanh(ky_{0})\equiv\frac{k V_{1}}{|\Lambda|}$ and
write the solution in the form: 
\begin{equation}
a(y)=\frac{\cosh(k(y_{0}-|y|))}{\cosh(ky_{0})}
\end{equation}
from which it is clear that the warp factor has a minimum at
$y=y_{0}$. From this point we can see the role of the
$AdS_{4}$ on the branes, \textit{i.e.} the role of the condition
$\frac{|\Lambda|}{k}>V_{1}$. This condition allows us to have the bounce
form of the warp factor and the second brane to have
positive tension and give us, as we will see shortly, a
 phenomenology quite similar to the $^{\prime \prime
}+-+^{\prime \prime }$ bigravity model \cite{Kogan:2000wc}.
This can be easily seen from the (\ref{ten}) which relates the brane
tensions and the distance between the branes. From this equation we indeed
see that by placing the second brane
after the minimum of the warp factor we can make the tension of
the second brane positive and thus both branes that appear in the model
have positive tension avoiding the problems associated with 
negative tension branes. 
In fact it is clear that the present model mimics the characteristics
of the $^{\prime \prime
}+-+^{\prime \prime }$ bigravity model since what we effectively do is to reproduce the
effect of the presence of a negative tension brane, \textit{i.e.} the bounce form
of the warp factor, with another mechanism allowing a negative four
dimensional cosmological constant on the brane. Note that in the
limit that ${V_{1}}\rightarrow\frac{|\Lambda|}{k}$ (flat limit) the
minimum of the warp factor tends to infinity and if we wish to
have a second brane at some finite distance, it will necessarily have negative
tension.

The relationship between the four dimensional effective fundamental scale  $M_{*}$ and the five dimensional
fundamental scale $M$ can be easily found by dimensional reduction to be:
\begin{equation}
M_{\rm
*}^2=\frac{M^3}{k\cosh^{2}(ky_{0})}\left[kL+\sinh(kL)\cosh(k(L-2y_{0}))\right]
\label{plank}
\end{equation}

The above formula tells us that for finite $L$ the compactification
volume is finite and thus the zero mode is normalizable. In the case
where we send the second brane to infinity, the compactification
volume becomes infinite which indicates that the zero mode becomes
non-normalizable. Note that $M_{*}$ is not
necessarily equal to the four dimensional Planck scale $M_{{\rm{Pl}}}$ since as will see shortly, at least for
a sector of the parameter space of our model, gravity is the result not
only of the massless graviton but also of an ultralight KK state.

The warp factor renormalizes the physical scales of the theory as
 in \cite{Randall:1999ee}. Thus, all 
mass parameters $m_0$ on a brane placed at the point $y_{br}$ are rescaled as:
\begin{equation}
m=a(y_{br})m_{0}
\end{equation}

Hence, one can choose  a distance between the
two branes such that this rescaling  leads to the creation of a desired mass
hierarchy. Let us note at this point that this interbrane distance,
and thus the resulting hierarchy, is dictated
by the relative strength of the tensions of the two branes. This is
not the case as the flat brane models where the hierarchy is to be
fixed by some stabilization mechanism. The  $AdS$ brane  system is
self-stabilized and this can be easily seen by the fact that the
dilaton in this case is massive. Indeed from (\ref{V}) we obtain
$m^2=4H^2>0$ which means that the system is truly stabilized.

However, since we consider non-flat solutions on the branes, we
have to make sure that the four dimensional effective cosmological
constant does not contradict  present experimental and
observational bounds. Recent  experimental data favour a positive
cosmological constant, nevertheless since
zero cosmological constant is not completely ruled out it can be argued that also
a tiny negative cosmological constant can be acceptable within the experimental
uncertainties. The effective cosmological constant on the two branes
is:
\begin{equation}
\Lambda_{4d}=-12H^{2}M_{*}^2=-\frac{12}{\cosh^2(ky_{0})}k^{2}M_{*}^{2}
\label{cosmolcon}
\end{equation}
   
From the above formula we can see that we can make the cosmological
constant small enough $|\Lambda_{4d}| \la 10^{-120} M_{\rm Pl}^4$ if we
choose large enough $ky_{0}$, \textit{i.e.} $ky_{0}\ga {135}$. This
however will make observable deviations from Newtonian gravity at
ultra-large scales impossible as we will see later.

To determine the phenomenology of the model we need to know the KK
spectrum that follows from the dimensional reduction. This is
determined by considering the linear fluctuations of the metric
around the vacuum solution that we found above. We can write the
metric perturbation in the form:
\begin{equation}
ds^2=\left[a(y)^{2}g^{AdS}_{~\mu\nu} +h_{\mu\nu}
(x,y)\right]dx^\mu dx^\nu +dy^2
\end{equation}
where $g^{AdS}_{~\mu\nu}$ is the four dimensional $AdS_4$ metric. Here
we have ignored the dilaton mode that has been calculated in the
previous chapter. We
expand the field $h_{\mu\nu}(x,y)$ into graviton and KK plane waves:
\begin{equation}
h_{\mu\nu}(x,y)=\sum_{n=0}^{\infty}h_{\mu\nu}^{(n)}(x)\Psi^{(n)}(y)
\end{equation}
where we demand
$\left(\nabla_\kappa\nabla^\kappa +2 H^2-m_n^2\right)h_{\mu\nu}^{(n)}=0$
and additionally 
\mbox{$\nabla^{\alpha}h_{\alpha\beta}^{(n)}=h_{\phantom{-}\alpha}^{(n)\alpha}=0$}.
The function $\Psi^{(n)}(y)$ will obey the second order differential
equation:
\begin{equation}
\left\{-{1 \over 2}\de_y^2+\left({a' \over a}\right)^2+ {a'' \over
a}\right\}\Psi^{(n)}(y)=\frac{m_n^2}{2}{1 \over a^2}\Psi^{(n)}(y)
\end{equation}

After a change of variables and a redefinition of the wavefunction the 
above equation reduces to an ordinary
Schr\"{o}dinger-type equation:
\begin{equation}
\left\{-
\frac{1}{2}\partial_z^2+V(z)\right\}\hat{\Psi}^{(n)}(z)=\frac{m_n^2}{2}\hat{\Psi
}^{(n)}(z)
\end{equation}
where the potential is given by:
\begin{eqnarray}
V(z)=&-&
\frac{9\tilde{k}^{2}}{8}~+~\frac{15\tilde{k}^2}{8}\frac{1}{\cos^{2}\left(\tilde{k}(|z|-z_{0})\right)}\cr      &-&\frac{3k}{2}\left[ \tanh(ky_{0})\delta(z)+\frac{\sinh(k(L-y_{0}))}{\cosh(ky_{0})}
\delta(z-z_{1})\right] 
\end{eqnarray}
with $\tilde{k}$  defined as
$\tilde{k}\equiv{\frac{k}{\cosh(ky_{0})}}$. The new variables  and the redefinition of the wavefunction are
related with the old ones by:
\begin{equation}
z\equiv {\rm sgn}(y)\frac{2}{\tilde{k}}\left[\arctan\left(\tanh(\frac{k(|y|-y_{0})}{2})\right)+\arctan\left(\tanh(\frac{ky_{0}}{2})\right)\right]
\
\end{equation}
\begin{equation}
\hat{\Psi}^{(n)}(z)\equiv\frac{1}{\sqrt{a(y)}}\Psi^{(n)}(y)
\end{equation}

Thus, in terms of the new coordinates, the branes are  at $z_{L_{0}}=0$
 and $z_{L}$, with the minimum of the potential  at $z_{0}={2 \over \tilde{k}}\arctan\left(\tanh(\frac{ky_{0}}{2})\right)$. Also note
that with this transformation the point $y=\infty$ is mapped to the
finite point $z_{\infty}={2 \over \tilde{k}}\left[{\pi \over 4} + \arctan\left(\tanh(\frac{ky_{0}}{2})\right)\right]$.

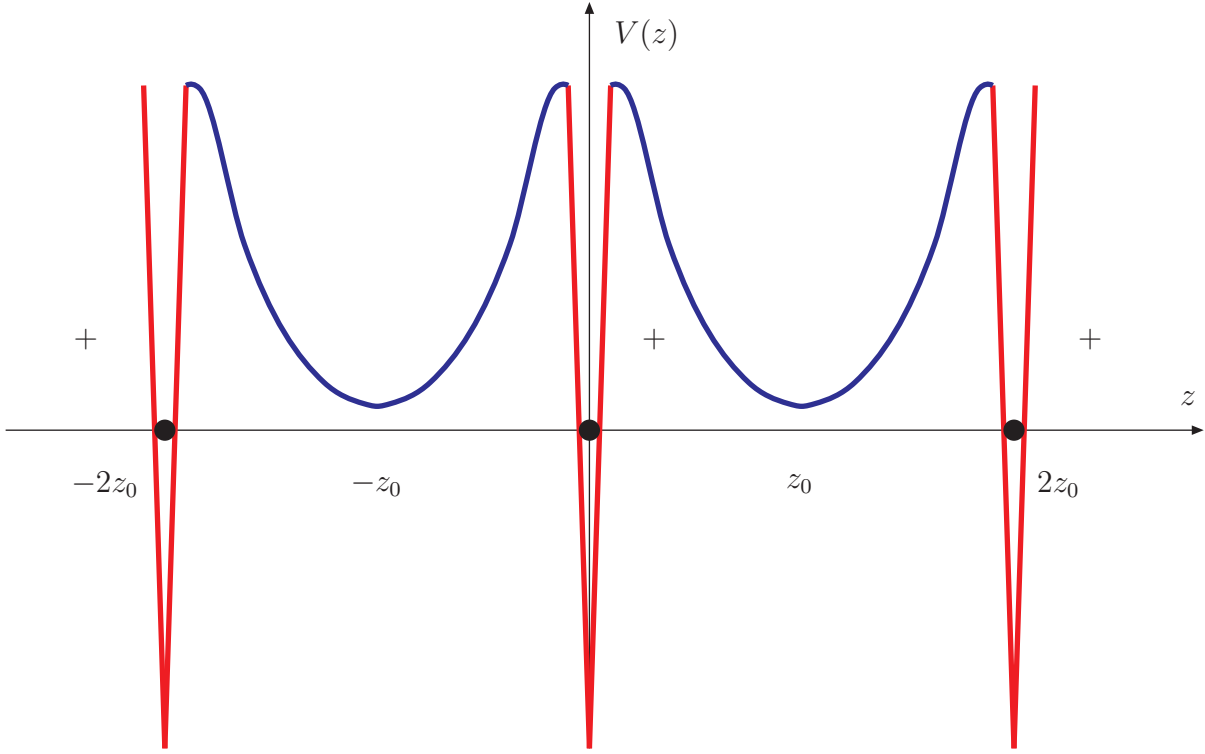
\begin{figure}[t]
\begin{center}
\begin{picture}(300,250)(0,50)
\LongArrow(140,0)(140,280)
\LongArrow(-80,120)(370,120)

\SetWidth{2}
\SetColor{Red}
\Line(140,0)(148,250)
\Line(140,0)(132,250)
\SetColor{Black}
\Vertex(140,120){4}

\Text(220,100)[c]{$z_0$}
\Text(60,100)[c]{$-z_0$}

\SetColor{Red}
\Line(-12,250)(-20,0)
\Line(-28,250)(-20,0)
\Line(308,250)(300,0)
\Line(292,250)(300,0)
\Text(310,100)[l]{$2z_0$}
\Text(-30,100)[r]{$-2z_0$}
\SetColor{Black}
\Vertex(-20,120){4}
\Vertex(300,120){4}

\Text(150,270)[l]{$V(z)$}
\Text(370,130)[rb]{$z$}

\Text(-50,155)[c]{$+$}
\Text(330,155)[c]{$+$}
\Text(165,155)[c]{$+$}

\SetColor{Blue}
\Curve{(148,250)(153,249)(170,190)(200,138)(215,130)(220,129)(225,130)(240,138)(270,190)(287,249)(292,250)}


\Curve{(-12,250)(-7,249)(10,190)(40,138)(55,130)(60,129)(65,130)(80,138)(110,190)(127,249)(132,250)}


\end{picture}
\end{center}
\vskip15mm
\caption{The potential $V(z)$ in the $''++''$ bigravity model for the
symmetric configuration. The
$\delta$-function wells correspond to $''+''$ branes.}
\label{++pot}
\end{figure}

From now on we restrict ourselves to the symmetric configuration of the two
branes with respect to the minimum  $z_{0}$ (\textit{i.e.} the first
brane at 0 and the
second at  $2z_{0}$ ), since the important characteristics of the model
appear independently of the details of the configuration.
Thus, the model has been reduced to a quantum mechanical problem
with $\delta$-function potentials wells of
the same weight and an extra smoothing term in-between (due to the $AdS$
geometry). This  gives the potential a double ``volcano'' form (see Fig.\ref{++pot}). An
interesting characteristic of this Schr\"{o}dinger equation  is that, as in the
flat brane case, it has a 
``supersymmetric'' form (see \cite{Miemiec:2000eq}) which guarantees
that it has no tachyonic modes.

Since the system is compact, \textit{i.e.} for $z_{L}<z_{\infty}$, 
it gives rise to a normalizable massless zero mode, something that is
expected since the volume of the extra dimension is finite.  The zero
mode wavefunction is given by:
\begin{equation}
\hat{\Psi}^{(0)}(z)=\frac{A}{[\cos(\tilde{k}(z_{0}-|z|))]^{3/2}}
\end{equation}
where the normalization factor $A$ is determined by the requirement 
$\displaystyle{\int_{-z_{L}}^{z_{L}}
dz\left[\hat{\Psi}^{(0)}(z)\right]^2=1}$, chosen so that we get the standard 
form 
of the Fierz-Pauli Lagrangian.

\begin{figure}[t]
\begin{center}
\begin{picture}(300,220)(0,50)

\Text(-10,250)[c]{$+$}
\Text(310,250)[c]{$+$}
\Text(160,130)[c]{$y_0$}
\Text(310,130)[c]{$2y_0$}

\SetWidth{2}

\SetColor{Red}
\Curve{(10,240)(50,192)(65,181)(80,173)(95,167)(110,161)(130,155)(150,150)}
\Curve{(150,150)(170,145)(190,139)(205,133)(220,127)(235,119)(250,108)(290,60)}
\Text(235,85)[c]{\Red{1st-KK}}

\SetColor{Blue}
\Curve{(10,240)(50,192)(65,181)(80,173)(95,167)(110,162)(130,157)(150,155)}
\Curve{(150,155)(170,157)(190,162)(205,167)(220,173)(235,181)(250,192)(290,240)}
\Text(235,220)[c]{\Blue{0-mode}}

\SetColor{Green}
\Curve{(10,145)(15,149)(20,150)(40,153)(65,158)(140,209)(150,210)}
\Curve{(150,210)(160,209)(235,158)(260,153)(280,150)(285,149)(290,145)}
\Text(110,220)[c]{\Green{2nd-KK}}


\SetWidth{.5}
\SetColor{Black}
\Line(10,150)(290,150)
\Line(10,50)(10,250)
\Line(290,50)(290,250)

\Text(370,220)[c]{$m$}
\LongArrow(350,50)(350,220)
\SetColor{Green}
\Vertex(350,163){3}
\Vertex(350,177){3}
\Vertex(350,188){3}
\SetColor{Red}
\Vertex(350,80){3}
\SetColor{Blue}
\Vertex(350,50){3}

\end{picture}
\end{center}
\caption{The graviton (blue line), first (red line) and second
(green line) KK states wavefunctions in the symmetric $''++''$
model. All wavefunctions are continuous at the position of the bounce. On the right the pattern of the
spectrum is sketched.}
\label{++wfunct}
\end{figure}
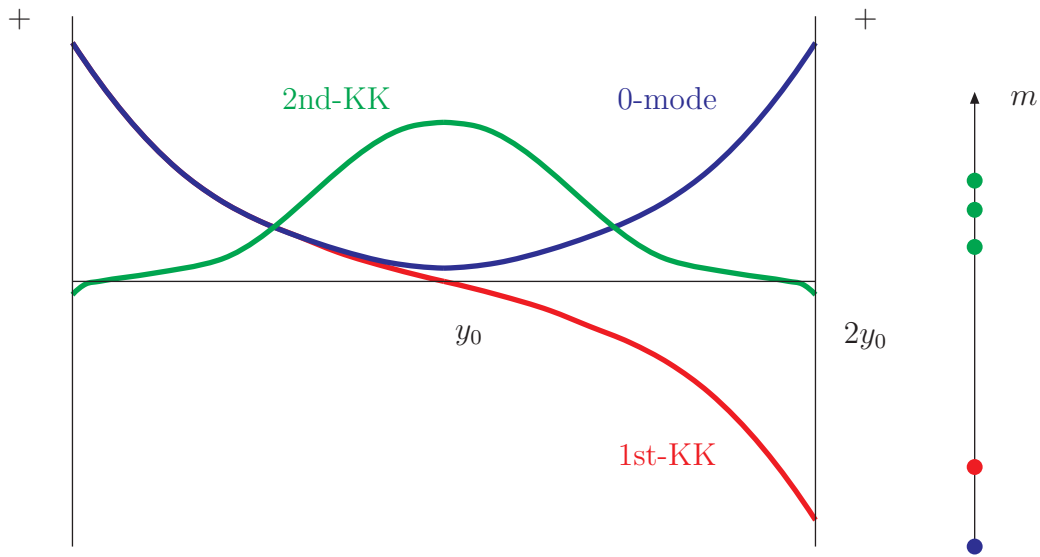

The form of the zero mode resembles the one  of the zero mode of the $''+-+''$
model, \textit{i.e.} it has a bounce form with the turning at $z_{0}$ 
(see Fig.\ref{++wfunct}). In the
case of the $''+-+''$ model the cause for this was the presence of the
$''-''$ brane. In the present model considering
$AdS_4$ spacetime on the branes we get the same effect.

In the case that we send the second brane to infinity
(\textit{i.e.} $z\rightarrow {z_{\infty}}$) the zero mode fails to be normalizable
due to singularity of the wavefunction exactly at that point. This can
be also seen from (\ref{plank}) which implies that at this limit $M_{*}$
becomes infinite (\textit{i.e.} the coupling of the zero mode becomes zero). Thus in this limit the
model has no zero mode and all gravitational interactions must be
produced by the ultralight first KK mode. The spectrum in this
case was discussed by Randall and Karch in \cite{Karch:2001ct}. 

Considering the Schr\"{o}dinger equation for $m\ne0$ we can determine
the wavefunctions of the KK tower. It turns out that the differential
equation can be brought to a hypergeometric form, and hence  the
general solution  is given in terms two hypergeometric functions: 
\begin{equation}
\renewcommand{\arraystretch}{1.5}
\begin{array}{c}\hat{\Psi}^{(n)}=\cos^{5/2}(\tilde{k}(|z|-z_{0}))\left[C_{1}~F(\tilde{a}_{n},\tilde{b}_{n},\frac{1}{2};\sin^{2}(\tilde{k}(|z|-z_{0})))~~~~~~~~\right.
\\ \left.  ~~~~~~~~~~~~~~~~~+C_{2}~|\sin(\tilde{k}(|z|-z_{0}))|~F(\tilde{a}_{n}+\frac{1}{2},\tilde{b}_{n}+\frac{1}{2},\frac{3}{2};\sin^{2}(\tilde{k}(|z|-z_{0})))\right]
\end{array}
\end{equation}
where
\begin{eqnarray}
\tilde{a}_{n}=\frac{5}{4}+\frac{1}{2}\sqrt{\left(\frac{m_{n}}{\tilde{k}}\right)^2+\frac{9}{4}}
\cr \tilde{b}_{n}=\frac{5}{4}-\frac{1}{2}\sqrt{\left(\frac{m_{n}}{\tilde{k}}\right)^2+\frac{9}{4}}
\end{eqnarray}
with the same normalization convention as the zero mode. The boundary conditions (\textit{i.e.} the jump of the wave function at the points
$z=0$, $z_{L}$) result in a
$2\times2$ homogeneous linear system which, in order to have a
non-trivial solution, should have vanishing determinant. In the
symmetric configuration which we consider, this procedure can be
simplified by considering even and odd functions with respect to the
minimum of the potential $z_{0}$.

In more detail, the odd eigenfunctions obeying
the  boundary condition  $\hat{\Psi}^{(n)}(z_{0})=0$  will  have $C_1=0$ and thus the form:
\begin{equation}
\hat{\Psi}^{(n)}=C_{2}\cos^{5/2}(\tilde{k}(|z|-z_{0}))|\sin(\tilde{k}(|z|-z_{0}))|~F(\tilde{a}_{n}+\frac{1}{2},\tilde{b}_{n}+\frac{1}{2},\frac{3}{2};\sin^{2}(\tilde{k}(|z|-z_{0})))
\end{equation}

On the other hand, the even eigenfunctions obeying
the  boundary condition  $\hat{\Psi}^{(n)}~'(z_{0})=0$  will have $C_2=0$ and thus the form:
\begin{equation}
\hat{\Psi}^{(n)}=C_{1}\cos^{5/2}(\tilde{k}(|z|-z_{0}))F(\tilde{a}_{n},\tilde{b}_{n},\frac{1}{2};\sin^{2}(\tilde{k}(|z|-z_{0})))
\end{equation}

The remaining boundary condition is given by:
\begin{equation}
\hat{\Psi}^{(n)}~'(0)+\frac{3k}{2}\tanh(ky_{0})\hat{\Psi}^{(n)}(0)=0
\end{equation}
and determines the mass spectrum of the KK states. From this
quantization condition we get that the KK spectrum has a special first
mode similar to the one of the $^{\prime \prime
}+-+^{\prime \prime }$ bigravity model. For $ky_0 \ga 5$  the  mass 
of the first mode is given by the approximate relation:
\begin{equation}
m_1=4\sqrt{3}~k~e^{-2ky_{0}}
\label{mass1}
\end{equation}

In contrast, the masses of the next levels, if we  put together the
results for even and odd wavefunctions, are given by the formula:
\begin{equation}
m_{n+1}=2\sqrt{n(n+3)}~k~e^{-ky_{0}}
\label{mr}
\end{equation}
with $n=1,2,...$.

We note that the first KK state has a different scaling law with respect
to the position of the minimum of the warp factor compared
to the rest of the KK tower, since it scales as $ e^{-2ky_{0}}$ while
the rest of the tower scales as $e^{-ky_{0}}$. Thus, the first
KK state is generally much lighter than the rest of the tower. It is clear that
this mass spectrum resembles the one of the $^{\prime \prime
}+-+^{\prime \prime }$ bigravity model. The deeper
reason for this is again the common form of the warp factor. In both
cases the warp factor has a minimum due to its bounce
form. The graviton wave function  follows
the form of the warp factor, \textit{i.e.} it is symmetric with respect to $z_{0}$, while
 the wavefunction of the first KK state is antisymmetric in
respect to $z_{0}$ (see Fig.\ref{++wfunct}). The absolute values of the two wavefunctions  are almost identical in all
regions except near $z_{0}$ (the symmetric is nonzero while the
antisymmetric is zero at $z_{0}$). The graviton
wavefunction is suppressed by the factor
$\cosh^{-2}(ky_{0})$ at $z_{0}$
which brings its value close to zero for reasonable values of
$ky_{0}$. Thus, the mass
difference which is determined by the wavefunction near  $z_{0}$ is
expected  to be generally very small, a fact which appears as
the extra suppression factor $e^{-ky_{0}}$ in the formula of $m_{1}$
in comparison with the rest of the KK tower.

In the case that we consider an asymmetric brane configuration,
for example
$z_{L}>2z_{0}$, the spectrum is effectively independent of the position
of the second brane $z_{L}$ (considering $ky_{0}\ga 5$). Thus, even in the case that we place the
second brane at $z_{\infty}$, \textit{i.e.} the point which corresponds to infinity in the
$y$-coordinates, the spectrum is given essentially by the same
formulas. This decompactification limit is exactly what has been
studied in \cite{Karch:2001ct}.

The coupling also of the first  KK state to matter on the branes is
 very different from the one of the remaining tower. We can easily
calculate it by the same procedure presented in chapter 2
considering the a minimally coupled to gravity matter Lagrangian. In
the case of scale equivalent positive tension branes, the gravitons
will couple to matter on them ($z=z_{br}=0~{\rm{or}}~L$) with the following couplings: 
\begin{equation} 
a_n=\frac{\hat{\Psi}^{(n)}\left(z_{br}\right)}{ 2\sqrt{2M^3}}
\label{coupling++}
\end{equation}   

In more detail we find that the zero mode and the first ultralight KK
state couple with the same strength on the positive tension branes:
\begin{equation} 
a_0=a_1=\frac{1}{ 2\sqrt{2}M_*}
\label{++coupl}
\end{equation}   
while the coupling of the remaining tower is exponentially
suppressed (it scales as $e^{-ky_{0}}$). The reason for the strong coupling of the first KK state in comparison 
with the remaining tower is readily understood because, being
dominantly a bound  state of the volcano potential on the positive tension brane, it is largely localized on it.

Exploiting the different mass scaling of the first KK relative to the
rest we can ask whether it is possible to realize a bigravity
scenario similar to that in $^{\prime \prime
}+-+^{\prime \prime }$ bigravity model. 
 In that model by appropriately choosing the position of the negative
tension brane, it was possible to make the first KK state have mass such that the
corresponding wavelength is  of the order of the cosmological scale that gravity
has been tested and at the same time have the rest of the KK tower wavelengths
below 0.2mm. In
this scenario the gravitational interactions are due to the net effect of the massless graviton and the first ultralight KK
state. A similar situation can be realized in the $''++''$ model as
well. From (\ref{++coupl}) it can be understood that in the symmetric 
$''++''$
configuration the massless graviton and the special KK state
contribute by the same amount to the gravitational interactions. In other words:
\begin{equation}
\frac{1}{M_{{\rm{Pl}}}^2}=\frac{1}{M_{*}^2}+\frac{1}{M_{*}^2}~~~ \Rightarrow~~~ M_{{\rm{Pl}}}=\frac{M_{*}}{\sqrt{2}}
\end{equation}

However, if our aim is to see modifications of gravity at
ultralarge distances, this is impossible because the Compton wavelength of our ultralight
graviton will be $e^{ky_0}$ times bigger than the horizon $H^{-1}$ of the
$AdS_4$ space on our brane due to (\ref{cosmolcon}), (\ref{mass1}). The ``Hubble''
parameter follows $m_2$ rather than $m_1$. Moreover, in the present model, the fact that the effective four dimensional
cosmological constant should be set very close to zero, requires that
the  warp factor is constrained by  \mbox{$ky_{0}\geq{135}$} and thus,
in this case, the spectrum of the remaining KK states will be very dense (tending
to continuum) bringing more states close to zero mass. The KK states
that have masses which correspond to wavelengths larger than $1$mm have
sufficiently small coupling so that there is no conflict with 
phenomenology (the situation is exactly similar to the RS2 case
where the coupling of the  KK states is proportional to their mass and thus
it is decreasing for lighter KK states).  The fact that the spectrum
tends to a continuum shadows
the special role of the first KK state.

It is interesting to note that
at the limit where the minimum of the warp factor is sent to infinity ($z_{\infty}$)
 the special behaviour of the first KK persists and does not catch the
other levels (by changing its scaling law) as
was the case in $''+--+''$ model of chapter 2. This means that the limit  $z\rightarrow
z_{\infty}$ will indeed
be identical to \textit{two}  RS2, but on the other hand we should
emphasize that what we call graviton in the RS2 limit is actually the
combination of a massless graviton
\textit{and} the ``massless'' limit of the special first KK
state. This ``massless'' limit exists as we will see in the next
section and  ensures that locality is respected by the
model, since physics on the brane does not get affected from the
boundary condition at infinity.

Finally, let us note that one can build  a crystalline model as in the 
flat brane case but without the need to introduce negative tension
branes. In this case one expects again to get a band structure for the 
KK continuum with the width of the first band (starting from zero)
much smaller than the width of the first forbidden zone. However, as
in the bigravity case the width of the first band which will mediate
normal four dimensional gravity at intermediate distances will be
linked to the remnant cosmological constant on the branes and quite
probably one will get modifications of gravity for distances much
bigger than the $AdS$ horizon. The analogue, however, of the quasi-localized model cannot be realized
in the $AdS$ brane case as it can be seen from  (\ref{ad2}),
(\ref{add}) since an asymptotically constant warp factor is forbidden.

\section{Graviton propagator in $dS_4$ and $AdS_4$ space}

In this section we will demonstrate how the van Dam-Veltman-Zakharov
no-go theorem for the non-decoupling of the extra polarization states
of the massive graviton can be evaded in $dS_4$ or $AdS_4$ space. For
the case of $dS_4$ there has been a proof in
\cite{Higuchi:1987py,Higuchi:1989gz}, but we will present an
alternative one in this section together with the case of $AdS_4$ for
the sake of completeness. In more
detail, as we have discussed at the end of chapter 2, in flat spacetime the $m\rightarrow0$ limit
of the massive graviton propagator does not give the massless one due
to the non-decoupling of the additional longitudinal components. This
generates  the well known discontinuity between massive and massless states. Considering the massive graviton
propagator in $dS_{4}$ or $AdS_{4}$ spacetime we can show that this result
persists if $m/H \rightarrow \infty$ where $H$ is the  ``Hubble'' parameter,  \textit{i.e.} the
discontinuity is still present in the  $m\rightarrow 0$ limit if it
happens that 
$m/H \rightarrow \infty$. However,
in the case that  $m/H \rightarrow 0$, we will explicitly show that
the $m \rightarrow 0$ limit is smooth. This
 is an important result since it gives us the possibility to circumvent
the van Dam - Veltman - Zakharov no go theorem about the
non-decoupling of the  extra graviton
polarizations. Thus, in the limit that $m/H \rightarrow 0$ 
all the extra polarizations of the graviton decouple, giving an
effective theory of a  massless graviton with just two polarization
states.

Our aim is to find the form of the massless and massive
graviton propagators in the case of $dS_4$ and $AdS_4$ spacetime with arbitrary
``Hubble'' parameter $H$ and graviton mass $m$. We wish to examine the behaviour of
these propagators in the limit where $H\rightarrow0$ and the limit
where the mass of the graviton tends to zero.
 For simplicity we will do our calculations in Euclidean $dS_4$ or
$AdS_4$ space. We can use for metric the one of the stereographic
projection of the sphere or the hyperboloid\footnote{Note that \cite{D'Hoker:1999jc} and
\cite{Naqvi:1999va} whose results we use in the following have a different
metric convention, but this makes no difference for our calculations.}:
\begin{equation}
ds^2= \frac{\delta_{\mu\nu}}{\left(1 \mp {H^2 x^2 \over4}\right)^2}dx^{\mu}dx^{\nu}\equiv g^{0}_{\mu\nu}dx^{\mu}dx^{\nu}
\end{equation}
where $x^2=\delta_{\mu\nu}x^{\mu}x^{\nu}$ and the scalar curvature is $R=\mp 12~H^2$. From now on, the upper
sign corresponds to $AdS_4$ space while the lower to $dS_4$ space.
The fundamental invariant in these spaces is the geodesic distance $\mu(x,y)$
between two points $x$ and $y$. For convenience, we will introduce
another invariant $u$ which is related with the geodesic distance 
by the relation $u=\cosh (H \mu)-1$ for $AdS_4$ ~($u\in[0,\infty)$)
and the relation $u=\cos (H \mu)-1$ for $dS_4$ ~($u\in[-2,0]$).  In
the small distance limit $u\sim \pm {\mu^2 H^2 \over 2}$.

This background metric is taken by the the Einstein-Hilbert action:
\begin{equation}
{\cal S}_G=\int d^4x \sqrt{g}\left(2M_4^2 R-\Lambda\right)
\end{equation}
where the cosmological constant is
$\Lambda=\mp 12~H^2 M_4^2$ and $M_4$ the four dimensional fundamental scale. The spin-2 massless 
graviton field can be obtained by the linear metric
fluctuations $ds^2=\left(g^0_{\mu\nu}+h_{\mu\nu}\right)dx^{\mu}dx^{\nu}$. This
procedure gives us the analogue of the Pauli-Fierz graviton action in
curved space:
\begin{eqnarray}
\frac{{\cal S}_0}{2 M_4^2}= \int d^4x \sqrt{g^0} \left\{-{1\over 4}h \bo^0 h +{1\over
2}h^{\mu \nu}\nabla^0_{\mu}\nabla^0_{\nu}h +{1\over 4}h^{\mu \nu} \bo^0
h_{\mu \nu} - {1\over
2} h^{\mu \nu}\nabla^0_{\nu}\nabla^0_{\kappa}
h^{\kappa}_{\mu}\right.\cr \pm \left. {1\over
2}H^2\left({1\over 2}h^2 +h_{\mu \nu}h^{\mu \nu}\right)\right\} 
\end{eqnarray}
The above action is invariant under the gauge transformation $\delta
h_{\mu\nu}=\nabla^0_{\mu}\xi_{\nu}+\nabla^0_{\nu}\xi_{\mu}$ which
guarantees that the graviton has only two physical degrees of
freedom. This is precisely the definition of masslessness in $dS_4$ or 
$AdS_4$
space (for example see \cite{Buchbinder:2000fy} and references therein).

The propagator of the above spin-2 massless field can be written in the form:
\begin{equation}
G^{0}_{\mu\nu;\mu'\nu'}(x,y)=(\partial_{\mu}\partial_{\mu'}u\partial_{\nu}\partial_{\nu'}u+\partial_{\mu}\partial_{\nu'}u\partial_{\nu}\partial_{\mu'}u)G^{0}(u)+g_{\mu\nu}g_{\mu'\nu'}E^{0}(u)+D[\cdots]
\label{mlesspr}
\end{equation}
where $\partial_{\mu}={\partial \over \partial x^{\mu}}$,
$\partial_{\mu'}={\partial \over \partial y^{\mu'}}$. The  last term, denoted $D[\cdots]$, is a total derivative and drops out of the
calculation when integrated with a conserved energy momentum tensor. Thus, all physical information is encoded in the first two terms.

The process of finding the functions $G^{0}$ and $E^{0}$ is quite
complicated and is the result of solving a system of six coupled differential
equations \cite{D'Hoker:1999jc}. We will  present here only the differential equation
that $G^{0}$ satisfies to show the difference between $AdS_4$ and
$dS_4$ space. This equation results from various integrations and has
the general form:
\begin{equation}
u(u+2)G^{0}(u)''+4(u+1)G^{0}(u)'=C_1 +C_2 u
\label{diff}
\end{equation}
where the constants $C_1$ and $C_2$ are to be fixed by the boundary
conditions. For the case of the $AdS_4$ space \cite{D'Hoker:1999jc}, these
constants were set to zero so that the  $G^{0}$ function vanishes at
the boundary  at infinity ($u \rightarrow \infty$). Using the same condition also for the  $E^{0}$ function, the 
exact form of them was found to be:
\begin{eqnarray}
G^{0}(u)&=&\frac{1}{8 \pi^2 H^2}\left[\frac{2(u+1)}{u(u+2)}-\log \frac{u+2}{u}\right]\cr
E^{0}(u)&=&-\frac{ H^2}{8
\pi^2}\left[\frac{2(u+1)}{u(u+2)}+4(u+1)-2(u+1)^2\log \frac{u+2}{u}\right]
\end{eqnarray}

For the case of the $dS_4$ space we iterated the procedure of
\cite{D'Hoker:1999jc} imposing the condition \cite{Allen:1986wd} that the  $G^{0}$ and $E^{0}$
functions should be non-singular at the antipodal point
($u=-2$). The constants $C_1$ and $C_2$ were kept non-zero and played
a crucial role in finding a consistent solution. It is straightforward 
to find the full expression of these functions, but we only need to know their short distance behaviour. Then with this accuracy the answer is:
\begin{eqnarray}
G^{0}(u)&=&-\frac{1}{8 \pi^2 H^2}\left[\frac{1}{u}+\log (-u)\right] +
\cdots \cr
E^{0}(u)&=&\frac{H^2}{8
\pi^2}\left[\frac{1}{u}+2(u+1)^2\log (-u)\right]+\cdots 
\end{eqnarray}

If we define $\Pi^{0}(u)=\frac{1}{H^4}\frac{E^{0}(u)}{G^{0}(u)}$, then 
for short distances ($H^2x^2 \ll 1$) where $u \rightarrow 0$ we get:
\begin{eqnarray}
g_{\mu\nu}g_{\mu'\nu'}&\rightarrow & \delta_{\mu\nu} \delta_{\mu'\nu'} 
\cr
\partial_{\mu}\partial_{\nu'}u &\rightarrow& \mp H^2 \delta_{\mu \nu'} \cr G^0(u) &\rightarrow& {1 \over 4 \pi^2 H^4 \mu^2} \cr
\Pi^{0}(u) &\rightarrow& -1
\end{eqnarray}
and so we recover
the short distance limit of the massless flat Euclidean space propagator:
\begin{equation}
G^{0}_{\mu\nu;\mu'\nu'}(x,y)=\frac{1}{4 \pi^2
\mu^2}(\delta_{\mu\mu'}\delta_{\nu\nu'}+\delta_{\mu\nu'}\delta_{\nu\mu'}-\delta_{\mu\nu}\delta_{\mu'\nu'})+\cdots
\label{mlessprop}
\end{equation}
Of course this is just as expected.

In order to describe a spin-2 massive field it is necessary to add to the above 
action  a Pauli-Fierz mass term:
\begin{equation}
\frac{{\cal S}_m}{2 M_4^2}=\frac{{\cal S}_0}{2 M_4^2}- \frac{m^2}{4}\int d^4 x \sqrt{g^0}(h_{\mu \nu}h^{\mu \nu}-h^2)
\end{equation}
By adding this term we immediately lose the gauge invariance
associated with the $dS_4$ or $AdS_4$ symmetry group and the
graviton acquires five degrees of freedom and becomes massive.

The propagator of this massive spin-2 field  can again be 
written in the form:
\begin{equation}
G^{m}_{\mu\nu;\mu'\nu'}(x,y)=(\partial_{\mu}\partial_{\mu'}u\partial_{\nu}\partial_{\nu'}u+\partial_{\mu}\partial_{\nu'}u\partial_{\nu}\partial_{\mu'}u)G^{m}(u)+g_{\mu\nu}g_{\mu'\nu'}E^{m}(u)+D[\cdots]
\label{mivepr}
\end{equation}
The last term
of the propagator in (\ref{mivepr}), denoted $D[\cdots]$, is again a total derivative and thus
drops out of the calculation when integrated with a conserved $T_{\mu
\nu}$.

At his point we should emphasize that in case of an arbitrary massive
spin-2 field, the absence of
gauge invariance means that there is no guarantee that the field will
couple to a conserved current. However, in the context of a higher
dimensional theory  whose symmetry group is spontaneously broken by
some choice of vacuum metric,  the massive spin-2 graviton KK states 
couple to a conserved $T_{\mu \nu}$. One can understand this by the
following example. Consider the case of the most simple KK
theory, the one with one compact extra dimension. By the time 
we choose a vacuum metric \textit{e.g.} $g^0_{MN}={\rm diag}\left(\eta_{\mu \nu},1\right)$, the higher dimensional
symmetry is broken. If we denote the
graviton fluctuations around
the background metric by $h_{\mu \nu}$, $h_{\mu 5}$ and $h_{55}$, there is still the gauge freedom:
\begin{eqnarray}
\delta h_{\mu \nu}&=&\partial_{\mu}\xi_{\nu}+\partial_{\nu}\xi_{\mu}\nonumber\\
\delta h_{\mu 5}&=&\partial_{\mu}\xi_{5}+\partial_{5}\xi_{\mu}\label{gauge}\\
\delta h_{5 5}&=&2\partial_{5}\xi_{5}\nonumber
\end{eqnarray}

If we Fourier decompose these fields, their $n$-th Fourier mode acquires
a mass $m_n \propto n$ with $n=0,1,2,\dots$, but there is mixing
between them. This means for example that $h^{(n)}_{\mu \nu}$ is not a 
massive spin-2 eigenstate \textit{etc.}. However, we can exploit the gauge
transformations (\ref{gauge}) to gauge away the massive $h^{(n)}_{\mu
5}$ and  $h^{(n)}_{55}$ and construct a pure spin-2 field (see for
example \cite{Cho:1992rq} and references therein). For a comprehensive account of KK theories see \cite{Appelquist:1987nr}. The new massive spin-2 field $\rho^{(n)}_{\mu \nu}$ is
invariant under (\ref{gauge}) and so its Lagrangian does not exhibit a
gauge invariance of the form $\delta \rho_{\mu
\nu}=\partial_{\mu}\chi_{\nu}+\partial_{\nu}\chi_{\mu}$. However,
since is originates from a Lagrangian that has the gauge invariance
(\ref{gauge}), it is bound to couple to a conserved $T_{\mu
\nu}$. The argument goes on for more complicated choices of vacuum
metric as for example warped metrics.

Again the functions $G^{m}$ and $E^{m}$ result from a complicated system of differential
equations \cite{Naqvi:1999va}. In that case, the differential equation
that $G^{m}$ satisfies is:
\begin{equation}
u(u+2)G^{m}(u)''+4(u+1)G^{m}(u)'\mp \left({m\over H}\right)^2 G^{m}(u)=C_1 +C_2 u
\label{diffm}
\end{equation}
where the constants $C_1$ and $C_2$ are to be fixed by the boundary
conditions. For the case of the $AdS_4$ space \cite{Naqvi:1999va}, these
constants were set to zero so that the  $G^{0}$ function vanishes at
the boundary  at infinity. Imposing 
additionally  the condition of fastest falloff at infinity ($u
\rightarrow \infty$) \cite{Allen:1986wd}, the exact form of the  $G^{m}$ and  $E^{0}$
function was found to be:
\begin{eqnarray}
G^{m}(u)&=&\frac{\Gamma(\Delta)\Gamma(\Delta-1)}{16 \pi^2
\Gamma(2\Delta-2)
H^2}\left(\frac{2}{u}\right)^{\Delta}F(\Delta,\Delta-1,2\Delta-2,-{2\over 
u})\cr
E^{m}(u)&=&-\frac{2}{3}~\frac{\Gamma(\Delta-1) H^2}{16
\pi^2 \Gamma(2\Delta
-2)[2+(m/H)^2]}\left(\frac{2}{u}\right)^{\Delta}\times\cr
& \times & \left\{ \renewcommand{\arraystretch}{1.5} \begin{array}{l} \phantom{-} 3[2+(m/H)^2]
\Gamma(\Delta-2) u^2 F(\Delta-1,\Delta-2,2\Delta-2,-{2\over u})
\\-3(u+1)u F(\Delta-1,\Delta-1,2\Delta-2,-{2\over u})\\+[3+(m/H)^2]\Gamma(\Delta)F(\Delta,\Delta-1,2\Delta-2,-{2\over u}) \end{array} \right\}\phantom{aa}
\end{eqnarray}
where $\Delta={3\over 2}+{1\over 2}\sqrt{9+4(m/H)^2}$. 

For the case of the $dS_4$ space we iterated the procedure of
\cite{Naqvi:1999va} imposing the condition \cite{Allen:1986wd} that the  $G^{m}$ and $E^{m}$
functions should be non-singular at the antipodal point
($u=-2$) and also finite as $m \rightarrow 0$. Again we kept the
constants $C_1$ and $C_2$  non-zero to obtain a consistent solution. It is straightforward to find the full expression of these functions, but we only need to know their short distance behaviour. Then with this accuracy the answer is:
\begin{eqnarray}
G^{m}(u)&=&\frac{\Gamma(\Delta)\Gamma(3-\Delta)}{16 \pi^2
H^2}\left[F(\Delta,3-\Delta,2,{u+2\over 2})-1\right]+\cdots \cr
E^{m}(u)&=&-\frac{2}{3}~\frac{\Gamma(\Delta)\Gamma(3-\Delta) H^2}{4
\pi^2 [2-(m/H)^2]}~\times\cr
& \times & \left\{ \renewcommand{\arraystretch}{1.5} \begin{array}{l} -3[2-(m/H)^2]\left[{2(u+2)\over(\Delta-1)(\Delta-2)}F(\Delta-1,2-\Delta,2,{u+2\over 2})+{u(u+2)\over2}\right]
\\-3(u+1)\left[{2\over(\Delta-1)(\Delta-2)} F(\Delta-1,2-\Delta,1,{u+2\over 2})+(u+1)\right]
\\+[3-(m/H)^2]\left[F(\Delta,3-\Delta,2,{u+2\over 2})-1\right]
\end{array} \right\}+\cdots \phantom{aaaAA}
\end{eqnarray}
where $\Delta={3\over 2}+{1\over 2}\sqrt{9-4(m/H)^2}$. 

If we define
$\Pi^{m}(u)=\frac{1}{H^4}\frac{E^{m}(u)}{G^{m}(u)}$, then 
for short distances ($H^2x^2 \ll 1$) where $u \rightarrow 0$ we get:
\begin{eqnarray}
G^m(u) &\rightarrow& {1 \over 4 \pi^2 H^4 \mu^2} \cr
\Pi^m(u) &\rightarrow& -{2\over 3}~\frac{~~3 \pm \left( {m \over H}
\right)^2}{~~2 \pm \left( {m \over H} \right)^2}
\label{ratio}
\end{eqnarray}

It is interesting to consider two massless flat limits. In the first
one  $m \rightarrow 0$ and $H \rightarrow 0$  while ~$m/H \rightarrow
\infty$. In this case, from
(\ref{ratio}) we see that we
recover the Euclidean propagator for a massive graviton in flat space:
\begin{equation}
G^{m}_{\mu\nu;\mu'\nu'}(x,y)=\frac{1}{4 \pi^2
\mu^2}(\delta_{\mu\mu'}\delta_{\nu\nu'}+\delta_{\mu\nu'}\delta_{\nu\mu'}-{2\over 
3}\delta_{\mu\nu}\delta_{\mu'\nu'})+\cdots
\end{equation}
This is in agreement with the van Dam - Veltman - Zakharov
theorem. The second limit has $m \rightarrow 0$ and $H \rightarrow 0$
but $m/H \rightarrow 0$. In this case the propagator passes smoothly
to the one of the flat massless
case (\ref{mlessprop}):
\begin{equation}
G^{0}_{\mu\nu;\mu'\nu'}(x,y)=\frac{1}{4 \pi^2
\mu^2}(\delta_{\mu\mu'}\delta_{\nu\nu'}+\delta_{\mu\nu'}\delta_{\nu\mu'}-\delta_{\mu\nu}\delta_{\mu'\nu'})+\cdots
\end{equation}
This is in contrary to the van Dam - Veltman  - 
Zakharov discontinuity in flat
space.

In general, we may consider the limit with $m/H$ finite. Then for
small   $m/H$ the contribution to the $\delta_{\mu\nu}\delta_{\mu'\nu'}$ 
structure is  $ - 1  \pm m^2/6 H^2$.  Since observations agree to $0.02\%$ 
accuracy with the prediction of Einstein gravitational theory for the
bending of light by the sun, we obtain the limit ${m
\over H} \la 0.04$. However, we should point out that the above results are meaningful only for
the $AdS$ case because in the $dS$ there are no unitary spin-2
representations in the region $0<m/H<2$ as it was shown in \cite{Higuchi:1987py,Higuchi:1989gz}.

\section{Discussion}

Let us now go back to the $''++''$ bigravity model and see how the
extra polarization states problem is evaded in the absence of a ghost
radion mode. In this model  we have a clear realization of the case
where $m/H \to 0$. Indeed the massive graviton, which in the symmetric
configuration is responsible for half of the gravitational
interactions that we feel in intermediate distances, has this
characteristic ratio to be  $m_1/H \sim e^{-k y_0}$. This
easily accommodates the bound that we presented earlier.  Then, the
Euclidean propagator (in configuration space) of the massive KK states   for relatively large $y_0$ will be given by:
\begin{equation}
G^{m}_{\mu\nu;\mu'\nu'}(x,x')=\frac{1}{4 \pi^2
\mu^2}(\delta_{\mu\mu'}\delta_{\nu\nu'}+\delta_{\mu\nu'}\delta_{\nu\mu'}-\left(
1-{1\over 6}e^{-2 ky_0}\right)\delta_{\mu\nu}\delta_{\mu'\nu'})
\end{equation}
where $\mu$ is the geodesic distance between two points. In the above, 
we have omitted terms that do not contribute when integrated with 
a conserved $T_{\mu \nu}$. For $ky_0 \ga 2.3$ there is no problem with the bending of
light since the extra polarization states are practically decoupled.
Additionally, as we discussed earlier the large scale  modifications
of gravity are  impossible to be observed because the Compton
wavelength of our ultralight graviton is exponentially larger than the horizon $H^{-1}$ of the
$AdS_4$ space on our brane. The facts that there is no
observable deviation from normal four dimensional Einsteinian gravity
in this model and that there is a cosmological constant
of negative sign present, make this model phenomenologically disfavoured.

Let us now make some further comments regarding the (dis)continuity of 
the graviton propagator. It was shown in
\cite{Dilkes:2001av,Duff:2001zz}  that the extra polarizations are
resurrected in quantum loops and thus the discontinuity reappears in
the quantum level. However, purely four dimensional theory with
massive gravitons where the  mass of the graviton is added by hand is
not well-defined as it was shown in \cite{Boulware:1972my}. If instead
one starts  from a higher dimensional theory in which the mass of the
gravitons is generated dynamically, as we are doing in the models we
have presented so far, and if one takes into account the whole KK
graviton tower in the quantum loops, this quantum discontinuity is
expected to be absent. It is worth noting that in \cite{Karch:2001jb}
it was suggested that the extra polarization states of the massive
graviton in the one $AdS_4$ brane model \cite{Karch:2001ct} can be
gauged away and thus the discontinuity is absent.

Moreover, one expects that the smoothness
of the limit $m \rightarrow  0$ 
  is not only a property of the $AdS_4$ space but holds 
for any background where the characteristic curvature invariants are  non-zero. For physical processes taking place 
 in some region of a curved space with some non-zero characteristic average curvature, 
the effect of graviton mass is controlled by positive powers of the ratio 
 $m^2/R^2$  where $R^2$ is a characteristic curvature invariant (made
from Riemann and Ricci tensors or scalar curvature).

Finally,  the
conjecture that  there is a smooth limit for phenomenologically 
 observable amplitudes in brane gravity  with ultralight gravitons 
 is also supported by results based on a very interesting paper   \cite{Vain}.
  In that paper it  was shown that the argument of van Dam, Veltman
and Zakharov that no mass for the graviton is allowed in flat
background is not entirely robust. In more detail, is was shown  that there is a smooth
limit for a   metric  around a spherically symmetric  source with a
mass  $M$   in a theory with
massive graviton  with  mass $m$ for distances  smaller than a
critical one which is $m^{-1}(mM/M_{{\rm{Pl}}}^2)^{1/5}$. This critical distance 
for a graviton of wavelength of the size of the observable universe
and for a massive body of the mass of the Sun, is of the order of the solar system size. In 
order to see this smooth limit one should perform a non-perturbative
analysis \cite{Deffayet:2001uk} since in this regime the lowest order
approximation breaks down. The discontinuity reappears for
distances larger than this critical distance where the linear
approximation is valid. This breakdown of perturbativity, is due to
the fact that the lowest order
approximation calculation does not take into account the
characteristic mass scale of the problem which in this case is the
mass of the Sun. Since the bending of the light is observed in the
first non-perturbative region, according to \cite{Deffayet:2001uk} it
is impossible to rule out the massive graviton proposal. 

However,  the
non-perturbative  solution discussed in \cite{Vain,Deffayet:2001uk} was found in a limited range of distance from
the center and it is still unclear if it can be
smoothly continued to spatial infinity (this problem was stressed in 
 \cite{Boulware:1972my}). Existence of this smooth
continuation depends on the full  nonlinear structure of the
theory. If one adds a mass term by hand the smooth asymptotic at
infinity may not exit. However, it 
 seems  plausible   that 
 in all cases when modification of gravity at large distances comes
from consistent higher-dimensional models, the global  smooth solution 
can exist because in this case there is 
a unique non-linear structure related to the mass term  which is
dictated by the  underlying 
higher-dimensional theory.

\chapter{Multigravity in six dimensions}

 As we have
discussed, in five dimensions it is
impossible to have flat brane multigravity models without negative
tension branes. For this reason, we will consider models in six dimensions. The
literature on six dimensional constructions is already quite rich 
\cite{Chodos:1999zt,Cohen:1999ia,Gregory:2000gv,Chacko:2000eb,Arkani-Hamed:2000dz,Chen:2000at,Olasagasti:2000gx,Gherghetta:2000qi,Chodos:2000tf,Gherghetta:2000jf,Ponton:2001gi,Moon:2001hn,Charmousis:2001hg,Corradini:2001su,Collins:2001ni,Kim:2001rm,Kanti:2001vb,Hayakawa:2001ke,Leblond:2001xr}.
In this chapter we explicitly show that the difficulties of the five
dimensional models can be evaded in six dimensions. It is possible to construct
multigravity models
with only flat positive tension branes.  This is due to the fact that in six dimensions the internal space 
is not trivial and its curvature allows bounce configurations with the 
above feature. The branes which localize gravity in
this setup are four-branes, but one of their dimensions is compact,
unwarped and of Planck length. Thus, in the low energy limit the
spacetime on the brane appears three dimensional. In order that these
constructions are realized it is crucial that the tensions and/or the bulk cosmological constant are anisotropic. The five dimensional constructions in this
setup come into two types. One of them involves conical singularities 
at finite distance from the four branes. These conical singularities
support three-branes which can be of
positive tension if one has an
angle deficit, of zero tension if one has no angle deficit and of
negative tension if one has an angle excess. The other type of the
multigravity models has no conical singularity at all. It is crucial
to note that there is no five dimensional effective theory for these
constructions, otherwise one would get all the problems faced in the
five dimensional constructions. The low energy effective theory is
directly four dimensional. These constructions give for the first time a theoretically and
phenomenologically viable realization of multigravity.

In the following section we present
the cylindrically symmetric single four-brane warped models and show the cases 
where gravity is localized on the four-branes. In section two we paste two 
of the single four-brane solutions and obtain the double four-brane bigravity
models. In section three we generalize the quasi-localized and
crystalline constructions in six dimensions. Finally, we  discuss 
the weaker energy condition in these six dimensional configurations.

\section{Single brane models in six dimensions}

At first we will discuss the single brane solutions in six dimensions to
get an insight on the more complicated multigravity
configurations. In the following subsection  we will review the
minimal model where the bulk cosmological constant is
isotropic and then a more generalized model with arbitrary
cosmological constant components along the four dimensional and the
extra dimensional directions. In the following we will implicitly
assume orbifolding around the four-brane.

\subsection{The minimal single brane model}

The simplest single brane model consists of a four-brane embedded
in six dimensional $AdS$ space \cite{Gherghetta:2000qi}. One of the longitudinal dimensions of
the four brane is compactified to a Planck length radius $R$ while the dimensions
transverse to the four-brane compact dimension are infinite.  We consider the following ansatz where  four dimensional Poincar\'e invariance is respected\footnote{Note
that in this thesis we use different metric signature and different
definition of the fundamental scale from \cite{Gherghetta:2000qi}.}:
\begin{equation}
ds^2=\sigma(\rho)\eta_{\mu \nu}dx^{\mu}dx^{\nu}+d\rho^2+\gamma(\rho)d\theta^2
\end{equation}
where $\theta$ is the compactified dimension with range $[0,2\pi]$ and
$\rho$ is the infinite radial dimension.

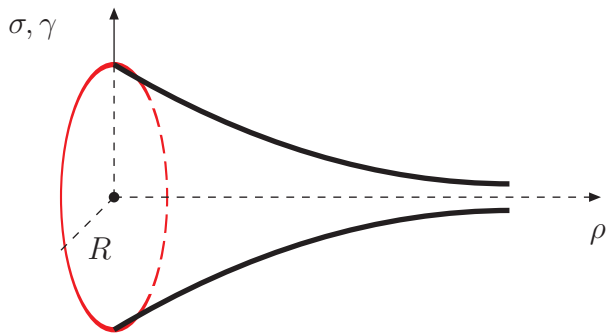
\begin{figure}[t]
\vskip10mm
\begin{center}
\begin{picture}(200,100)(0,50)

\SetWidth{2}
\SetColor{Red}
\Oval(0,100)(50,20)(0)
\SetColor{Black}
\Curve{(0,150)(150,105)(300,150)}
\Curve{(0,50)(150,95)(300,50)}
\Vertex(0,100){2}
\Text(-40,160)[lb]{$\sigma,\gamma$}

\CBox(150,50)(300,150){White}{White}
\SetWidth{.5}
\DashLine(0,100)(180,100){3}
\LongArrow(180,100)(182,100)

\LongArrow(0,150)(0,170)
\DashLine(0,100)(0,150){3}
\DashLine(0,100)(-20,80){3}

\SetWidth{3}
\SetColor{White}
\Line(10,135)(20,135)
\Line(10,125)(20,125)
\Line(10,115)(20,115)
\Line(10,105)(25,105)
\Line(10,95)(25,95)
\Line(10,85)(20,85)
\Line(10,75)(20,75)
\Line(10,65)(20,65)

\SetWidth{2}
\SetColor{Black}
\Text(180,90)[lt]{$\rho$}
\Text(-10,85)[lt]{$R$}

\end{picture}
\end{center}

\caption{The minimal single four-brane model warp factors $\sigma(\rho)$
and $\gamma(\rho)$.}
\label{1min}
\end{figure}

It is straightforward to solve the Einstein equations for the bulk
energy momentum tensor $T^{(B)~N}_{~~~M}=-\Lambda \delta_M^N$ with the brane contribution:
\begin{equation}
T^{(br)~N}_{~~~M}=-\delta(\rho)\left(\begin{array}{ccc}V
~\delta_{\mu}^{\nu}&~&~\\~&0&~\\~&~&V \end{array}\right)
\end{equation}

From the form of the Einstein equations given in  Appendix C we see
that the solution for the two warp factors is: 
\begin{equation}
\sigma(\rho)=e^{-k \rho} ~~~,~~~ \gamma(\rho)=R^2 e^{-k \rho}
\end{equation}
with $k^2=-{\Lambda \over 10M^4}$, where $M$ is the six dimensional
fundamental scale and  the arbitrary integration constant $R$ is just the radius of
the four-brane (see Fig.\ref{1min}). The Einstein equations require the usual fine tuning
between the bulk cosmological constant and the tension of the four brane:
\begin{equation}
V=-{8\Lambda \over 5k}
\end{equation}
Let us note at this point that in \cite{Gherghetta:2000qi} this fine tuning was absent
because a smooth local defect was considered instead of a four-brane. In this case, the fine tuning emerges between the different
components of the defect energy momentum tensor. The physics of the
four-brane idealization and the one of the defect model is the same.

The four dimensional KK decomposition can be carried out as
usual by considering the following graviton perturbations:
\begin{equation}
ds^2=\sigma(\rho)\left[\eta_{\mu \nu}+h_{\mu \nu}(\rho,\theta,x)\right]dx^{\mu}dx^{\nu}+d\rho^2+\gamma(\rho)d\theta^2
\end{equation}

Here, as well as throughout this chapter, we have ignored the modulus
associated to the radius $R$ of the four-brane. This will be a
massless scalar and in order not to give rise to an additional long range
force it should be given mass through some stabilization mechanism.

We expand the graviton perturbations in a complete set of radial
eigenfunctions and Fourier angular modes:  
\begin{equation}
h_{\mu \nu}(\rho,\theta,x)=\sum_{n,l} \phi_{(n,l)}(\rho)e^{il\theta} h^{(n,l)}_{\mu \nu}(x)
\end{equation}

The differential equation for the radial wavefunctions $\phi$ is:
\begin{equation}
\phi''-{5 \over 2}k\phi'+\left(m^2-{l^2 \over R^2}\right)e^{k\rho}\phi=0
\end{equation}
with normalization  $\int_0^\infty d\rho \sigma \sqrt{\gamma} \phi_m^*
\phi_n=\delta_{mn}$. We can convert this equation to a two dimensional 
Schr\"{o}dinger-like equation by the following redefinitions:
\begin{equation}
z={2 \over k}\left(e^{{k \over 2}\rho}-1\right)~~~,~~~\hat{\Psi}=\sigma^{3/4}\phi
\end{equation}
so that 
\begin{equation}
-{1 \over 2 \sqrt{\gamma}} \de_z\left(\sqrt{\gamma}\de_z\hat{\Psi}\right)+V_{eff}\hat{\Psi}={m^2 \over 2}\hat{\Psi}~~~,~~~V_{eff}(z)=\frac{21k^2}{32\left({kz \over
2}+1\right)^2}+{l^2 \over 2R^2}-{3k \over 4}\delta(z)
\end{equation}

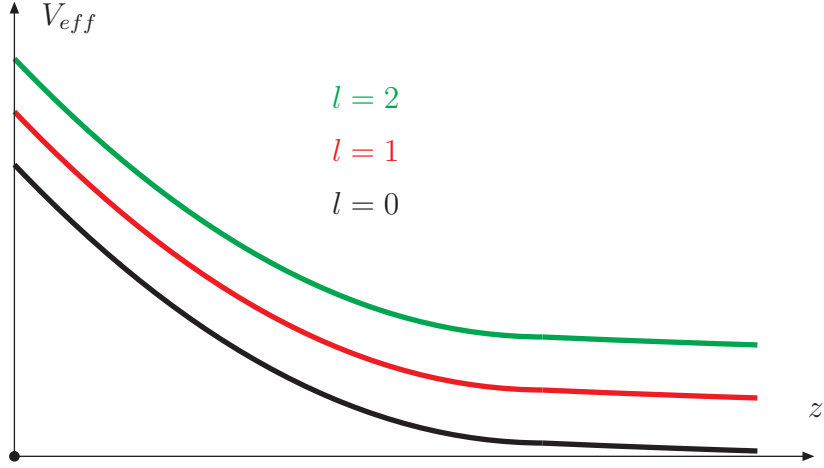
\begin{figure}[t]
\vskip10mm
\begin{center}
\begin{picture}(200,100)(0,50)

\SetWidth{2}
\SetColor{Black}
\Curve{(-50,110)(150,5)(350,110)}
\SetColor{Red}
\Curve{(-50,130)(150,25)(350,130)}
\SetColor{Green}
\Curve{(-50,150)(150,45)(350,150)}
\CBox(150,5)(350,150){White}{White}
\SetColor{Black}
\Curve{(149,5)(200,3)(230,2)}
\SetColor{Red}
\Curve{(149,25)(200,23)(230,22)}
\SetColor{Green}
\Curve{(149,45)(200,43)(230,42)}
\SetColor{Black}
\Vertex(-50,0){2}

\SetWidth{.5}
\LongArrow(-50,0)(250,0)
\LongArrow(-50,0)(-50,170)

\SetWidth{2}
\SetColor{Black}
\Text(250,20)[lt]{$z$}
\Text(-40,160)[lb]{$V_{eff}$}
\Text(70,100)[lt]{$l=0$}
\Text(70,120)[lt]{${\Red{l=1}}$}
\Text(70,140)[lt]{${\Green{l=2}}$}
\end{picture}
\end{center}
\vskip15mm
\caption{The form of the effective potential $V_{eff}$ for different angular
quantum numbers $l$ for the minimal single four-brane model. The $\delta$-function in the origin is omitted.}
\label{pot1min}
\end{figure}

From the form of the potential (see Fig.\ref{pot1min}) we can easily deduce that the angular
excitations spectrum will consist of continuum starting from a gap of
the order ${l^2 \over R^2}$ and thus can be safely ignored. For the s-wave ($l=0$) there is a normalizable zero mode which is a
constant in the $\rho$ coordinate, \textit{i.e.} $\phi_0=const.$ The
KK tower for the s-waves will again form a continuum but this time
gapless with wavefunctions given by \cite{Gherghetta:2000qi}:
\begin{equation}
\phi_m=N_m e^{{5 \over 4}k \rho}\left[J_{3/2}\left({2m \over k}\right)Y_{5/2}\left({2m \over k}e^{{k \over 2} \rho}\right)-Y_{3/2}\left({2m \over k}\right)J_{5/2}\left({2m \over k}e^{{k \over 2} \rho}\right)\right]
\end{equation}

The correction to Newton's law due to the s-modes can be easily calculated and one
finds that it is more suppressed than the RS2 model \cite{Gherghetta:2000qi,Randall:1999vf}. In six dimensions the 
correction reads:
\begin{equation}
\Delta V=-{1 \over {\mathcal{O}}(M_{{\rm{Pl}}}^5)}{1 \over r^4}
\end{equation}

\subsection{The generalized single brane model}

We can now try to find more general single four-brane solutions by relaxing
the requirement of isotropy of the bulk and brane energy momentum
tensors. In contrast to
the five dimensional case,  this is possible in six dimensions because there are two independent functions
in the metric (see Appendix C). We will consider the following anisotropic  bulk energy-momentum tensor: 
\begin{equation}
T^{(B)~N}_{~~~M}=-\left(\begin{array}{ccc}\Lambda_0
~\delta_{\mu}^{\nu}&~&~\\~&\Lambda_{\rho}&~\\~&~&\Lambda_{\theta}\end{array}\right)
\label{bulkT}
\end{equation}
and allow for an anisotropic brane tension of the form: 
\begin{equation}
T^{(br)~N}_{~~~M}=-\delta(\rho)\left(\begin{array}{ccc}V_0~
\delta_{\mu}^{\nu}&~&~\\~&0&~\\~&~&V_{\theta}\end{array}\right)
\label{braneT}
\end{equation}
This  anisotropy can be due to different contributions to the
Casimir energy in the different directions
\cite{Kogan:1983fp,Candelas:1984ae}  (see also \cite{Appelquist:1987nr}) or due to a background
three-form gauge field with non-zero field strength \cite{Chen:2000at}.

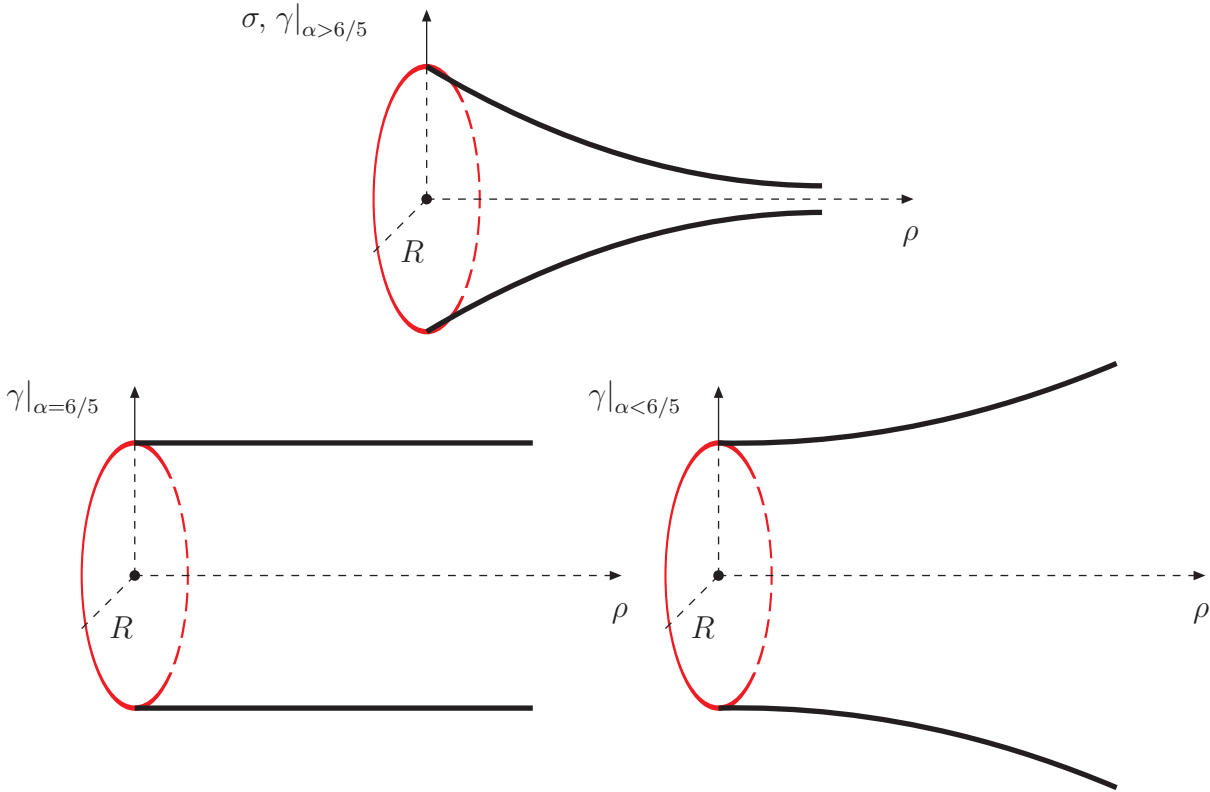
\begin{figure}[b!]
\vskip10mm
\begin{center}
\begin{picture}(200,100)(0,50)

\SetWidth{2}
\SetColor{Red}
\Oval(0,100)(50,20)(0)
\SetColor{Black}
\Curve{(0,150)(150,105)(300,150)}
\Curve{(0,50)(150,95)(300,50)}
\Vertex(0,100){2}
\Text(-70,160)[lb]{$\sigma,\left.\gamma\right|_{\alpha>6/5}$}

\CBox(150,50)(300,150){White}{White}
\SetWidth{.5}
\DashLine(0,100)(180,100){3}
\LongArrow(180,100)(182,100)

\LongArrow(0,150)(0,170)
\DashLine(0,100)(0,150){3}
\DashLine(0,100)(-20,80){3}

\SetWidth{3}
\SetColor{White}
\Line(10,135)(20,135)
\Line(10,125)(20,125)
\Line(10,115)(20,115)
\Line(10,105)(25,105)
\Line(10,95)(25,95)
\Line(10,85)(20,85)
\Line(10,75)(20,75)
\Line(10,65)(20,65)

\SetWidth{2}
\SetColor{Black}
\Text(180,90)[lt]{$\rho$}
\Text(-10,85)[lt]{$R$}

\end{picture}
\end{center}

\vskip5mm

\vskip5mm
\begin{center}
\begin{picture}(200,100)(0,50)

\SetWidth{2}
\SetColor{Red}
\Oval(-110,100)(50,20)(0)
\SetColor{Black}
\Line(-110,150)(40,150)
\Line(-110,50)(40,50)
\Vertex(-110,100){2}
\Text(-160,160)[lb]{$\left.\gamma\right|_{\alpha=6/5}$}

\SetWidth{.5}
\DashLine(-110,100)(70,100){3}
\LongArrow(70,100)(72,100)

\LongArrow(-110,150)(-110,170)
\DashLine(-110,100)(-110,150){3}
\DashLine(-110,100)(-130,80){3}

\SetWidth{3}
\SetColor{White}
\Line(-100,135)(-90,135)
\Line(-100,125)(-90,125)
\Line(-100,115)(-90,115)
\Line(-100,105)(-85,105)
\Line(-100,95)(-85,95)
\Line(-100,85)(-90,85)
\Line(-100,75)(-90,75)
\Line(-100,65)(-90,65)

\SetWidth{2}
\SetColor{Black}
\Text(70,90)[lt]{$\rho$}
\Text(-120,85)[lt]{$R$}

\SetWidth{2}
\SetColor{Red}
\Oval(110,100)(50,20)(0)
\SetColor{Black}
\Curve{(110,150)(200,160)(260,180)}
\Curve{(110,50)(200,40)(260,20)}
\Vertex(110,100){2}
\Text(60,160)[lb]{$\left.\gamma\right|_{\alpha<6/5}$}

\SetWidth{.5}
\DashLine(110,100)(290,100){3}
\LongArrow(290,100)(292,100)

\LongArrow(110,150)(110,170)
\DashLine(110,100)(110,150){3}
\DashLine(110,100)(90,80){3}

\SetWidth{3}
\SetColor{White}
\Line(120,135)(130,135)
\Line(120,125)(130,125)
\Line(120,115)(130,115)
\Line(120,105)(135,105)
\Line(120,95)(135,95)
\Line(120,85)(130,85)
\Line(120,75)(130,75)
\Line(120,65)(130,65)

\SetWidth{2}
\SetColor{Black}
\Text(290,90)[lt]{$\rho$}
\Text(100,85)[lt]{$R$}

\end{picture}
\end{center}

\vskip5mm
\caption{The generalized single four-brane warp factors $\sigma(\rho)$
and $\gamma(\rho)$ for different values of the parameter $\alpha$.}
\label{1gen}
\end{figure}

If we define the parameter $\alpha={2\Lambda_{\rho} \over
\Lambda_{\theta}}$, then the one four-brane solutions for the warp factors are:
\begin{equation}
\sigma(\rho)=e^{-k \rho} ~~~,~~~\gamma(\rho)=R^2 e^{-{k \over 4}(5\alpha-6) \rho}
\end{equation}
with $k^2=-{\Lambda_{\theta} \over 10M^4}$, where again $R$ is the
radius of the four-brane (see Fig.\ref{1gen}). The flatness of the
brane is achieved by the following fine tunings of the the parameters
of the bulk and brane energy momentum tensors:
\begin{equation}
V_{\theta}=-{8\Lambda_{\theta} \over 5k} ~~~,~~~V_0={5\alpha+6 \over
16}V_{\theta}~~~,~~~\Lambda_0={5\alpha^2+12 \over 32}\Lambda_{\theta}
\end{equation}

\begin{figure}[t]
\vskip10mm
\begin{center}
\begin{picture}(200,100)(0,50)

\SetWidth{2}
\SetColor{Black}
\Curve{(-50,110)(150,5)(350,110)}
\CBox(150,5)(350,150){White}{White}
\SetColor{Black}
\Curve{(149,5)(200,3)(230,2)}
\SetColor{Red}
\Curve{(-50,130)(110,35)(270,130)}
\SetColor{Green}
\Curve{(-50,150)(90,60)(230,180)}
\CBox(230,70)(270,190){White}{White}
\SetColor{Black}
\Vertex(-50,0){2}

\SetWidth{.5}
\LongArrow(-50,0)(250,0)
\LongArrow(-50,0)(-50,170)

\SetWidth{2}
\SetColor{Black}
\Text(250,20)[lt]{$z$}
\Text(-40,160)[lb]{$V_{eff}$}
\Text(70,100)[lt]{$l=0$}
\Text(70,120)[lt]{${\Red{l=1}}$}
\Text(70,140)[lt]{${\Green{l=2}}$}

\end{picture}
\end{center}

\vskip15mm

\caption{The form of the effective potential $V_{eff}$ for $\alpha>2$ for different angular
quantum numbers $l$. The $\delta$-function in the origin is omitted.}
\label{pot1gen1}
\end{figure}
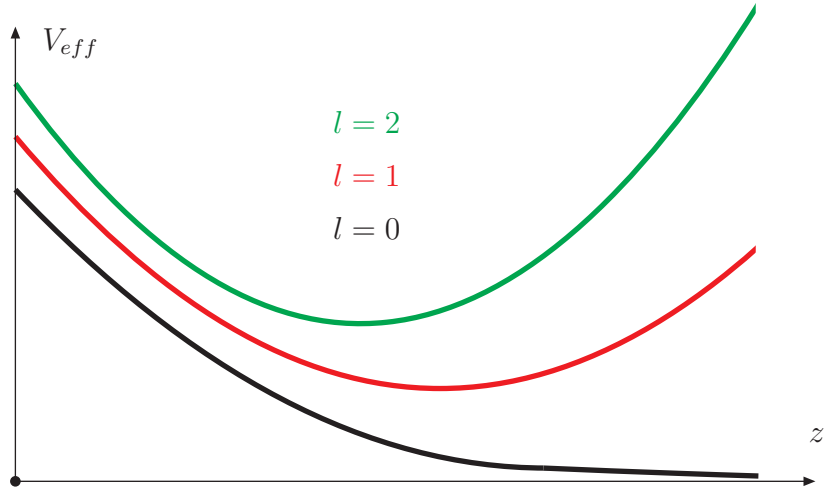

\begin{figure}[b!]
\vskip10mm
\begin{center}
\begin{picture}(200,100)(0,50)

\SetWidth{2}
\SetColor{Black}
\Curve{(-50,110)(150,5)(350,110)}
\SetColor{Red}
\Curve{(-50,130)(150,5)(350,130)}
\SetColor{Green}
\Curve{(-50,150)(150,5)(350,150)}
\CBox(150,5)(350,150){White}{White}
\SetColor{Black}
\Curve{(149,5)(200,3)(230,2)}
\SetColor{Red}
\Curve{(149,5)(200,3)(230,2)}
\SetColor{Green}
\Curve{(149,5)(200,3)(230,2)}
\SetColor{Black}
\Vertex(-50,0){2}

\SetWidth{.5}
\LongArrow(-50,0)(250,0)
\LongArrow(-50,0)(-50,170)

\SetWidth{2}
\SetColor{Black}
\Text(250,20)[lt]{$z$}
\Text(-40,160)[lb]{$V_{eff}$}
\Text(70,100)[lt]{$l=0$}
\Text(70,120)[lt]{${\Red{l=1}}$}
\Text(70,140)[lt]{${\Green{l=2}}$}

\end{picture}
\end{center}

\vskip15mm
\caption{The form of the effective potential $V_{eff}$ for $\alpha<2$ for different angular
quantum numbers $l$. The $\delta$-function in the origin is omitted.}
\label{pot1gen2}
\end{figure}
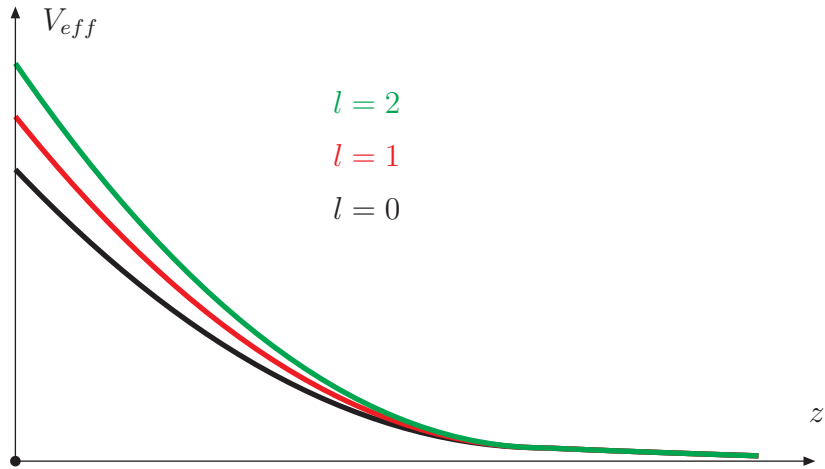

Let us note that the two components of the brane tension have 
the same sign as long as $\alpha>-{6 \over 5}$. For $\alpha>{6 \over
5}$ the internal space is shrinking as in \cite{Gherghetta:2000qi},  for $\alpha<{6 \over 5}$ the internal space is growing and
for $\alpha={6 \over 5}$ the internal space is a cylinder (see Fig.\ref{1gen}).

Proceeding with the usual KK decomposition we find the following
equation for the  radial wavefunction $\phi$:
\begin{equation}
\phi''-{5\alpha +10 \over 8}k\phi'+\left(m^2-{l^2 \over R^2}e^{{5
\over 4}(\alpha-2)k \rho}\right)e^{k\rho}\phi=0
\end{equation}
with normalization  $\int_0^\infty d\rho \sigma \sqrt{\gamma} \phi_m
\phi_n=\delta_{mn}$. With the wavefunction redefinition and the
coordinate change discussed in the previous subsection, we obtain the
following effective potential for the two dimensional Schr\"{o}dinger-like equation:
\begin{equation}
V_{eff}(z)=\frac{3 (5\alpha+4)k^2}{64\left({kz \over
2}+1\right)^2}+{l^2 \over 2R^2}\left({kz \over 2}+1\right)^{{5 \over 2}(\alpha-2)}-{3k \over 4}\delta(z)
\end{equation}

For the s-wave there is a normalizable zero mode when $\alpha>-{2
\over 5}$ which is a
constant in the $\rho$ coordinate, \textit{i.e.} $\phi_0=const.$ The
KK tower for the s-waves will again form a continuum and their
wavefunctions are given by:
\begin{equation}
\phi_m=N_m e^{{\nu \over 2}k \rho}\left[J_{\nu-1}\left({2m \over k}\right)Y_{\nu}\left({2m \over k}e^{{k \over 2} \rho}\right)-Y_{\nu-1}\left({2m \over k}\right)J_{\nu}\left({2m \over k}e^{{k \over 2} \rho}\right)\right]
\end{equation}
where $\nu={5 \over 8}(\alpha+2)$. The correction to the Newton's law due to the s-modes can be easily calculated and one
finds that it is negligible as long as $\alpha \gtsim -{2 \over 5}$:
\begin{equation}
\Delta V=-{1 \over {\mathcal{O}}(M_{{\rm{Pl}}}^{2\nu})}{1 \over r^{2\nu-1}}
\end{equation}

As far as the angular excitations $l\neq0$ are concerned, in the case where
$\alpha>2$ the potential diverges at infinity (see Fig. \ref{pot1gen1}), so the spectrum will be 
discrete starting from a scale bigger than ${l^2 \over R^2}$ and thus
can be safely ignored. On the other hand, the potential for the case
$\alpha<2$ asymptotically vanishes (see Fig. \ref{pot1gen2}) and so the spectrum will be gapless
continuum. One should examine the contribution of the angular excitations to
the Newtonian potential in this case.

\section{Bigravity in six dimensions}

The flat single four-brane models considered in the previous sections have the
characteristic that gravity is localized on the brane in the same way
as in the five dimensional analogue \cite{Randall:1999vf}. In this section we will
show how we can construct realistic multi-localization scenarios for
gravity by sewing two single four-brane solutions together. We will
present two explicit examples of compact double four-brane models where bigravity is realized. Firstly, we will consider a
double four-brane bigravity model which in general
contains an additional three-brane associated with the existence of a conical
singularity. This can be done by allowing for an anisotropic four-brane 
tension. In the following subsection we demonstrate how one can avoid the conical singularity by also allowing for anisotropic bulk
cosmological constant. In the following we will consider the
 orbifold geometry $S^1/Z_2$ in which the two four-branes lie on the
fixed points. 

\subsection{The conifold model}

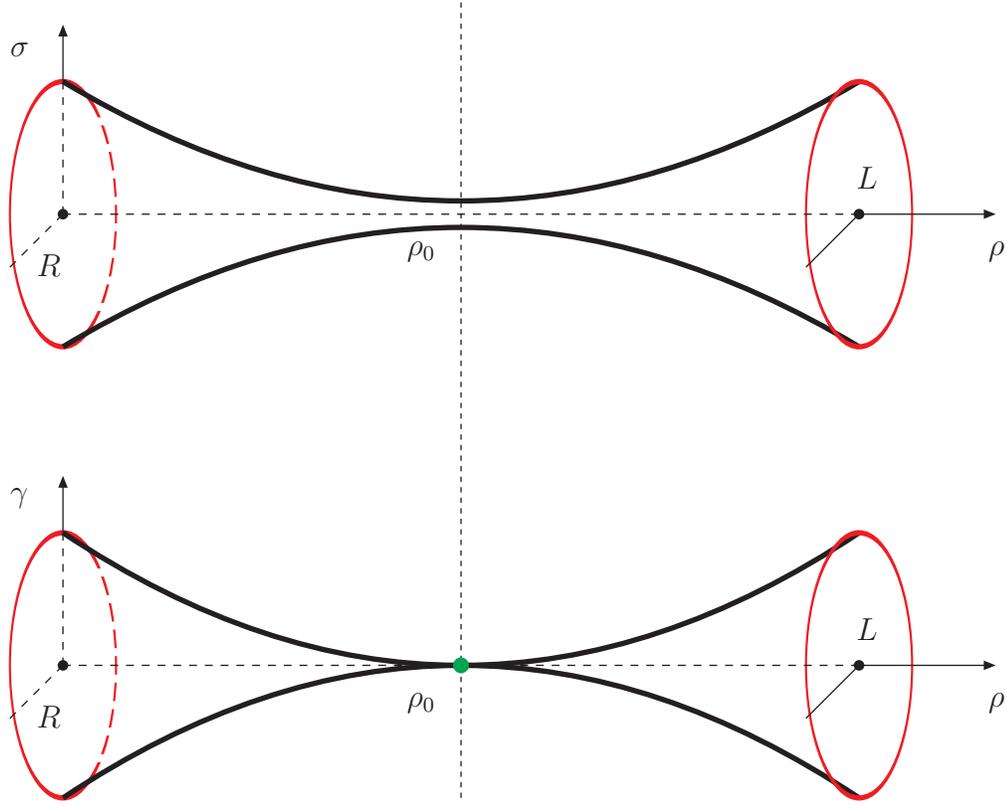
\begin{figure}[t]
\vskip10mm
\begin{center}
\begin{picture}(200,100)(0,50)

\SetWidth{2}
\SetColor{Red}
\Oval(-50,100)(50,20)(0)
\SetColor{Black}
\Curve{(-50,150)(100,105)(250,150)}
\Curve{(-50,50)(100,95)(250,50)}
\Vertex(-50,100){2}
\Vertex(250,100){2}

\SetColor{Red}
\Oval(250,100)(50,20)(0)

\SetColor{Black}
\SetWidth{.5}
\LongArrow(-50,150)(-50,170)
\DashLine(-50,100)(-50,150){3}
\DashLine(-50,100)(-70,80){3}
\Line(250,100)(230,80)

\SetColor{Black}
\SetWidth{.5}
\DashLine(-50,100)(250,100){3}
\LongArrow(250,100)(300,100)

\SetWidth{3}
\SetColor{White}
\Line(-40,135)(-30,135)
\Line(-40,125)(-30,125)
\Line(-40,115)(-30,115)
\Line(-40,105)(-25,105)
\Line(-40,95)(-25,95)
\Line(-40,85)(-30,85)
\Line(-40,75)(-30,75)
\Line(-40,65)(-30,65)

\SetWidth{2}
\SetColor{Black}
\Text(300,90)[lt]{$\rho$}
\Text(80,90)[lt]{$\rho_0$}
\Text(250,110)[lb]{$L$}
\Text(-60,85)[lt]{$R$}
\Text(-70,160)[lb]{$\sigma$}

\end{picture}
\end{center}

\vskip5mm

\vskip15mm
\begin{center}
\begin{picture}(200,100)(0,50)

\SetWidth{2}
\SetColor{Red}
\Oval(-50,100)(50,20)(0)
\SetColor{Black}
\Curve{(-50,150)(100,100)(250,150)}
\Curve{(-50,50)(100,100)(250,50)}
\Vertex(-50,100){2}
\Vertex(250,100){2}

\SetColor{Red}
\Oval(250,100)(50,20)(0)

\SetColor{Black}
\SetWidth{.5}
\LongArrow(-50,150)(-50,170)
\DashLine(-50,100)(-50,150){3}
\DashLine(-50,100)(-70,80){3}
\Line(250,100)(230,80)

\SetColor{Black}
\SetWidth{.5}
\DashLine(-50,100)(250,100){3}
\LongArrow(250,100)(300,100)
\SetColor{Black}
\SetWidth{.5}
\DashLine(100,50)(100,350){2}

\SetWidth{2}
\SetColor{Green}
\Vertex(100,100){3}

\SetWidth{3}
\SetColor{White}
\Line(-40,135)(-30,135)
\Line(-40,125)(-30,125)
\Line(-40,115)(-30,115)
\Line(-40,105)(-25,105)
\Line(-40,95)(-25,95)
\Line(-40,85)(-30,85)
\Line(-40,75)(-30,75)
\Line(-40,65)(-30,65)

\SetWidth{2}
\SetColor{Black}
\Text(300,90)[lt]{$\rho$}
\Text(80,90)[lt]{$\rho_0$}
\Text(250,110)[lb]{$L$}
\Text(-60,85)[lt]{$R$}
\Text(-70,160)[lb]{$\gamma$}

\end{picture}
\end{center}

\caption{The conifold bigravity model warp factors $\sigma(\rho)$ and
$\gamma(\rho)$ for the symmetric configuration where $L=2\rho_0$. Both 
the $\sigma(\rho)$ and $\gamma(\rho)$ have a bounce at
$\rho_0$. However,  $\gamma(\rho)$ vanishes at this point. This point in general corresponds to a conical singularity.}
\label{bigrcon}
\end{figure}

We are interested in a double four-brane generalization of the minimal
single four-brane model. In order to achieve this,  we still assume an isotropic bulk energy momentum tensor $T^{(B)~N}_{~~~M}=-\Lambda \delta_M^N$ but we do not impose
any constraints on the four-brane tension: 
\begin{equation}
T^{(br)~N}_{~~~M}=-\delta(\rho)\left(\begin{array}{ccc}V_0~
\delta_{\mu}^{\nu}&~&~\\~&0&~\\~&~&V_{\theta}\end{array}\right)
\end{equation}
The Einstein equations for $\sigma(\rho)$ and $\gamma(\rho)$ in this
case  give the following solutions for the warp factors:
\begin{equation}
\sigma(\rho)=\frac{\cosh^{4/5}\left[{5 \over 4}k (\rho-\rho_0) \right]}{\cosh^{4/5}\left[{5 \over 4}k\rho_0 \right]} ~~~,~~~\gamma(\rho)=R^2\frac{\cosh^{6/5}\left[{5 \over 4}k \rho_0 \right]}{\sinh^2\left[{5 \over 4}k \rho_0 \right]}\frac{\sinh^2\left[{5 \over 4}k  (\rho-\rho_0) \right]}{\cosh^{6/5}\left[{5 \over 4}k  (\rho-\rho_0) \right]}
\end{equation}
with $k^2=-{\Lambda \over 10M^4}$, where we have normalized
$\sigma(0)=1$ and $\gamma(0)=R^2$ (see Fig. \ref{bigrcon}). From the above relations it is obvious  that both $\sigma(\rho)$ and
$\gamma(\rho)$ have a bounce form. However, we note that
$\gamma(\rho_{0})=0$, that is,  $\gamma(\rho)$ is vanishing at the
minimum of the warp factor $\sigma(\rho)$. This is a general
characteristic of the solutions even when the branes are non-flat. From (\ref{tt}) and (\ref{rr}) we can easily show that
$\gamma(\rho)=C\frac{(\sigma'(\rho))^2}{\sigma(\rho)}$ (where C is an
integration constant) which implies
that whenever we have a bounce in the warp factor the function $\gamma(\rho)$
will develop a zero.

In order to examine the nature of this singularity we study the form of the metric at the vicinity of the point
$\rho=\rho_{0}$ along the lines of \cite{Deser:1984tn,Deser:1984dr}. Taking in account that in this limit we have 
$\sigma(\rho) \rightarrow b^2$ and $\gamma(\rho) \rightarrow \beta^2 (\rho-\rho_{0})^2$
the metric becomes:
\begin{eqnarray}
ds^2=&b^2~\eta_{\mu \nu}dx^{\mu}dx^{\nu}+d\rho^2 + \beta^2
(\rho-\rho_{0})^2 d\theta^2\\
&{\rm where}~~~b^2=\cosh^{-4/5}\left[{5 \over 4}k\rho_0 \right]\nonumber\\&{\rm and}~~~\beta^2 \equiv \frac{25 k^2 R^2}{16}
\frac{\cosh^{6/5}(\frac{5}{4}k\rho_{0})}{\sinh^{2}(\frac{5}{4}k\rho_{0})}\nonumber
\end{eqnarray}

From the form of the metric, it is clear  that for general values of
the $\beta$ parameter there will be a conical singularity with a
corresponding 
deficit angle $\delta=2\pi(1-\beta)$. The existence of this conifold singularity is
connected to the presence of a 3-brane at
$\rho=\rho_{0}$. In order
to find its tension one has to carefully examine the Einstein tensor
at the vicinity of the singularity. For this reason we write the metric for 
the internal manifold in the conformally flat form:
\begin{equation}
ds^2=b^2\eta_{\mu \nu}dx^{\mu}dx^{\nu}+f(r)(dr^2 + r^2 d\theta^2)
\end{equation}
with $f(r)=r^{2(\beta-1)}$ and $\rho-\rho_{0}=\beta^{-1}r^{\beta}$. In these coordinates it is easy to see how the three brane appears at
the conifold point. The Einstein tensor can be calculated for $\rho
\rightarrow \rho_{0}$:
\begin{equation}
R_{MN}-\frac{1}{2}G_{MN}R=\left(\begin{array}{ccc}\frac{\nabla^{2}\log(f(r))}{2f(r)}~b^2~
\eta_{\mu \nu}&~&~\\~&0&~\\~&~&0\end{array}\right)
\end{equation}
where $\nabla^{2}$ is the flat two dimensional Laplacian. Now, given that $\nabla^{2}\log(r)=2 \pi \delta^{(2)}({\bf{r}})$ and by comparing with:
\begin{equation}
R_{MN}-\frac{1}{2}G_{MN}R=-\frac{V_3}{4M^{4}}\frac{\sqrt{-\hat{G}}}{\sqrt{-G}}\hat{G}_{\mu\nu}\delta^{\mu}_{M}\delta^{\nu}_{N}\delta(r)
\end{equation}
where $\delta^{(2)}({\bf{r}})={\delta(r) \over 2\pi r}$ (see Appendix
of \cite{Leblond:2001xr}), we find that the tension of the 3-brane is:
\begin{equation}
V_{3}=4(1-\beta)M^{4}={2M^4 \over \pi}\delta
\end{equation}

Thus, if there is angle deficit $\delta>0$ ($\beta<1$) the tension of
the brane is  positive, whereas if there is angle excess $\delta<0$
($\beta>1$) the tension of the brane is negative. At the critical
value $\beta=0$ there is no conical singularity at all and we have a
situation where two locally flat spaces touch each other at one point.


In the following we will concentrate on the symmetric case, that is,
the four-branes will be considered placed at symmetric points with respect to 
the minimum of the warp factor $\rho_{0}$, \textit{i.e.}
$L=2\rho_0$. For the previous solution to be consistent the tensions of the
four-branes for the symmetric configuration must satisfy: 
\begin{equation}
V_{\theta}=-{8\Lambda \over 5k}\tanh\left[{5 \over 2}k
\rho_0\right]~~~,~~~V_0={3 \over 8}V_{\theta}+{8\Lambda^2 \over
5k^2}{1 \over V_{\theta}}
\end{equation}

The brane tensions for both branes are identical. Thus, the above construction consists of two positive tension four-branes placed at the end of the compact space and an intermediate three-brane due to the conifold singularity with tension depending on the
parameters of the model. In the limit $\rho_0 \rightarrow \infty$ we
correctly obtain two identical minimal single four-brane models for the case
where $\delta \to 2\pi$ ($\beta \to 0$).

\begin{figure}[t]
\vskip20mm
\begin{center}
\begin{picture}(200,100)(0,50)

\SetWidth{2}
\SetColor{Black}
\Curve{(-70,110)(75,-10)(220,110)}
\SetColor{Red}
\Curve{(-70,130)(-20,65)(10,35)(30,35)(70,150)}
\Curve{(80,150)(120,35)(140,35)(170,65)(220,130)}
\SetColor{Green}
\Curve{(-70,150)(-20,95)(10,65)(30,65)(70,200)}
\Curve{(80,200)(120,65)(140,65)(170,95)(220,150)}
\SetColor{Black}
\Vertex(-70,-20){2}

\SetWidth{.5}
\LongArrow(-70,-20)(250,-20)
\LongArrow(-70,-20)(-70,170)
\DashLine(220,-20)(220,170){3}
\DashLine(75,-20)(75,170){3}

\SetWidth{2}
\SetColor{Black}
\Text(250,0)[lt]{$z$}
\Text(-110,160)[lb]{$V_{eff}$}
\Text(250,100)[lt]{$l=0$}
\Text(250,120)[lt]{${\Red{l=1}}$}
\Text(250,140)[lt]{${\Green{l=2}}$}
\Text(75,-30)[lt]{$z_0$}
\Text(220,-30)[lt]{$2z_0$}

\end{picture}
\end{center}

\vskip25mm
\caption{The form of the effective potential $V_{eff}$ for different angular
quantum numbers $l$ for the conifold bigravity model in the symmetric configuration. The
$\delta$-functions in the positions of the four-branes are omitted.}
\label{potbigrcon}
\end{figure}
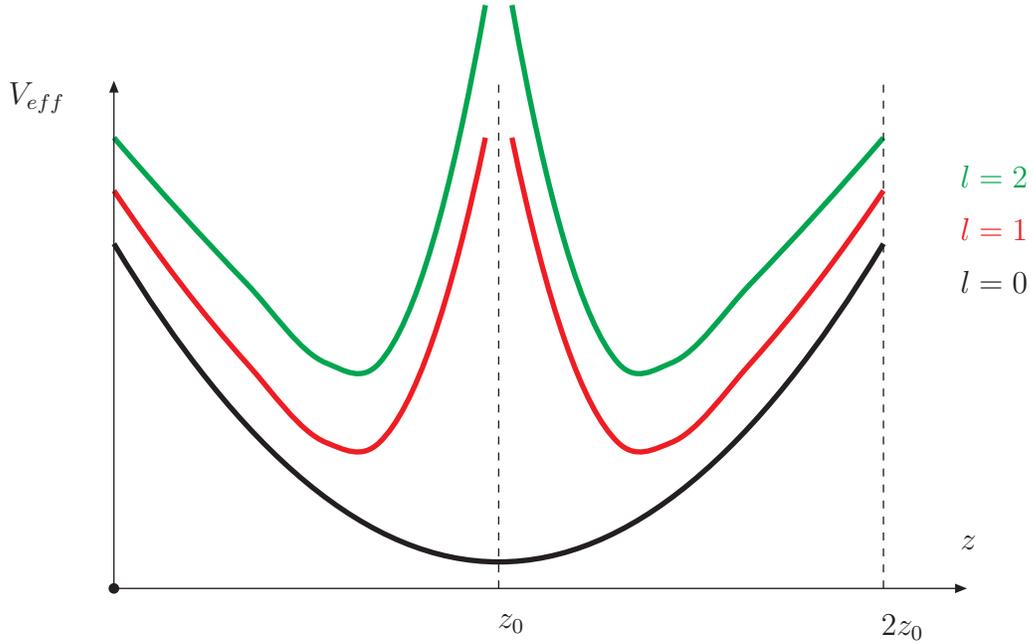

The differential equation for the radial wavefunction $\phi$ of 
the graviton excitations is:
\begin{equation} 
\phi''+2\left({\sigma' \over \sigma}+{\gamma' \over
4\gamma}\right)\phi'+\left({m^2 \over \sigma}-{l^2  \over \gamma}\right)\phi=0
\label{differ}
\end{equation}
with normalization $\int_0^{2\rho_0} d\rho \sigma \sqrt{\gamma} \phi_m
\phi_n=\delta_{mn}$.

There is an obvious normalizable zero mode with $\phi_0=const.$ and a
tower of discrete KK states. It is easy to demonstrate that for
$\delta>0$ ($\beta<1$) this model has an ultralight state, leading to a
theory of bigravity without negative tension branes. This follows considering the limit $\rho_0 \to \infty$ when the theory
describes two identical single four-brane models each of which has a
massless graviton. For finite $\rho_0$ only their symmetric combination
remains massless but their antisymmetric combination acquires an
anomalously small mass, because the difference in the modulus of the
wavefunctions between the massive and the massless states is
significant only near $\rho_0$ where the wavefunctions are
exponentially small\footnote{ There is, however, a 
mismatch of the degrees of freedom of the massive and massless
graviton, which may not be phenomenologically dangerous, but still raises the
issue of locality in the infinite four-brane separation limit.}  (see \cite{Kogan:2001wp} for discussion of this phenomenon for fields of all spin).

We can transform this differential equation to a two
dimensional Schr\"{o}dinger-like one by the coordinate change ${dz
\over d\rho}=\sigma^{-1/2}\equiv g(z)$ and the usual wavefunction
redefinition. Then the effective potential reads:
\begin{equation}
V_{eff}={15 \over 8}\left({\de_z g \over g}\right)^2-{3 \over 4}{\de_z^2 g \over
g}-{3 \over 8}{\de_z g \de_z \gamma \over g \gamma}+{l^2 \over 2\gamma 
g^2}
\label{pot}
\end{equation}

We cannot write an explicit analytic formula for of the above
potential in the $z$-coordinates because the coordinate transformation 
is not invertible analytically. However, since the transformation is
monotonic, we can easily sketch the form of the potential (see Fig.\ref{potbigrcon}) by
calculating it in the $\rho$ coordinates. From this procedure we see
that the potential for the s-wave is finite at the conical singularity 
$\rho_0$ but has a divergence for all the angular excitations with
$l\neq0$. This means that only the s-wave excitations with communicate
the two parts of the  conifold and the other excitations will be
confined in the two semicones. The spectrum of KK states in this case will be discrete. 

The construction may be readily modified to obtain a six-dimensional
analogue of the locally localized model \cite{Karch:2001ct}. This is
done by considering the asymmetric situation in which one of the fixed 
point four-branes is moved to infinity (in the asymmetric case the tensions of
the two four-branes are different). Then gravity on
the four-brane at $\rho=0$ will be mediated by only the ultralight
state since the  graviton zero mode will not be normalizable.

\subsection{The non-singular model}

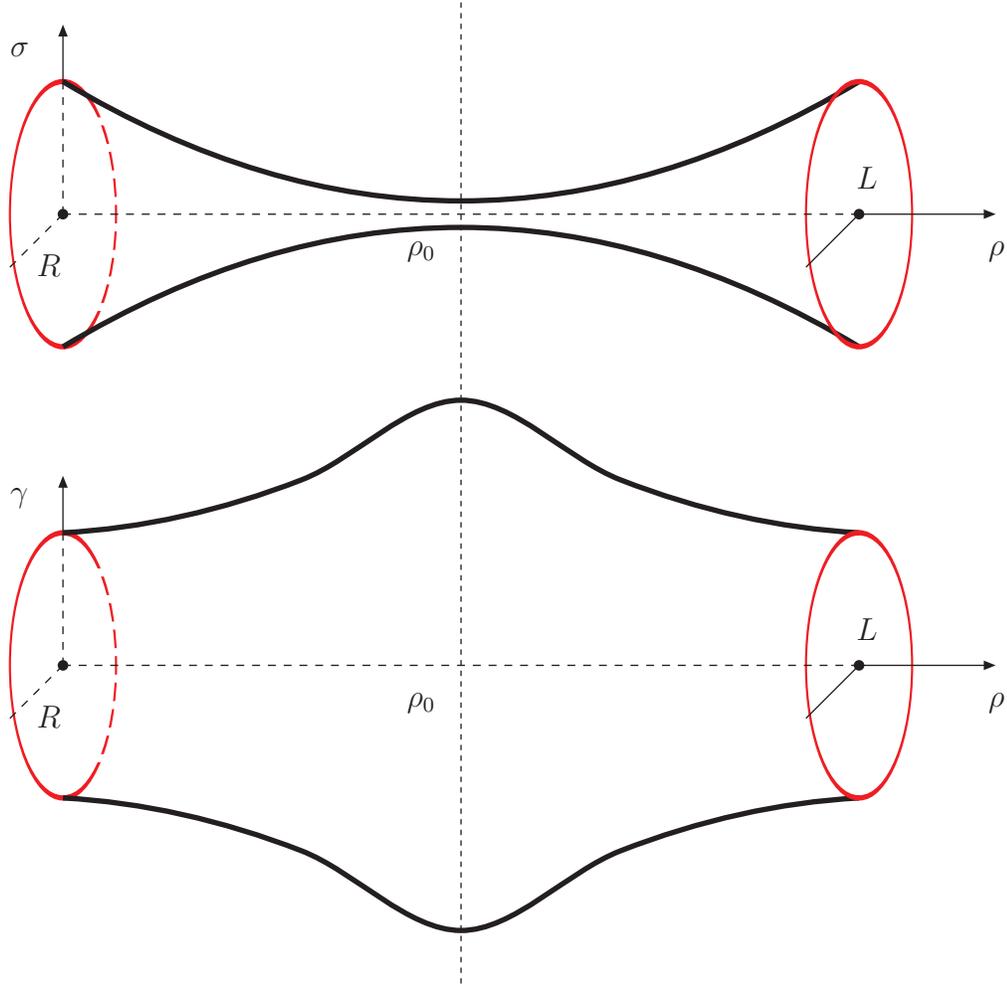
\begin{figure}[t]
\vskip10mm
\begin{center}
\begin{picture}(200,100)(0,50)

\SetWidth{2}
\SetColor{Red}
\Oval(-50,100)(50,20)(0)
\SetColor{Black}
\Curve{(-50,150)(100,105)(250,150)}
\Curve{(-50,50)(100,95)(250,50)}
\Vertex(-50,100){2}
\Vertex(250,100){2}

\SetColor{Red}
\Oval(250,100)(50,20)(0)

\SetColor{Black}
\SetWidth{.5}
\LongArrow(-50,150)(-50,170)
\DashLine(-50,100)(-50,150){3}
\DashLine(-50,100)(-70,80){3}
\Line(250,100)(230,80)

\SetColor{Black}
\SetWidth{.5}
\DashLine(-50,100)(250,100){3}
\LongArrow(250,100)(300,100)

\SetWidth{3}
\SetColor{White}
\Line(-40,135)(-30,135)
\Line(-40,125)(-30,125)
\Line(-40,115)(-30,115)
\Line(-40,105)(-25,105)
\Line(-40,95)(-25,95)
\Line(-40,85)(-30,85)
\Line(-40,75)(-30,75)
\Line(-40,65)(-30,65)

\SetWidth{2}
\SetColor{Black}
\Text(300,90)[lt]{$\rho$}
\Text(80,90)[lt]{$\rho_0$}
\Text(250,110)[lb]{$L$}
\Text(-60,85)[lt]{$R$}
\Text(-70,160)[lb]{$\sigma$}

\end{picture}
\end{center}

\vskip5mm

\vskip15mm
\begin{center}
\begin{picture}(200,100)(0,50)

\SetWidth{2}
\SetColor{Red}
\Oval(-50,100)(50,20)(0)
\SetColor{Black}
\Curve{(-50,150)(40,170)(100,200)(160,170)(250,150)}
\Curve{(-50,50)(40,30)(100,0)(160,30)(250,50)}
\Vertex(-50,100){2}
\Vertex(250,100){2}

\SetColor{Red}
\Oval(250,100)(50,20)(0)

\SetColor{Black}
\SetWidth{.5}
\LongArrow(-50,150)(-50,170)
\DashLine(-50,100)(-50,150){3}
\DashLine(-50,100)(-70,80){3}
\Line(250,100)(230,80)

\SetColor{Black}
\SetWidth{.5}
\DashLine(-50,100)(250,100){3}
\LongArrow(250,100)(300,100)

\SetWidth{3}
\SetColor{White}
\Line(-40,135)(-30,135)
\Line(-40,125)(-30,125)
\Line(-40,115)(-30,115)
\Line(-40,105)(-25,105)
\Line(-40,95)(-25,95)
\Line(-40,85)(-30,85)
\Line(-40,75)(-30,75)
\Line(-40,65)(-30,65)

\SetWidth{2}
\SetColor{Black}
\Text(300,90)[lt]{$\rho$}
\Text(80,90)[lt]{$\rho_0$}
\Text(250,110)[lb]{$L$}
\Text(-60,85)[lt]{$R$}
\Text(-70,160)[lb]{$\gamma$}

\SetColor{Black}
\SetWidth{.5}
\DashLine(100,-20)(100,350){2}

\end{picture}
\end{center}

\vskip20mm

\caption{The non-singular bigravity model warp factors $\sigma(\rho)$ and
$\gamma(\rho)$ for the symmetric configuration where $L=2\rho_0$. The
$\sigma(\rho)$ warp factor has exactly the same form as in the
conifold model. The $\gamma(\rho)$ warp factor however now has an
inverse bounce form.  In this case $\gamma(\rho)$ does not vanish anywhere and thus the model is free of singularities.}
\label{bigrnosing}
\end{figure}

The appearance of a conifold singularity can be avoided if  instead 
of using the minimal single four-brane solution to build the double four-brane
solution, one uses the generalized single four-brane model. By
allowing for an isotropic bulk
cosmological constant one can arrange that the function $\gamma(\rho)$ 
does not develop a zero. In this case the $(\theta,\theta)$ and
$(\rho,\rho)$ part of Einstein equations in the bulk, (\ref{tt}) and (\ref{rr}), have solutions: 
\begin{equation}
\sigma(\rho)=\frac{\cosh^{4/5}\left[{5 \over 4}k (\rho-\rho_0) \right]}{\cosh^{4/5}\left[{5 \over 4}k\rho_0 \right]} ~~~,~~~\gamma(\rho)=A\frac{\sinh^{\alpha}\left[{5 \over 4}k  (\rho-\rho_0) \right]}{\cosh^{6/5}\left[{5 \over 4}k  (\rho-\rho_0) \right]}
\end{equation}
where $A$ is a constant and $\alpha \equiv \frac{2
\Lambda_{\rho}}{\Lambda_{\theta}}$.  The $(\mu,\nu)$ component of the Einstein equations (\ref{mne}), however, restricts
the possible values of $\alpha$ to $\alpha=2$ or  $\alpha=0$. The case 
of $\alpha=2$ corresponds to the previous conifold model. For the 
case $\alpha=0$, which corresponds  to the choice $\Lambda_{\rho}=0$,
the solution is:  
\begin{equation}
\sigma(\rho)=\frac{\cosh^{4/5}\left[{5 \over 4}k (\rho-\rho_0) \right]}{\cosh^{4/5}\left[{5 \over 4}k\rho_0 \right]} ~~~,~~~\gamma(\rho)=R^2\frac{\cosh^{6/5}\left[{5 \over 4}k \rho_0 \right]}{\cosh^{6/5}\left[{5 \over 4}k  (\rho-\rho_0) \right]}
\end{equation}
with $k^2=-{\Lambda_{\theta} \over 10M^4}$, where we have normalized
$\sigma(0)=1$ and $\gamma(0)=R^2$ (see Fig. \ref{bigrnosing}). From the above the absence of any singularity is obvious since
$\gamma(\rho)$ does not vanish at any finite value of $\rho$.

The four branes will appear as $\delta$-function singularities at the
points where we will cut the space in the $\rho$
direction. In the following we will concentrate on the symmetric case, that is,
the branes will be considered placed in symmetric points in respect to 
the extremum of the warp factor $\rho_{0}$, \textit{i.e.} $L=2\rho_0$.
In order for the above configuration to be realized, the components of the four-brane tensions and
the bulk cosmological constant must be tuned to give: 
\begin{equation}
V_{\theta}=-{8\Lambda \over 5k}\tanh\left[{5 \over 2}k
\rho_0\right]~~~,~~~V_0={3 \over 8}V_{\theta}~~~,~~~\Lambda_0={3 \over 8}\Lambda_{\theta}
\end{equation}

The brane tensions for both branes are identical. Thus, we have constructed a compact model with two positive tension
flat branes with a bounce in the warp factors. Note the extra
condition relating the different components of the bulk cosmological
constant; this anisotropy in the bulk cosmological constant
 is what allows us to have a non-singular solution. 
In the limit $\rho_0 \rightarrow \infty$ we  obtain two identical
generalized single four-brane $\alpha=0$ models.

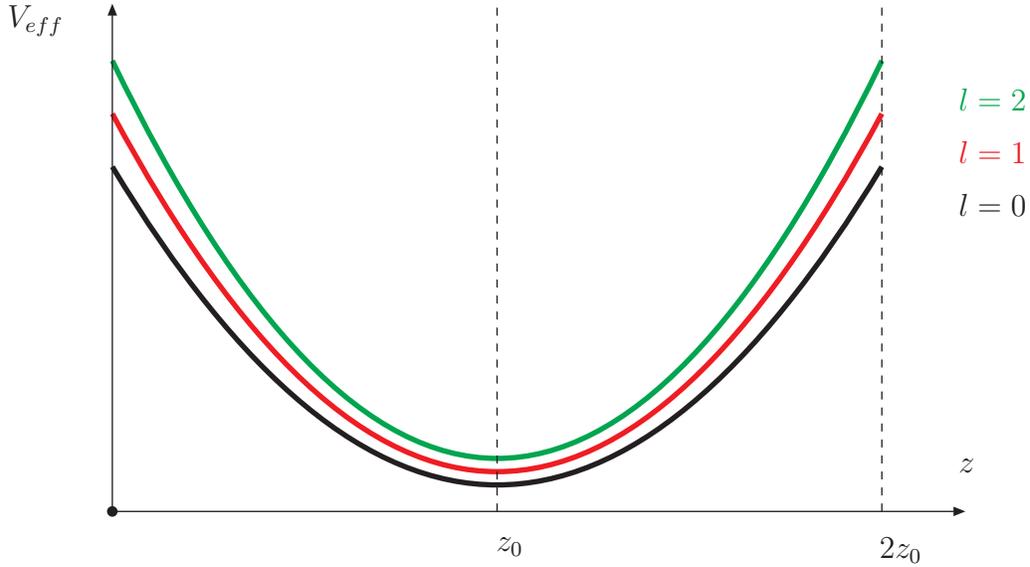
\begin{figure}[t]
\vskip20mm
\begin{center}
\begin{picture}(200,100)(0,50)

\SetWidth{2}
\SetColor{Black}
\Curve{(-70,110)(75,-10)(220,110)}
\SetColor{Red}
\Curve{(-70,130)(75,-5)(220,130)}
\SetColor{Green}
\Curve{(-70,150)(75,0)(220,150)}
\SetColor{Black}
\Vertex(-70,-20){2}

\SetWidth{.5}
\LongArrow(-70,-20)(250,-20)
\LongArrow(-70,-20)(-70,170)
\DashLine(220,-20)(220,170){3}
\DashLine(75,-20)(75,170){3}

\SetWidth{2}
\SetColor{Black}
\Text(250,0)[lt]{$z$}
\Text(-110,160)[lb]{$V_{eff}$}
\Text(250,100)[lt]{$l=0$}
\Text(250,120)[lt]{${\Red{l=1}}$}
\Text(250,140)[lt]{${\Green{l=2}}$}
\Text(75,-30)[lt]{$z_0$}
\Text(220,-30)[lt]{$2z_0$}

\end{picture}
\end{center}

\vskip25mm
\caption{The form of the effective potential $V_{eff}$ for different angular
quantum numbers $l$ for the non-singular bigravity model in the symmetric configuration. The
$\delta$-functions in the positions of the four-branes are omitted.}
\label{potbigrnosing}
\end{figure}

 The differential equation for the radial wavefunction $\phi$ of 
the graviton excitations is given by (\ref{differ}) for the relevant
$\sigma(\rho)$ and $\gamma(\rho)$  functions.
Although it is difficult to find an  analytical solution 
of the corresponding differential equation for the KK states,
the existence of a light special KK is assured from the locality
arguments discussed above. Since at the infinite brane
separation limit we recover two identical generalized single four-brane
models (for $\alpha=0$) where each of them supports a massless zero mode, at a finite
separation configuration the one of the zero modes will become the
 special light KK state.

By making an appropriate change of coordinates and a redefinition of
the wavefunction we can bring the previous differential equation in a
two dimensional Schr\"{o}dinger form with the effective potential (\ref{pot}) that is 
plotted in Fig.\ref{potbigrnosing}. Note that in this case the angular
excitations feel the whole region
between the two positive branes and thus communicate physics between them.
The spectrum of KK states in this case will again be discrete.

\section{Multigravity in six dimensions}
  
In this section we will show how we can obtain the analogues of the
GRS model of quasi-localized gravity and  of the crystal
universe in six dimensions, in both cases  without the need of introducing moving
negative tension branes.

\begin{figure}[t]
\vskip10mm
\begin{center}
\begin{picture}(200,100)(0,50)

\SetWidth{2}
\SetColor{Red}
\Oval(-50,100)(50,20)(0)
\SetColor{Black}
\Curve{(-50,150)(100,105)(250,150)}
\Curve{(-50,50)(100,95)(250,50)}
\Vertex(-50,100){2}

\SetColor{Black}
\SetWidth{.5}
\LongArrow(-50,150)(-50,170)
\DashLine(-50,100)(-50,150){3}
\DashLine(-50,100)(-70,80){3}
\Line(250,100)(230,80)

\CBox(151,0)(250,170){White}{White}
\SetColor{Black}
\SetWidth{.5}
\DashLine(-50,100)(230,100){3}
\LongArrow(230,100)(232,100)

\SetWidth{2}
\Line(150,110)(220,110)
\Line(150,90)(220,90)

\SetWidth{3}
\SetColor{White}
\Line(-40,135)(-30,135)
\Line(-40,125)(-30,125)
\Line(-40,115)(-30,115)
\Line(-40,105)(-25,105)
\Line(-40,95)(-25,95)
\Line(-40,85)(-30,85)
\Line(-40,75)(-30,75)
\Line(-40,65)(-30,65)

\SetWidth{2}
\SetColor{Black}
\Text(240,90)[lt]{$\rho$}
\Text(80,90)[lt]{$\rho_0$}
\Text(150,80)[lt]{$\rho_b$}
\Text(-60,85)[lt]{$R$}
\Text(-70,160)[lb]{$\sigma$}

\SetWidth{2}
\SetColor{Red}
\Oval(150,100)(10,3)(0)

\SetWidth{3}
\SetColor{White}
\Line(165,105)(150,105)
\Line(165,95)(150,95)

\end{picture}
\end{center}

\vskip5mm

\vskip5mm
\begin{center}
\begin{picture}(200,100)(0,50)

\SetWidth{2}
\SetColor{Red}
\Oval(-50,100)(50,20)(0)
\SetColor{Black}
\Curve{(-50,150)(40,170)(100,200)(160,170)(250,150)}
\Curve{(-50,50)(40,30)(100,0)(160,30)(250,50)}
\Vertex(-50,100){2}

\SetColor{Black}
\SetWidth{.5}
\LongArrow(-50,150)(-50,170)
\DashLine(-50,100)(-50,150){3}
\DashLine(-50,100)(-70,80){3}
\Line(250,100)(230,80)

\CBox(151,0)(250,200){White}{White}
\SetColor{Black}
\SetWidth{.5}
\DashLine(-50,100)(230,100){3}
\LongArrow(230,100)(232,100)

\SetWidth{2}
\SetColor{Black}
\Line(150,175)(220,175)
\Line(150,25)(220,25)

\SetWidth{3}
\SetColor{White}
\Line(-40,135)(-30,135)
\Line(-40,125)(-30,125)
\Line(-40,115)(-30,115)
\Line(-40,105)(-25,105)
\Line(-40,95)(-25,95)
\Line(-40,85)(-30,85)
\Line(-40,75)(-30,75)
\Line(-40,65)(-30,65)

\SetWidth{2}
\SetColor{Black}
\Text(240,90)[lt]{$\rho$}
\Text(80,90)[lt]{$\rho_0$}
\Text(150,90)[lt]{$\rho_b$}
\Text(-60,85)[lt]{$R$}
\Text(-70,160)[lb]{$\gamma$}

\SetColor{Black}
\SetWidth{.5}
\DashLine(100,-10)(100,320){2}

\SetWidth{2}
\SetColor{Red}
\Oval(150,100)(75,20)(0)

\SetWidth{3}
\SetColor{White}
\Line(170,165)(155,165)
\Line(170,155)(160,155)
\Line(170,145)(160,145)
\Line(170,135)(160,135)
\Line(170,125)(160,125)
\Line(175,115)(160,115)
\Line(175,105)(160,105)
\Line(175,95)(160,95)
\Line(175,85)(160,85)
\Line(170,75)(160,75)
\Line(170,65)(160,65)
\Line(170,55)(160,55)
\Line(170,45)(160,45)
\Line(170,35)(155,35)

\end{picture}
\end{center}

\vskip15mm
\caption{The $\alpha=0$ quasi-localized model warp factors $\sigma(\rho)$ and
$\gamma(\rho)$. The bulk region for $\rho>\rho_b$ is flat.}
\label{quasi1}
\end{figure}
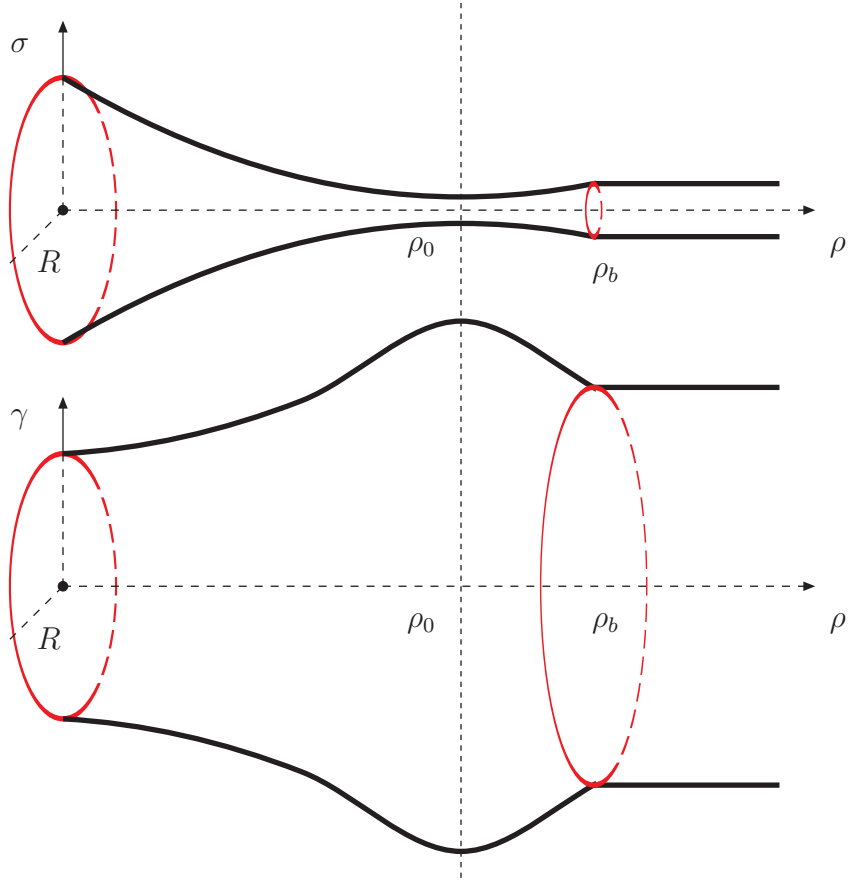

\subsection{Quasi-localized gravity}

As in the bigravity models we have two possible ways of realizing a
quasi-localized scenario in six dimensions. The first model can be built
from the $\alpha=0$ bigravity model, by cutting the space at a point
$\rho_b$ on the right 
of the position of the bounce and sewing it to flat space on the right (see
Fig.\ref{quasi1}). In that case we will have a system of two positive tension
branes and by appropriately tuning the different components of the
brane tension of the second brane we can find solutions where the warp factor is constant
in the flat bulk region. The fine tunings which achieve this are:
\begin{equation}
V^{(2)}_{\theta}=-{4\Lambda \over 5k}\tanh\left[{5 \over 2}k (\rho_b-\rho_0)\right]~~~,~~~V^{(2)}_0={3 \over 8}V^{(2)}_{\theta}
\end{equation}

From the above formulas it is obvious that if we paste the flat space
at the position of the bounce, \textit{i.e} $\rho_b=\rho_0$ then the
second brane is tensionless. If we relax these fine tunings we can find 
solutions where the warp factors in the flat bulk region are non
trivial, but we will not discuss this possibility here.

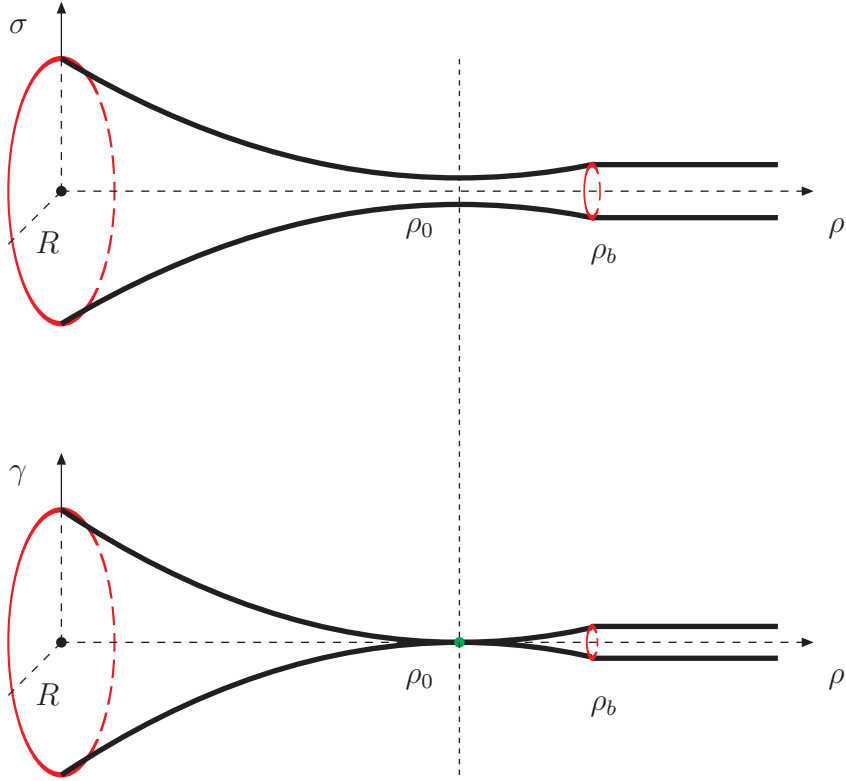
\begin{figure}[t]
\vskip10mm
\begin{center}
\begin{picture}(200,100)(0,50)

\SetWidth{2}
\SetColor{Red}
\Oval(-50,100)(50,20)(0)
\SetColor{Black}
\Curve{(-50,150)(100,105)(250,150)}
\Curve{(-50,50)(100,95)(250,50)}
\Vertex(-50,100){2}

\SetColor{Black}
\SetWidth{.5}
\LongArrow(-50,150)(-50,170)
\DashLine(-50,100)(-50,150){3}
\DashLine(-50,100)(-70,80){3}
\Line(250,100)(230,80)

\CBox(151,0)(250,170){White}{White}
\SetColor{Black}
\SetWidth{.5}
\DashLine(-50,100)(230,100){3}
\LongArrow(230,100)(232,100)

\SetWidth{2}
\Line(150,110)(220,110)
\Line(150,90)(220,90)

\SetWidth{3}
\SetColor{White}
\Line(-40,135)(-30,135)
\Line(-40,125)(-30,125)
\Line(-40,115)(-30,115)
\Line(-40,105)(-25,105)
\Line(-40,95)(-25,95)
\Line(-40,85)(-30,85)
\Line(-40,75)(-30,75)
\Line(-40,65)(-30,65)

\SetWidth{2}
\SetColor{Black}
\Text(240,90)[lt]{$\rho$}
\Text(80,90)[lt]{$\rho_0$}
\Text(150,80)[lt]{$\rho_b$}
\Text(-60,85)[lt]{$R$}
\Text(-70,160)[lb]{$\sigma$}

\SetWidth{2}
\SetColor{Red}
\Oval(150,100)(10,3)(0)

\SetWidth{3}
\SetColor{White}
\Line(165,105)(150,105)
\Line(165,95)(150,95)

\end{picture}
\end{center}

\vskip5mm

\vskip15mm
\begin{center}
\begin{picture}(200,100)(0,50)

\SetWidth{2}
\SetColor{Red}
\Oval(-50,100)(50,20)(0)
\SetColor{Black}
\Curve{(-50,150)(100,100)(250,150)}
\Curve{(-50,50)(100,100)(250,50)}
\Vertex(-50,100){2}

\CBox(151,0)(250,200){White}{White}
\SetColor{Black}
\SetWidth{.5}
\LongArrow(-50,150)(-50,170)
\DashLine(-50,100)(-50,150){3}
\DashLine(-50,100)(-70,80){3}
\Line(250,100)(230,80)

\CBox(151,0)(250,200){White}{White}
\SetColor{Black}
\SetWidth{.5}
\DashLine(-50,100)(230,100){3}
\LongArrow(230,100)(232,100)

\SetWidth{2}
\Line(150,106)(220,106)
\Line(150,94)(220,94)

\SetWidth{3}
\SetColor{White}
\Line(-40,135)(-30,135)
\Line(-40,125)(-30,125)
\Line(-40,115)(-30,115)
\Line(-40,105)(-25,105)
\Line(-40,95)(-25,95)
\Line(-40,85)(-30,85)
\Line(-40,75)(-30,75)
\Line(-40,65)(-30,65)

\SetWidth{2}
\SetColor{Black}
\Text(240,90)[lt]{$\rho$}
\Text(80,90)[lt]{$\rho_0$}
\Text(150,80)[lt]{$\rho_b$}
\Text(-60,85)[lt]{$R$}
\Text(-70,160)[lb]{$\gamma$}

\SetWidth{2}
\SetColor{Green}
\Vertex(100,100){2}

\SetWidth{2}
\SetColor{Red}
\Oval(150,100)(6,2)(0)

\SetWidth{2}
\SetColor{White}
\Line(165,102.5)(150,102.5)
\Line(165,97.5)(150,97.5)

\SetColor{Black}
\SetWidth{.5}
\DashLine(100,50)(100,320){2}

\end{picture}
\end{center}

\caption{The $\alpha=2$ quasi-localized model warp factors $\sigma(\rho)$ and
$\gamma(\rho)$. The bulk region for $\rho>\rho_b$ is flat.}
\label{quasi2}
\end{figure}

The above construction will obviously have no normalizable zero mode
due to the infinite volume of the system and the KK spectrum will be
continuous. However, as in the five dimensional case we expect that
the low part of the KK tower will have a ``resonance''-like coupling
to matter on the $''+''$  brane at $\rho=0$. This will mediate normal 
gravity at intermediate distances but will change the nature of the
gravitational law at ultralarge scales.

\begin{figure}[t]
\vskip10mm
\begin{center}
\begin{picture}(200,100)(0,50)

\SetWidth{2}
\SetColor{Red}
\Oval(-50,100)(50,20)(0)
\SetColor{Black}
\Curve{(-50,150)(25,105)(100,150)}
\Curve{(100,150)(175,105)(250,150)}
\Curve{(250,150)(325,105)(400,150)}
\Curve{(-200,150)(-125,105)(-50,150)}

\Curve{(-200,50)(-125,95)(-50,50)}
\Curve{(250,50)(325,95)(400,50)}
\Curve{(-50,50)(25,95)(100,50)}
\Curve{(100,50)(175,95)(250,50)}
\Vertex(100,100){2}

\SetColor{Red}
\Oval(250,100)(50,20)(0)
\Oval(100,100)(50,20)(0)

\CBox(-200,0)(-100,200){White}{White}
\CBox(300,0)(400,200){White}{White}
\SetColor{Black}
\SetWidth{.5}
\LongArrow(100,150)(100,170)
\DashLine(100,100)(100,150){3}
\DashLine(-50,100)(-70,80){3}

\SetColor{Black}
\SetWidth{.5}
\DashLine(-100,100)(300,100){3}
\LongArrow(300,100)(301,100)

\SetWidth{3}
\SetColor{White}
\Line(-40,125)(-30,125)
\Line(-40,115)(-30,115)
\Line(-40,105)(-25,105)
\Line(-40,95)(-25,95)
\Line(-40,85)(-30,85)
\Line(-40,75)(-30,75)

\Line(110,125)(120,125)
\Line(110,115)(120,115)
\Line(110,105)(125,105)
\Line(110,95)(125,95)
\Line(110,85)(120,85)
\Line(110,75)(120,75)

\Line(260,125)(270,125)
\Line(260,115)(270,115)
\Line(260,105)(275,105)
\Line(260,95)(275,95)
\Line(260,85)(270,85)
\Line(260,75)(270,75)

\SetWidth{2}
\SetColor{Black}
\Text(300,90)[lt]{$\rho$}
\Text(-60,85)[lt]{$R$}
\Text(80,160)[lb]{$\sigma$}
\Text(175,90)[lt]{$\rho_0$}
\Text(15,90)[lt]{$-\rho_0$}
\Text(310,110)[lt]{${\Huge{\bf{\cdots}}}$}
\Text(-110,110)[rt]{${\Huge{\bf{\cdots}}}$}

\end{picture}
\end{center}

\vskip5mm

\vskip15mm
\begin{center}
\begin{picture}(200,100)(0,50)

\SetWidth{2}
\SetColor{Red}
\Oval(-50,100)(50,20)(0)
\SetColor{Black}
\Curve{(-50,150)(-10,158)(25,175)(60,158)(100,150)}
\Curve{(100,150)(140,158)(175,175)(210,158)(250,150)}
\Curve{(250,150)(290,158)(325,175)(360,158)(400,150)}
\Curve{(-200,150)(-160,158)(-125,175)(-90,158)(-50,150)}

\Curve{(-200,50)(-160,42)(-125,25)(-90,42)(-50,50)}
\Curve{(250,50)(290,42)(325,25)(360,42)(400,50)}
\Curve{(-50,50)(-10,42)(25,25)(60,42)(100,50)}
\Curve{(100,50)(140,42)(175,25)(210,42)(250,50)}
\Vertex(100,100){2}

\SetColor{Red}
\Oval(250,100)(50,20)(0)
\Oval(100,100)(50,20)(0)

\CBox(-200,0)(-100,200){White}{White}
\CBox(300,0)(400,200){White}{White}
\SetColor{Black}
\SetWidth{.5}
\LongArrow(100,150)(100,170)
\DashLine(100,100)(100,150){3}
\DashLine(-50,100)(-70,80){3}

\SetColor{Black}
\SetWidth{.5}
\DashLine(-100,100)(300,100){3}
\LongArrow(300,100)(301,100)

\SetWidth{3}
\SetColor{White}
\Line(-40,135)(-30,135)
\Line(-40,125)(-30,125)
\Line(-40,115)(-30,115)
\Line(-40,105)(-25,105)
\Line(-40,95)(-25,95)
\Line(-40,85)(-30,85)
\Line(-40,75)(-30,75)
\Line(-40,65)(-30,65)

\Line(110,135)(120,135)
\Line(110,125)(120,125)
\Line(110,115)(120,115)
\Line(110,105)(125,105)
\Line(110,95)(125,95)
\Line(110,85)(120,85)
\Line(110,75)(120,75)
\Line(110,65)(120,65)

\Line(260,135)(270,135)
\Line(260,125)(270,125)
\Line(260,115)(270,115)
\Line(260,105)(275,105)
\Line(260,95)(275,95)
\Line(260,85)(270,85)
\Line(260,75)(270,75)
\Line(260,65)(270,65)

\SetWidth{2}
\SetColor{Black}
\Text(300,90)[lt]{$\rho$}
\Text(-60,85)[lt]{$R$}
\Text(80,160)[lb]{$\gamma$}
\Text(175,90)[lt]{$\rho_0$}
\Text(15,90)[lt]{$-\rho_0$}
\Text(310,110)[lt]{${\Huge{\bf{\cdots}}}$}
\Text(-110,110)[rt]{${\Huge{\bf{\cdots}}}$}

\end{picture}
\end{center}

\vskip5mm
\caption{The $\alpha=0$ crystalline model warp factors $\sigma(\rho)$ and
$\gamma(\rho)$.}
\label{crystal1}
\end{figure}

The second model of quasi-localized gravity can be built in a similar
fashion from the $\alpha=2$ bigravity model (see Fig.\ref{quasi2}). Again we obtain a system
of two positive tension branes which will have a constant warp factor
in the flat bulk region when the following fine tunings are demanded:
\begin{equation}
V^{(2)}_{\theta}=-{4\Lambda \over 5k}\tanh\left[{5 \over 2}k (\rho_b-\rho_0)\right]~~~,~~~V^{(2)}_0={3 \over 8}V^{(2)}_{\theta}+{8\Lambda^2 \over 5k^2}{1 \over V^{(2)}_{\theta}}
\end{equation}

From the above formulas it is obvious that if we paste the flat space
at the position of the bounce, \textit{i.e} $\rho_b=\rho_0$ then the
tension $V^{(2)}_{\theta}$ vanishes whereas the tension $V^{(2)}_0$
diverges.  This limit is not well behaved since the internal space
becomes singular in a whole line. At this limit obviously classical
gravity breaks down and one would expect that quantum gravity
corrections would resolve the singularity. However, one should check if these quantum gravity corrections affect the
solution in the vicinity of the four-brane at $\rho=0$, \textit{e.g.} by examining the
$R^2$ correction to the Einstein action.

The above construction will again have no normalizable zero mode and
the KK spectrum will be continuous. A ``resonance''-like coupling of
the KK states to matter on the $''+''$  brane at $\rho=0$ is again
expected which will give normal  four dimensional gravity at
intermediate distances but will change the nature of the gravitational law at ultralarge scales.

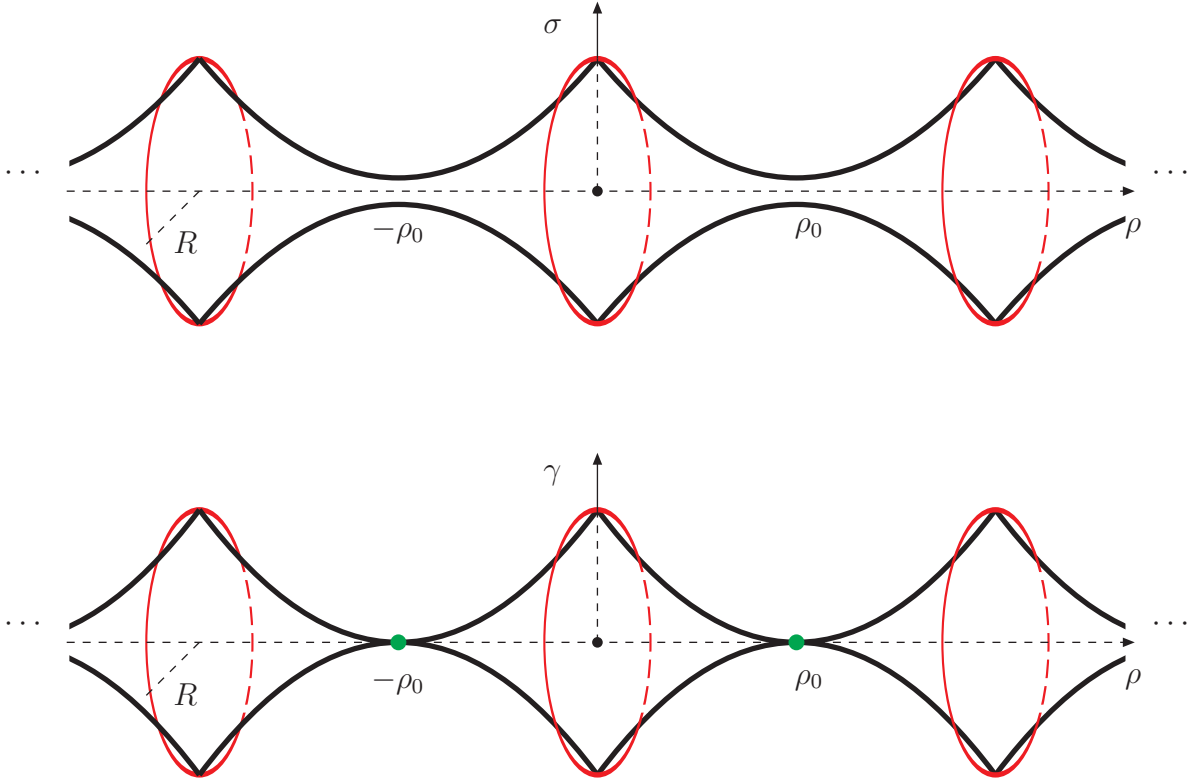
\begin{figure}[t]
\vskip10mm
\begin{center}
\begin{picture}(200,100)(0,50)

\SetWidth{2}
\SetColor{Red}
\Oval(-50,100)(50,20)(0)
\SetColor{Black}
\Curve{(-50,150)(25,105)(100,150)}
\Curve{(100,150)(175,105)(250,150)}
\Curve{(250,150)(325,105)(400,150)}
\Curve{(-200,150)(-125,105)(-50,150)}

\Curve{(-200,50)(-125,95)(-50,50)}
\Curve{(250,50)(325,95)(400,50)}
\Curve{(-50,50)(25,95)(100,50)}
\Curve{(100,50)(175,95)(250,50)}
\Vertex(100,100){2}

\SetColor{Red}
\Oval(250,100)(50,20)(0)
\Oval(100,100)(50,20)(0)

\CBox(-200,0)(-100,200){White}{White}
\CBox(300,0)(400,200){White}{White}
\SetColor{Black}
\SetWidth{.5}
\LongArrow(100,150)(100,170)
\DashLine(100,100)(100,150){3}
\DashLine(-50,100)(-70,80){3}

\SetColor{Black}
\SetWidth{.5}
\DashLine(-100,100)(300,100){3}
\LongArrow(300,100)(301,100)

\SetWidth{3}
\SetColor{White}
\Line(-40,125)(-30,125)
\Line(-40,115)(-30,115)
\Line(-40,105)(-25,105)
\Line(-40,95)(-25,95)
\Line(-40,85)(-30,85)
\Line(-40,75)(-30,75)

\Line(110,125)(119,125)
\Line(110,115)(120,115)
\Line(110,105)(125,105)
\Line(110,95)(125,95)
\Line(110,85)(120,85)
\Line(110,75)(119,75)

\Line(260,125)(269,125)
\Line(260,115)(270,115)
\Line(260,105)(275,105)
\Line(260,95)(275,95)
\Line(260,85)(270,85)
\Line(260,75)(269,75)

\SetWidth{2}
\SetColor{Black}
\Text(300,90)[lt]{$\rho$}
\Text(-60,85)[lt]{$R$}
\Text(80,160)[lb]{$\sigma$}
\Text(175,90)[lt]{$\rho_0$}
\Text(15,90)[lt]{$-\rho_0$}
\Text(310,110)[lt]{${\Huge{\bf{\cdots}}}$}
\Text(-110,110)[rt]{${\Huge{\bf{\cdots}}}$}

\end{picture}
\end{center}

\vskip5mm

\vskip15mm
\begin{center}
\begin{picture}(200,100)(0,50)

\SetWidth{2}
\SetColor{Red}
\Oval(-50,100)(50,20)(0)
\SetColor{Black}
\Curve{(-50,150)(25,100)(100,150)}
\Curve{(100,150)(175,100)(250,150)}
\Curve{(250,150)(325,100)(400,150)}
\Curve{(-200,150)(-125,100)(-50,150)}

\Curve{(-200,50)(-125,100)(-50,50)}
\Curve{(250,50)(325,100)(400,50)}
\Curve{(-50,50)(25,100)(100,50)}
\Curve{(100,50)(175,100)(250,50)}
\Vertex(100,100){2}

\SetColor{Red}
\Oval(250,100)(50,20)(0)
\Oval(100,100)(50,20)(0)

\CBox(-200,0)(-100,200){White}{White}
\CBox(300,0)(400,200){White}{White}
\SetColor{Black}
\SetWidth{.5}
\LongArrow(100,150)(100,170)
\DashLine(100,100)(100,150){3}
\DashLine(-50,100)(-70,80){3}

\SetColor{Black}
\SetWidth{.5}
\DashLine(-100,100)(300,100){3}
\LongArrow(300,100)(301,100)

\SetWidth{3}
\SetColor{White}
\Line(-40,125)(-31,125)
\Line(-40,115)(-30,115)
\Line(-40,105)(-25,105)
\Line(-40,95)(-25,95)
\Line(-40,85)(-30,85)
\Line(-40,75)(-31,75)

\Line(110,125)(119,125)
\Line(110,115)(120,115)
\Line(110,105)(125,105)
\Line(110,95)(125,95)
\Line(110,85)(120,85)
\Line(110,75)(119,75)

\Line(260,125)(269,125)
\Line(260,115)(270,115)
\Line(260,105)(275,105)
\Line(260,95)(275,95)
\Line(260,85)(270,85)
\Line(260,75)(269,75)

\SetWidth{2}
\SetColor{Black}
\Text(300,90)[lt]{$\rho$}
\Text(-60,85)[lt]{$R$}
\Text(80,160)[lb]{$\gamma$}
\Text(175,90)[lt]{$\rho_0$}
\Text(15,90)[lt]{$-\rho_0$}
\Text(310,110)[lt]{${\Huge{\bf{\cdots}}}$}
\Text(-110,110)[rt]{${\Huge{\bf{\cdots}}}$}

\SetWidth{2}
\SetColor{Green}
\Vertex(25,100){3}
\Vertex(175,100){3}

\end{picture}
\end{center}


\caption{The $\alpha=2$ crystalline model warp factors $\sigma(\rho)$ and
$\gamma(\rho)$.}
\label{crystal2}
\end{figure}

\subsection{The crystal universe model}

Another obvious generalization of the five dimensional multigravity
models is the one of the crystalline brane model. All that one has to
do is to paste an infinite array of bigravity models. We again take
this constructions in two copies. One for  $\alpha=0$ (Fig.\ref{crystal1}) and one for
$\alpha=2$ (Fig.\ref{crystal2}). A band structure is again expected as in the five
dimensional case and the width of the first band will be exponentially 
smaller than the one of the first forbidden zone. As far as the
phenomenology is concerned, these models will generate normal gravity
at intermediate scales as the first band will behave as an effective
zero mode, whereas at ultralarge scales will have the same change of
the gravitational law as the one of the corresponding quasi-localized models.

\section{The weaker energy condition}

In this chapter we have constructed, for the first time, flat brane theories
which can lead to multigravity models and their associated
modifications of gravity at large distances {\textit{without}}
introducing moving negative tension branes. The constructions are made
possible by going to a six dimensional theory. In five dimensions with 
flat branes, the presence of a bounce of the warp factor was linked
to the violation of the weaker energy condition. This is not true,
however, in six dimensions as we will show in this section. 
In five dimensions with the metric:
\begin{equation}
ds^2=e^{-A(\rho)}\eta_{\mu \nu}dx^{\mu}dx^{\nu}+d\rho^2
\end{equation}
one can readily show that the weaker energy condition requires that:
\begin{equation}
A''\geq 0
\end{equation}
which is violated at the position of the moving negative tension
branes. In the case that the branes were $AdS$ one could have a bounce (without
having negative tension branes) and still satisfy the weaker energy
condition because the above relation is modified to [19]:
\begin{equation}
A''\geq -2H^2 e^{A}
\end{equation}
However, such models do not lead to modifications of gravity at large
distances and moreover the remnant negative cosmological constant is
in conflict with current observations.

In the six dimensional case with metric:
\begin{equation}
ds^2=e^{-A(\rho)}\eta_{\mu \nu}dx^{\mu}dx^{\nu}+d\rho^2+e^{-B(\rho)}d\theta^2
\end{equation}
from the weaker energy condition one finds two inequalities which can be cast into the following relation:
\begin{equation}
{1 \over 6} (B')^2 -{1 \over 3}B''-{1 \over 6}A'B'\leq A'' \leq -{1
\over 2} (B')^2 +B''-{3 \over 2}A'B'+ 2(A')^2
\end{equation}

These inequalities hold for the $\alpha=2$ models as long as the
three-branes sitting on the conical singularities have positive
tension. It is violated, however, in the $\alpha=0$ models everywhere
in the bulk. This shows 
that in six dimensional flat brane models, the presence of bounces of the warp factor is not
necessarily linked to whether the weaker energy condition is satisfied
or not. It is not yet clear if the violation of the weaker energy
condition in the  $\alpha=0$ models is  a sign of a possible instability.

\chapter{Conclusions}

In thesis we have presented a comprehensive account of the
multigravity scenario and how it can be realized in warped brane-world 
models in five and six dimensions. It is clear that the possibility of 
having massive gravitons contributing significantly to intermediate
distance gravitational interactions is not excluded by current observations and
furthermore it offers phenomenological signatures which are in
principle testable. Moreover, taking into account that massive
gravitons necessarily arise as KK states from the dimensional
reduction of a higher dimensional theory, these modifications of
gravity can be a very unusual window to extra
dimensions. Multigravity models which involve brane configurations
with interbrane distances of the order of the Planck length, modify
gravity not only at distances of the order of the Planck length as
usually do ordinary KK theories, but also at ultralarge distances. The 
crossover scale of normal to modified gravity is related to the
interbrane distances and in that sense is generated by the geometry
itself. 

The first attempts to build such a model faced inconsistency
problems. The $''+-+''$ bigravity model, the quasi-localized GRS model and
the crystalline model, although they had promising phenomenology, they
all shared the characteristic of moving negative tension branes. As
we explicitly discussed in this thesis the moduli corresponding 
to the fluctuations of these branes are ghost fields and therefore
unacceptable. However, the general characteristics of these models
persist in more involved constructions that have less or no
problems. For example, the appearance of light states whenever we have 
more than one positive branes that localize gravitons is a generic
characteristic which persists in the $AdS$ brane models or the six
dimensional ones.

The $AdS$ models on the other hand, although they do not have the
ghost radion problem, do not give some interesting phenomenology. The
first reason is the very fact that the remnant cosmological constant
is of negative sign, in contradiction with observations. The second
reason is that they do not lead to any observable modifications of
gravity at any observable scale. The modifications of gravity at ultralarge
scales are hidden behind the $AdS$ horizon, while the modifications at 
all scales due to the different propagator structure of the massive
gravitons are enormously suppressed. It is, however, interesting how the 
$AdS$ brane models escape the van Dam-Veltman-Zakharov no-go theorem
about the non-decoupling of the additional polarization states of the
massive graviton. This in addition to the observation of
Vainshtein that the latter no-go theorem may not be valid even for flat
background, suggests that the massive graviton proposal does not face
any fundamental obstructions.

Certainly the only theoretically and
phenomenologically viable possibility of multigravity at the moment is
the one in six dimensional models. Due to the non-trivial curvature of 
the internal space one can have bounces of the warp factor with only
positive tension flat branes. There is no five dimensional effective regime in these
models, since otherwise one would face all the problems present in
five dimensions. The  low energy theory is directly four
dimensional. Moreover, the replication of the multigravity
models in two types each, one with conical singularities and one without
them, gives a wealth of possibilities that offer promising phenomenology.

There are a lot of open questions regarding multigravity at the
moment. The most important one is the construction of a cosmological
model of multigravity. So far the models we considered involved either 
flat or $AdS$ branes which are not valid descriptions of our Universe
at large scales where the modifications of gravity due to the massive
gravitons are expected to appear. A cosmological generalization of
these models would probably reveal corrections to the four dimensional 
Friedman equation at cosmological scales. This information is
essential if one wishes to put these models into observational test.

Furthermore, as far as model building is concerned, there  are a lot of 
issues to be investigated in the six dimensional models. A  careful
treatment of the angular excitations in the case of the non-singular
models  should be carried out to
examine the nature of gravity on the four-branes. Furthermore, one
should examine the moduli of the system, the  radions and the dilaton,
and calculate their mass. It is quite probable that the modulus
corresponding to $\rho_0$ is massive as in the five dimensional $AdS_4$
branes case and so the system is self-stabilized. Finally, the non-trivial
way that locality is preserved by these model should be carefully
studied.

\renewcommand{\chaptername}{Appendix} 
\appendix
\chapter{Effective action for graviton-radion-dilaton 
system}

It is convenient to define a new variable $z$ defined by:   
\begin{equation}   
\frac{1}{a(y)} = \frac{dz}{dy} 
\end{equation}   

In the coordinates $(x,z)$ the metric (\ref{pert}) is conformal to a flat    
perturbed metric $\tilde{G}_{AB}=   
\eta_{AB} + \tilde{H}_{AB}$:
\begin{equation}   
\begin{split}   
ds^2 &= a^2 \left[\tilde{G}_{\mu \nu}  \, dx^\mu dx^\nu + \tilde{G}_{zz} \, dz^2 \right] =    
a^2 \left[\eta_{MN} + \tilde{H}_{MN}  \, dx^M dx^N \right]  \\   
\tilde{H}_{\mu \nu} &= \varphi_1(x,z)  \,  \eta_{\mu \nu} + 2 \epsilon(z) \,    
\de_\mu \de_\nu f_2(x) +    
\tilde{h}_{\mu \nu}(x,z) \\   
\tilde{H}_{zz} &= \varphi_2(x,z)  \end{split}   
\label{pert1}   
\end{equation}   
where the fields $\varphi_1$ and $\varphi_2$ are:
\begin{equation}  
\begin{gathered}  
\varphi_1(x,z)  =  Q(z) f_1(x) + B(z) f_2(x)\\  
\varphi_2(x,z)  = q(z) f_1(x) + A(z) f_2(x)     
\end{gathered} 
\end{equation}

Inserting (\ref{pert1}) in (\ref{act}) and taking into account the equation   
of motion satisfied by $a$ one gets:   
\begin{equation}   
\begin{split}   
{\cal S}_{eff}& = \int d^4x \; {\cal L}_{eff} =  \int d^4x  \; 2 M^3\int dz     
\left[a^3  {\cal L}^{(0)}_{PF}(\tilde{h}) + \frac{a^3}{4}    
\left[(\de_z \tilde{h})^2 - \de_z \tilde{h}_{\mu \nu} \, \de_z \tilde{h}^{\mu \nu}\right]+ {\cal L}_{\varphi} +    
{\cal L}_{\tilde{h} \varphi}    
\right]  \\   
 {\cal L}_{\varphi} &={\cal L}_1  + {\cal L}_2 +  {\cal L}_{12} 
\end{split}   
\end{equation}   
where
\begin{equation}   
\begin{split}    
{\cal L}_{\tilde{h} \varphi} &= \left[a^3 \left(\varphi_1 + \frac{1}{2} \varphi_2    
\right)+ f_2 \frac{d}{dz}    
\left(\epsilon^\prime a^3 \right)    
\right] \partial_\mu \partial_\nu \tilde{h}^{\mu \nu} \\   
&- \left[a^3 \left(\varphi_1 +   
 \frac{1}{2}\varphi_2 \right) + f_2     
\frac{d}{dz} \left(\epsilon^\prime a^3 \right) \right] \Box \tilde{h}    
+\frac{3}{2} \frac{d}{dz} \left(a^2 a^\prime \varphi_2 - a^3    
\varphi_1^\prime \right) \tilde{h}  
\end{split}   
\end{equation}   
\begin{equation}   
{\cal L}_1 = \de_\mu f_1 \de^\mu f_1 \, \frac{3}{2} a^3 \left(Q^2 + Q q \right)   
+ f_1^2 \left(3 a^3 {Q^\prime}^2 + 3 {a^\prime}^2 a q^2 - 6 a^2 a^\prime    
Q^\prime q \right) 
\end{equation}   
\begin{equation}   
\begin{split}   
{\cal L}_2 &= \de_\mu f_2 \de^\mu f_2 \left(\frac{3}{2} a^3 B^2 + \frac{3}{2} a^3 AB   
-3 a^3 B^\prime \epsilon^\prime + 3 a^2 a^\prime \epsilon^\prime A \right)\\   
&+ f_2^2 \left(3 a^3 { B^\prime}^2 + 3 {a^\prime}^2 a A^2 - 6 a^2 a^\prime   
A B^\prime \right)   
\end{split}   
\end{equation}   
\begin{equation}   
\begin{split}   
{\cal L}_{12} &= \de_\mu f_1 \de^\mu f_2 \left(\frac{3}{2} a^3 B q +\frac{3}{2}a^3 A Q   
+3a^3B Q - 3 a^3 \epsilon^\prime Q^\prime + 3 a^2 a^\prime \epsilon^\prime q \right) \\   
&+ f_1 f_2 \left[6 a^3  B^\prime Q^\prime + 6 {a^\prime}^2 a A q - 6 a^2   
a^\prime \left( A Q^\prime + q  B^\prime \right) \right]    
\end{split}   
\end{equation}   
and ${\cal L}^{(0)}_{PF}(\tilde{h})$ is the four dimensional
Pauli-Fierz Lagrangian for $\tilde{h}$ without the mass term:   
\begin{equation}   
{\cal L}^{(0)}_{PF}(\tilde{h}) = \frac{1}{2} \de_\nu \tilde{h}_{\mu \alpha} \, \de^\alpha    
\tilde{h}^{\mu \nu} -    
\frac{1}{4} \de_\mu \tilde{h}_{\alpha \beta} \, \de^\mu \tilde{h}^{\alpha \beta}    
- \frac{1}{2} \de_\alpha \tilde{h} \, \de_\beta \tilde{h}^{\alpha \beta} +\frac{1}{4}    
\de_\alpha \tilde{h} \, \de^\alpha \tilde{h} 
\end{equation}  

All indices in the above expressions are raised and lowered by $\eta_{\mu\nu}$. Differentiation with respect of $z$ is denoted with a prime. The absence of mixing   
terms in ${\cal L}_{eff}$ yields the following constraints:   
\begin{eqnarray}   
&&A(z) = \frac{a B^\prime}{ a^\prime} \, , \qquad \frac{d}{dz}    
\left(B a^2 \right) + \frac{2 a^\prime}{a^2} \, \frac{d}{dz} \left(a^3    
\epsilon^\prime \right) = 0 \;  \label{c1}\\   
&& Q(z) = c \, a^{-2} \, , \qquad q(z) = -2 c \, a^{-2}, \quad c \text{ is a   
constant}  \label{c2}\\   
&& \int dz \, a A(z) = 0  \label{c3}   
\end{eqnarray}   

Equations (\ref{c1})-(\ref{c3}) give:   
\begin{equation}   
\begin{split}   
{\cal L}_{eff} &=  2 M^3 \int dz  \Big \{a^3  {\cal L}^{(0)}_{PF}(\tilde{h}) +    
\frac{a^3}{4} \left[(\de_z \tilde{h})^2 - \de_z\tilde{h}_{\mu \nu} \,
\de_z\tilde{h}^{\mu \nu}    
\right] \\   
&- \frac{3}{2}c^2  a^{-1} \de_\mu f_1 \de^\mu f_1 + \frac{3}{4}    
\frac{a^2}{a^\prime} \, \frac{d}{dz} \left(B^2 a^2 \right)  \, \de_\mu f_2   
\de^\mu f_2 \Big \}
\end{split}   
\label{eff}   
\end{equation}   

In particular the effective Lagrangian ${\cal L}_{Scal}$ for the dilaton    
$f_1$ and the radion   
$f_2$  is:   
\begin{equation}   
\begin{split}   
{\cal L}_{Scal} &= {\cal K}_1 \, f_1 \Box f_1 +  {\cal K}_2 \, f_2 \Box f_2 \\   
 {\cal K}_1 &=  2 M^3 \, \frac{3}{2}c^2 \int_{- L}^{L} a^{-2}    
\, dy \\   
 {\cal K}_2 &= - 2 M^3 \, \frac{3}{4} \int_{- L}^{L} a    
\left(\frac{da}{dy} \right)^{-1} \, \frac{d}{dy} \left( B^2 a^2 \right) \, dy      
\end{split}   
\end{equation}   
with:   
\begin{equation}   
\begin{split}   
&\frac{d}{dy} \left(B a^2 \right) + 2 a^{-1} \, \frac{da}{dy} \, \frac{d}{dy}    
\left(a^4 \de_y \epsilon \right) = 0  \\   
& \int_{- L}^{L} dy \; a \left(\frac{da}{dy} \right)^{-1} \,    
\frac{dB}{dy} = 0   
\end{split}   
\end{equation}   

As a result only $\de_y\epsilon(0)$, $\de_y\epsilon(r)$ and 
$\de_y\epsilon(L)$ enter the radion effective action.

\chapter{Ricci tensor and scalar for the dilaton perturbation}

In this Appendix we list the Ricci tensor components, the Ricci
scalar and the action obtained by the metric:
\begin{equation}
ds^2=e^{-a(y)^{-2}\gamma(x)}a(y)^2g_{\mu\nu}(x)dx^{\mu}dx^{\nu}+\left(1+a(y)^{-2}\gamma(x)\right)^2dy^2
\end{equation}

The spacetime components of the five dimensional Ricci tensor are:
\begin{eqnarray}
R_{\mu\nu}~=~R_{g~\mu\nu}&+&{g_{\mu\nu} \over 2}a^{-2}\bo\gamma -{g_{\mu\nu}
\over 2}{a^{-6}\gamma \over
(1+a^{-2}\gamma)}\gamma_{,\kappa}\gamma^{,\kappa}-{1 \over 2}\left({1-a^{-2}\gamma \over
1+a^{-2}\gamma}\right)a^{-4}\gamma_{,\mu}\gamma_{,\nu}\nonumber\\&+&{a^{-4}\gamma \over
(1+a^{-2}\gamma)}D_{\mu}\partial_{\nu}\gamma- g_{\mu\nu}
e^{-a^{-2}\gamma}(a')^2 \left({3+4a^{-2}\gamma \over
1+a^{-2}\gamma}\right)\nonumber\\&-&g_{\mu\nu}e^{-a^{-2}\gamma}{ aa'' \over (1+a^{-2}\gamma)}
\end{eqnarray}
and the (55) component:
\begin{eqnarray}
R_{55}&=&e^{a^{-2}\gamma}(1+a^{-2}\gamma)a^{-6}\gamma_{,\mu}\gamma^{,\mu}-e^{a^{-2}\gamma}(1+a^{-2}\gamma)a^{-4}\bo\gamma\nonumber\\&\phantom{=}&-4(1+a^{-2}\gamma)a^{-1}a''-4(1+a^{-2}\gamma)a^{-4}(a')^2\gamma
\end{eqnarray}

Finally the Ricci scalar is:
\begin{eqnarray}
R&=&e^{a^{-2}\gamma}a^{-2}R_g+e^{a^{-2}\gamma}\left({1+3a^{-2}\gamma \over
1+a^{-2}\gamma}\right)a^{-4}\bo\gamma+{1 \over 2}e^{a^{-2}\gamma}\left({1-3a^{-2}\gamma \over
1+a{-2}\gamma}\right)a^{-6}\gamma_{,\mu}\gamma^{,\mu}\nonumber\\&\phantom{=}&-8{
a^{-1}a'' \over (1+a^{-2}\gamma)}-{12a^{-2}(a')^2+20a^{-4}(a')^2\gamma
\over(1+a^{-2}\gamma) }
\end{eqnarray}

In the above expressions the indices are raised and lowered by
$g_{\mu\nu}$.

The action (\ref{action}) then becomes:
\begin{eqnarray}
{\cal S}&=&\int d^4xdy \sqrt{-g}
\left\{2M^3\left[e^{-a^{-2}\gamma}a^2(1+a^{-2}\gamma)R_g+e^{-a^{-2}\gamma} 
\left(1+{5\over 2}a^{-2}\gamma +{3 \over
2}a^{-4}\gamma^2\right)g^{\mu\nu}D_{\mu}\partial_{\nu} \gamma
\right.\phantom{\sum_{i}}
\right.\nonumber\\&\phantom{=}&~~~~~~~~~~~~~~~~~~~~~~~~~~~~~~~~~~~~~~~~~~~~~~~~~~~~\phantom{5 
\over 2}\left.+e^{-2a^{-2}\gamma}(-8a''a^3-12(a')^2a^2-20(a')^2\gamma)\right]\nonumber\\&\phantom{=}&~~~~~~~~~~~~-{1 
\over 2}{a^4 \over
(1+a^{-2}\gamma)}e^{-2a^{-2}\gamma}\Phi'\cdot\Phi'-e^{-2a^{-2}\gamma}a^4(1+a^{-2}\gamma)V(\Phi)\nonumber\\&\phantom{=}&~~~~~~~~~~~~\left.-e^{-2a^{-2}\gamma}a^4\sum_{i}\lambda_{i}(\Phi)\delta_(y-y_i)\right\}
\end{eqnarray}

\chapter{Equations of motion in six dimensions}

In this Appendix we provide the Einstein equations for the class of
models that we considered in chapter 5 where the metric is:
\begin{equation}
ds^2=\sigma(\rho)\eta_{\mu \nu}dx^{\mu}dx^{\nu}+d\rho^2+\gamma(\rho)d\theta^2
\end{equation}

Ignoring the three-branes located on the
conical singularities, the Einstein equations are generally written as:
\begin{equation}
G_{MN}={1 \over 4M^4}(T^{(B)}_{MN}+T^{(br)}_{MN})
\end{equation}
where $T^{(B)}_{MN}$ is the bulk energy momentum tensor of the general 
form (\ref{bulkT}) and
$T^{(br)}_{MN}$ the four-brane energy momentum tensor of the form (\ref{braneT}).

In the absence of a four dimensional cosmological constant, the $(\theta,\theta)$ component of the above equation is:
\begin{equation}
2{\sigma'' \over \sigma}+{1 \over 2}\left({\sigma' \over 
\sigma}\right)^2=-{\Lambda_{\theta} \over 4M^4}-{V^i_{\theta} \over
4M^4}\delta(\rho-\rho_i)
\label{tt}
\end{equation}
The $(\rho,\rho)$ component is:
\begin{equation}
{3 \over 2}\left({\sigma' \over \sigma}\right)^2+{\sigma'\gamma' \over 
\sigma \gamma}=-{\Lambda_{\rho} \over 4M^4}
\label{rr}
\end{equation}
Finally, the $(\mu,\nu)$ component is:
\begin{equation}
{3 \over 2}{\sigma'' \over \sigma}+{3 \over 4}{\sigma'\gamma' \over
\sigma \gamma}-{1 \over 4}\left({\gamma' \over \gamma}\right)^2+{1
\over 2}{\gamma'' \over \gamma}=-{\Lambda_0 \over 4M^4}-{V^i_0 \over 4M^4}\delta(\rho-\rho_i)
\label{mne}
\end{equation}

These equations may be compared with the ones of the five
dimensional case where the metric:
\begin{equation}
ds^2=\sigma(\rho)\eta_{\mu \nu}dx^{\mu}dx^{\nu}+d\rho^2
\end{equation}
gives rise to the $(\rho,\rho)$ component:
\begin{equation}
{3 \over 2}\left({\sigma' \over \sigma}\right)^2=-{\Lambda \over 4M^3}
\label{rr5}
\end{equation}
and the $(\mu,\nu)$ component:
\begin{equation}
{3 \over 2}{\sigma'' \over \sigma}=-{\Lambda \over 4M^3}-{V^i \over 4M^3}\delta(\rho-\rho_i)
\label{mn5}
\end{equation}

The extra freedom that we have in the six dimensional case is apparent.

\addcontentsline{toc}{chapter}{\protect\numberline{Bibliography\hspace{-96pt}}}

\end{document}